\documentclass[a4paper]{article}
\pdfoutput=1

\usepackage{macros}
\usepackage{booktabs}
\usepackage{multirow,makecell}
\usepackage{import}
\usepackage{tikz}
\usepackage{graphicx}
\usepackage{standalone}
\usepackage{asymptote}

\preprint{ }
\title{Shell formulas for instantons and gauge origami}

\author{Jiaqun Jiang}

\affiliation{Department of Physics and Center for Field Theory \& Particle Physics, Fudan University, 20005, Songhu Road, 200438 Shanghai, China}
\emailAdd{jiangjiaqun@gmail.com}
\abstract{We introduce the shell formula—a framework that unifies the description of partition functions whose pole structures are classified by Young diagrams of arbitrary dimension. The formalism yields explicit closed-form expressions and recursion relations for a wide range of physical systems, including instanton partition functions of 5d pure super Yang–Mills theory with classical gauge groups, as well as gauge origami configurations such as the magnificent four, tetrahedron instantons, spiked instantons, and Donaldson–Thomas invariants in $\C^3$ and $\C^4$.}

\begin{document}
\allowdisplaybreaks

\maketitle
\section{Introduction}

The study of supersymmetric gauge theories has revealed deep and unexpected
interrelations among quantum field theory, algebraic geometry, and
combinatorics. One of the most striking manifestations of this interplay is the
appearance of integer partitions ($2$d Young diagrams)---or more generally,
plane and solid partitions ($3$d and $4$d Young diagrams)---in exact partition
functions of supersymmetric systems. These combinatorial objects arise naturally
in supersymmetric localization computations, instanton counting
\cite{Nekrasov:2002qd,Nekrasov:2003rj,Nekrasov:2016qym,Nekrasov:2017cih,Pomoni:2021hkn},
and topological string amplitude calculations
\cite{Aganagic:2003db,Katz:1996fh,Leung:1997tw,Hayashi:2020hhb,Nawata:2021dlk},
where they encode the BPS spectra of brane configurations.

The story began with the application of supersymmetric localization to compute
the instanton partition function of $\U(N)$ supersymmetric Yang-Mills (SYM)
theory with 8 supercharges \cite{Nekrasov:2002qd,Nekrasov:2003rj,Pestun:2016zxk},
in which each box of a $2$d Young diagram corresponds to a fixed point of the
torus action on the instanton moduli space. Beyond reproducing the
Seiberg-Witten prepotential \cite{Seiberg:1994rs}, Nekrasov's program connected
instanton counting with a broad web of frameworks, including the BPS/CFT
correspondence
\cite{Gaiotto:2009we,Alday:2009aq,Nekrasov:2015wsu,Nekrasov:2016qym,Nekrasov:2016ydq,Nekrasov:2017rqy,Nekrasov:2017gzb},
topological vertex computations \cite{Aganagic:2003db,Iqbal:2007ii,Awata:2008ed},
and quantum algebras
\cite{Nekrasov:2015wsu,Kimura:2015rgi,Nawata:2023wnk,Maulik:2012wi,Awata:2011ce}.

Subsequent developments in gauge origami and topological vertex techniques
revealed that Young diagrams with asymptotic boundaries are intimately connected
to a wide class of physical systems. Intersecting D4-brane configurations give
rise to spiked instantons
\cite{Nekrasov:2016qym,Nekrasov:2016gud,Rapcak:2018nsl}, while D6-brane
configurations produce tetrahedron instantons
\cite{Pomoni:2021hkn,Pomoni:2023nlf,Fasola:2023ypx}; these correspond to 2d
and 3d Young diagrams growing along different directions, respectively. The
magnificent four
\cite{Nekrasov:2017cih,Nekrasov:2018xsb,Billo:2021xzh,Kool:2025qou} provides a
physical realization of D0-D8 systems, with 4d Young diagrams labeling bound
states. In the Donaldson-Thomas (DT) framework for Calabi-Yau (CY) threefolds
\cite{Maulik:2003rzb,Maulik:2004txy,Okounkov:2003sp}, 3d Young diagrams with
prescribed 2d asymptotic boundaries describe D0-D2-D6 bound states on $\C^3$.
This perspective lifts naturally to the DT4 setting on $\C^4$
\cite{Monavari_2022,Kimura:2025lig,Nekrasov:2023nai,Piazzalunga:2023qik},
where 4d Young diagrams carry asymptotic data of two distinct types: leg-type
asymptotics, arising from D2-branes ending on D8-branes, and surface-type
asymptotics, arising from D4-branes ending on D8-branes. This construction
furnishes a unifying geometric framework that lifts lower-dimensional
combinatorial structures into a single parent theory on $\C^4$.

In this paper, we introduce a universal formula---referred to as the \emph{shell
formula}---that provides a compact and systematic representation of the Witten
index for all classes of systems described above. The shell formula is built
from two geometric ingredients attached to a Young diagram of arbitrary
dimension $d$: its \emph{shell} (the set of boxes on the outer boundary) and a
\emph{charge} assigned to each shell box. The central object is the
$\mathcal{J}$-factor, defined as a product of $\mathrm{sh}$ functions over the
shell boxes, each raised to its charge. This construction has three concrete
advantages. 

First, the
Nekrasov factor is expressed in terms of arm and leg lengths, which are
intrinsic to 2d Young diagrams and do not extend naturally to $d \geq 3$,
whereas the $\mathcal{J}$-factor is defined uniformly for any $d$ via the shell
and charge data, making closed-form expressions for the tetrahedron instanton
(3d), magnificent four (4d), DT3, and DT4 partition functions directly
accessible. 

Second, the recursion relation~\eqref{recursion} expresses the ratio of partition function
  contributions between a Young diagram $Y_A$ and its one-box extension
  $Y_A \cup \{\boldsymbol{w}\}$ as a local product involving only the new box.
  It holds uniformly for $d = 2, 3, 4$ and is used throughout
  Secs.~\ref{5d pure SYM}--\ref{gauge origami} to derive the DT3 and DT4
  integrands.

Third, for Sp$(2N)$ SYM the partition function receives Young-diagram-dependent
BPS jumping coefficients $C^{\mathrm{Sp}}_{\vec\lambda,v}$ that must otherwise
be inserted by hand in each term of the sum \cite{Kim:2024vci,Nawata:2021dlk};
the shell formula in the unrefined limit $\epsilon_2 \to -\epsilon_1$
automatically absorbs these coefficients into the limiting procedure, as
demonstrated explicitly in Appendix~\ref{Sp2 example}.

The paper is organized as follows. In Sec.~\ref{YD and Shell}, we review the
definition of Young diagrams, introduce the shell and the charge of each shell
box, and define the $\mathcal{J}$-factor for Young diagrams of arbitrary
dimension. We also establish key algebraic properties, including expansion,
translation invariance, swapping, recursion, and splitting. In
Sec.~\ref{5d pure SYM}, we review the instanton moduli space and partition
functions of $5$d $\cN=1$ pure SYM theories with classical gauge groups, and
recast them in terms of the shell formula. In Sec.~\ref{gauge origami}, we
apply the shell formula to express the partition functions of the magnificent
four, tetrahedron instantons, spiked instantons, and DT3 and DT4 theories.
Appendix~\ref{charges of shellbox} collects the notation used throughout the
paper and provides illustrative examples of $\mathcal{J}$-factor computations.
Appendix~\ref{witten inedx and JK} gives a concise review of the Witten index
and the Jeffrey-Kirwan (JK) residue method. Appendix~\ref{examples} presents
detailed computational examples for several theories.

\section{Young diagrams and shell formulas}\label{YD and Shell}

In this section, we introduce Young diagrams in arbitrary dimensions, define the shell formula and its central ingredient—the $\mathcal{J}$-factor—and establish several algebraic properties that will be used throughout the paper.

\subsection{Young diagrams and poles}

A Young diagram is a combinatorial object defined by a simple monotonicity rule. In two dimensions, the distinct ways to write a positive integer as an ordered sum of non-increasing positive integers correspond bijectively to 2d Young diagrams, also known as integer partitions. This concept generalizes directly to higher dimensions: 3d Young diagrams (plane partitions) and 4d Young diagrams (solid partitions), and so on. The general definition is as follows.

\begin{definition}[Young diagram]\label{young diagram def}
    A $d$-dimensional \emph{Young diagram} is a finite subset $\mathbf{Y} \subseteq \mathbb{Z}^d$ for which there exists a point $\boldsymbol{z} = (z_1,\dots,z_d) \in \mathbb{Z}^d$, called the \emph{origin} of $\mathbf{Y}$, such that the following monotonicity condition holds:
    \begin{align*}
        &\text{if }\, \boldsymbol{x} = (x_1,\dots,x_d) \in \mathbf{Y} \,\text{ and }\, z_i \leq y_i \leq x_i \text{ for all } i = 1,\dots,d,\cr
        &\text{then }\, \boldsymbol{y} = (y_1,\dots,y_d) \in \mathbf{Y}.
    \end{align*}
    The empty set $\emptyset$ is also regarded as a Young diagram (with no origin specified).
\end{definition}

Throughout this paper, the origin of every Young diagram is taken to be $(1,1,1,\ldots)$ unless stated otherwise.

Young diagrams provide a natural language for the partition functions discussed in Sec.~\ref{5d pure SYM} and Sec.~\ref{gauge origami}. The key point is that the poles of an instanton partition function are in one-to-one correspondence with the box coordinates of Young diagrams. More precisely, given an integrand $\mathcal{I}(\boldsymbol{\phi})$ depending on integration variables $\boldsymbol{\phi}=(\phi_1,\ldots,\phi_k)$, the JK-residue (reviewed in Appendix~\ref{JK}) evaluates the integral as a sum over poles:
\begin{align}
    \oint_{\text{JK}}\prod^k_{I=1}\frac{d\phi_I}{2\pi i}\,\mathcal{I}(\boldsymbol{\phi})=\sum_{\boldsymbol{\phi}_*}\underset{\boldsymbol{\phi}=\boldsymbol{\phi}_*}{\operatorname{JK-Res}}(\eta)\,\mathcal{I}(\boldsymbol{\phi}).
\end{align}
In all cases considered here, the JK-residue selects poles indexed by a list of Young diagrams $\boldsymbol{\mathbf{Y}}=(\mathbf{Y}_{\mathcal{A}},\mathbf{Y}_{\mathcal{B}},\dots)$. Each Young diagram $\mathbf{Y}_{\mathcal{A}}$ carries a label $\mathcal{A}\equiv(A,\alpha)=(a_1a_2\ldots a_d,\alpha)$, specifying its basis directions $\boldsymbol{\e}_A=(\e_{a_1},\ldots,\e_{a_d})$ and a color $\alpha$. The pole corresponding to this Young diagram list is:
\begin{align}
    (\phi_{1*},\ldots,\phi_{k*})=(\mathcal{X}_\mathcal{A}(\boldsymbol{x}_{\mathcal{A},1}),\ldots,\mathcal{X}_\mathcal{A}(\boldsymbol{x}_{\mathcal{A},n_{\mathcal{A}}}),\,\mathcal{X}_\mathcal{B}(\boldsymbol{x}_{\mathcal{B},1}),\ldots,\mathcal{X}_\mathcal{B}(\boldsymbol{x}_{\mathcal{B},n_{\mathcal{B}}}),\ldots),
\end{align}
where $n_\mathcal{A}=|\mathbf{Y}_{\mathcal{A}}|$ is the number of boxes in $\mathbf{Y}_{\mathcal{A}}$, and the total box count satisfies $\sum_{\mathcal{A}}n_{\mathcal{A}}=k$. The coordinate function $\mathcal{X}_\mathcal{A}$ is defined in terms of the Coulomb branch parameter $v_\mathcal{A}$ by:
\begin{align}
    \mathcal{X}_{\mathcal{A}}(\boldsymbol{x})\equiv v_\mathcal{A}+(\boldsymbol{x}-\boldsymbol{1})\cdot\boldsymbol{\e}_{A}=v_{\mathcal{A}}+\sum_{i=1}^d(x_i-1)\e_{a_i}.
\end{align}

A crucial observation is that all partition functions considered in this paper share a common structure at each pole:
\begin{align}\label{index J expand}
    \underset{\boldsymbol{\phi}=\boldsymbol{\phi}_*}{\operatorname{JK-Res}}(\eta)\,\mathcal{I}(\boldsymbol{\phi})\supset\frac{1}{\sh(\phi_{i*}-\mathcal{X}_\mathcal{A}(\boldsymbol{1}))}\prod_{\boldsymbol{y}\in\mathbf{Y}_{\mathcal{A}}}\frac{\prod_{\boldsymbol{b}\in\mathbf{B}_{\text{even}}}\sh(\phi_{i*}-\mathcal{X}_{\mathcal{A}}(\boldsymbol{y})-\boldsymbol{b}\cdot\boldsymbol{\e}_A)}{\prod_{\boldsymbol{b}\in\mathbf{B}_{\text{odd}}}\sh(\phi_{i*}-\mathcal{X}_{\mathcal{A}}(\boldsymbol{y})-\boldsymbol{b}\cdot\boldsymbol{\e}_A)},
\end{align}
where $\sh(x)=e^{x/2}-e^{-x/2}$. Here $\mathbf{B}_d=\{\boldsymbol{b}=(b_1,\ldots,b_d)\mid b_i\in\{0,1\}\}$ is the set of all $d$-tuples of binary digits, while $\mathbf{B}_{\text{even}}\subset\mathbf{B}_d$ consists of those tuples with $|\boldsymbol{b}|=\sum_i b_i\in 2\mathbb{Z}$, and $\mathbf{B}_{\text{odd}}$ consists of those with $|\boldsymbol{b}|\in 2\mathbb{Z}+1$. It is precisely this universal structure that motivates the definition of the shell formula.

\subsection{Shell formulas and \texorpdfstring{$\mathcal{J}$}{J}-Factor}

The shell formula is built from two geometric ingredients attached to a Young diagram: its \emph{shell} (the set of boxes on its outer boundary) and a \emph{charge} assigned to each shell box. We define these in turn.

\begin{definition}[Shell of a Young diagram]
    Given a $d$-dimensional Young diagram $\mathbf{Y}$ and the set of binary tuples $\mathbf{B}_d$, the \emph{shell} $\mathcal{S}(\mathbf{Y})$ of $\mathbf{Y}$ is:
    \begin{align}\label{shell}
        \mathcal{S}(\mathbf{Y})\equiv (\mathbf{Y}+\mathbf{B}_d)\backslash\mathbf{Y}.
    \end{align}
    In words, $\mathcal{S}(\mathbf{Y})$ consists of all boxes that are not in $\mathbf{Y}$ but can be reached from some box of $\mathbf{Y}$ by a unit step in any combination of coordinate directions.
\end{definition}

\begin{definition}[Charge of a shell box]
    For each box $\boldsymbol{x}=(x_1,\ldots,x_d)$ in the shell $\mathcal{S}(\mathbf{Y})$, its \emph{charge} is defined by the inclusion-exclusion sum:
    \begin{align}\label{charge}
        \operatorname{Q}_{\mathbf{Y}}(\boldsymbol{x})=\sum_{\substack{\boldsymbol{b}\in\mathbf{B}_d\\\boldsymbol{x}-\boldsymbol{b}\in\mathbf{Y}}}(-1)^{|\boldsymbol{b}|},
    \end{align}
    where $|\boldsymbol{b}|\equiv\sum_{i=1}^d b_i$ counts the number of $1$'s in $\boldsymbol{b}$. Intuitively, the charge measures how many corners of the unit hypercube centered at $\boldsymbol{x}$ already belong to $\mathbf{Y}$, weighted by sign.
\end{definition}

Explicit examples of shells and charges are collected in Appendix~\ref{shell and J}. For a generic $d$-dimensional Young diagram, shell box charges take integer values between $-d$ and $d$.

With these ingredients, the central object of the shell formula can now be defined.

\begin{definition}[$\mathcal{J}$-factor]
    Given a $d$-dimensional Young diagram $\mathbf{Y}_{\mathcal{A}}$ with label $\mathcal{A}=(A,\alpha)=(a_1a_2\ldots a_d,\alpha)$, the \emph{$\mathcal{J}$-factor} is the product over shell boxes:
    \begin{align}\label{J-def}
        \mathcal{J}\big(x\big|\mathbf{Y}_{\mathcal{A}}\big)&\equiv \prod_{\boldsymbol{y}\in\mathcal{S}(\mathbf{Y}_{\mathcal{A}})}\sh\left(x-\mathcal{X}_{\mathcal{A}}(\boldsymbol{y})\right)^{\operatorname{Q}_{\mathbf{Y}_{\mathcal{A}}}(\boldsymbol{y})},\cr
        \mathcal{J}\big(x\big|\emptyset_{\mathcal{A}}\big)&\equiv \frac{1}{\sh(x-\mathcal{X}_\mathcal{A}(\boldsymbol{1}))}.
    \end{align}
    That is, each shell box $\boldsymbol{y}$ contributes a factor of $\sh(x - \mathcal{X}_\mathcal{A}(\boldsymbol{y}))$ raised to its charge.
\end{definition}

Detailed computational examples of the $\mathcal{J}$-factor are provided in Appendix~\ref{shell and J}.

\medskip

We now establish four algebraic properties of the $\mathcal{J}$-factor that will be used in subsequent sections.  This discussion is self-contained and may be skipped on a first reading.

\paragraph{Expansion.} The $\mathcal{J}$-factor admits a box-by-box expansion:
\begin{align}\label{J-expand}
    \mathcal{J}\big(x\big|\mathbf{Y}_{\mathcal{A}}\big)=\frac{1}{\sh(x-\mathcal{X}_\mathcal{A}(\boldsymbol{1}))}\prod_{\boldsymbol{y}\in\mathbf{Y}_{\mathcal{A}}}\frac{\prod_{\boldsymbol{b}\in\mathbf{B}_{\text{even}}}\sh(x-\mathcal{X}_{\mathcal{A}}(\boldsymbol{y})-\boldsymbol{b}\cdot\boldsymbol{\e}_A)}{\prod_{\boldsymbol{b}\in\mathbf{B}_{\text{odd}}}\sh(x-\mathcal{X}_{\mathcal{A}}(\boldsymbol{y})-\boldsymbol{b}\cdot\boldsymbol{\e}_A)}.
\end{align}
This expansion simplifies dramatically after cancellations: for any box $\boldsymbol{y}$ in the interior of $\mathbf{Y}$, the boxes $\boldsymbol{y}-\boldsymbol{b}$ (for all $\boldsymbol{b}\in\mathbf{B}_d$) all lie in $\mathbf{Y}$ by the Young diagram monotonicity condition, and their contributions cancel pairwise. More precisely:
\begin{align}
    \frac{\prod_{\boldsymbol{b}\in\mathbf{B}_{\text{even}}}\sh(x-\mathcal{X}_{\mathcal{A}}(\boldsymbol{y}-\boldsymbol{b})-\boldsymbol{b}\cdot\boldsymbol{\e}_A)}{\prod_{\boldsymbol{b}\in\mathbf{B}_{\text{odd}}}\sh(x-\mathcal{X}_{\mathcal{A}}(\boldsymbol{y}-\boldsymbol{b})-\boldsymbol{b}\cdot\boldsymbol{\e}_A)}=\frac{\prod_{\boldsymbol{b}\in\mathbf{B}_{\text{even}}}\sh(x-\mathcal{X}_{\mathcal{A}}(\boldsymbol{y}))}{\prod_{\boldsymbol{b}\in\mathbf{B}_{\text{odd}}}\sh(x-\mathcal{X}_{\mathcal{A}}(\boldsymbol{y}))}=1.
\end{align}
For boxes on the boundary of $\mathbf{Y}$, only half of the shifted boxes $\boldsymbol{y}-\boldsymbol{b}$ are present, but they still cancel mutually. After all cancellations, only the shell boxes survive, leaving exactly the definition~\eqref{J-def}. This also confirms that the expansion~\eqref{J-expand} coincides with the universal pole structure~\eqref{index J expand}, providing the precise justification for the definition of the $\mathcal{J}$-factor.

As a concrete illustration, for the 2d Young diagram $\lambda_{12,\alpha}$ with label $(12,\alpha)$:
\begin{align}
    \mathcal{J}\big(x\big|\lambda_{12,\alpha}\big)
    =&\frac{1}{\sh(x-\mathcal{X}_{12,\alpha}(\boldsymbol{1}))}\prod_{\boldsymbol{y}\in\lambda_{12,\alpha}}\frac{\sh(x-\mathcal{X}_{12,\alpha}(\boldsymbol{y})-(0,0)\cdot\boldsymbol{\e}_{12})\,\sh(x-\mathcal{X}_{12,\alpha}(\boldsymbol{y})-(1,1)\cdot\boldsymbol{\e}_{12})}{\sh(x-\mathcal{X}_{12,\alpha}(\boldsymbol{y})-(0,1)\cdot\boldsymbol{\e}_{12})\,\sh(x-\mathcal{X}_{12,\alpha}(\boldsymbol{y})-(1,0)\cdot\boldsymbol{\e}_{12})}\cr
    =&\frac{1}{\sh(x-v_{12,\alpha})}\prod_{\boldsymbol{y}\in\lambda_{12,\alpha}}\frac{\sh(x-\mathcal{X}_{12,\alpha}(\boldsymbol{y}))\,\sh(x-\mathcal{X}_{12,\alpha}(\boldsymbol{y})-\e_{12})}{\sh(x-\mathcal{X}_{12,\alpha}(\boldsymbol{y})-\e_1)\,\sh(x-\mathcal{X}_{12,\alpha}(\boldsymbol{y})-\e_2)}.
\end{align}

\paragraph{Translation invariance.} The $\mathcal{J}$-factor depends only on coordinate differences, so shifting both its argument and the Young diagram by the same vector $\boldsymbol{y}\in\mathbb{Z}^d$ leaves it unchanged:
\begin{align}\label{J-translation}
    \mathcal{J}\big(\mathcal{X}_{A,\alpha}(\boldsymbol{x})\big|\mathbf{Y}_{A,\beta}\big)=\mathcal{J}\big(\mathcal{X}_{A,\alpha}(\boldsymbol{x}+\boldsymbol{y})\big|\mathbf{Y}_{A,\beta}+\boldsymbol{y}\big),
\end{align}
where $\mathbf{Y}_{A,\beta}+\boldsymbol{y}$ denotes the Young diagram obtained by translating every box of $\mathbf{Y}_{A,\beta}$ by $\boldsymbol{y}$.

\paragraph{Swapping property.} For two $d$-dimensional Young diagrams $\mathbf{Y}_{A,\alpha}$ and $\mathbf{Y}_{A,\beta}$ sharing the same basis $A$, the following identity exchanges their roles:
\begin{align}\label{swapping prop}
    &\prod_{\boldsymbol{x}\in\mathbf{Y}_{A,\alpha}}\sh(\mathcal{X}_{A,\alpha}(\boldsymbol{x})-\mathcal{X}_{A,\beta}(\boldsymbol{1}))\,\mathcal{J}\big(\mathcal{X}_{A,\alpha}(\boldsymbol{x})\big|\mathbf{Y}_{A,\beta}\big)\cr
    =&\prod_{\boldsymbol{y}\in\mathbf{Y}_{A,\beta}+\boldsymbol{1}}\left(\sh(\mathcal{X}_{A,\beta}(\boldsymbol{y})-\mathcal{X}_{A,\alpha}(\boldsymbol{1}))\,\mathcal{J}\big(\mathcal{X}_{A,\beta}(\boldsymbol{y})\big|\mathbf{Y}_{A,\alpha}\big)\right)^{(-1)^d}.
\end{align}
The sign $(-1)^d$ reflects the parity of the dimension. In particular, for $d=2$ the right-hand side has the same sign as the left, while for $d=3$ it is inverted. Concretely, for two 2d Young diagrams $\lambda_{12,1}$ and $\lambda_{12,2}$:
\begin{align}
    &\prod_{\boldsymbol{x}\in\lambda_{12,1}}\sh(\mathcal{X}_{12,1}(\boldsymbol{x})-\mathcal{X}_{12,2}(\boldsymbol{1}))\,\mathcal{J}\big(\mathcal{X}_{12,1}(\boldsymbol{x})\big|\lambda_{12,2}\big)\cr
    =&\prod_{\boldsymbol{y}\in\lambda_{12,2}+\boldsymbol{1}}\sh(\mathcal{X}_{12,2}(\boldsymbol{y})-\mathcal{X}_{12,1}(\boldsymbol{1}))\,\mathcal{J}\big(\mathcal{X}_{12,2}(\boldsymbol{y})\big|\lambda_{12,1}\big),
\end{align}
while for two 3d Young diagrams $\pi_{123,1}$ and $\pi_{123,2}$ the inversion gives:
\begin{align}
    &\prod_{\boldsymbol{x}\in\pi_{123,1}}\sh(\mathcal{X}_{123,1}(\boldsymbol{x})-\mathcal{X}_{123,2}(\boldsymbol{1}))\,\mathcal{J}\big(\mathcal{X}_{123,1}(\boldsymbol{x})\big|\pi_{123,2}\big)\cr
    =&\prod_{\boldsymbol{y}\in\pi_{123,2}+\boldsymbol{1}}\frac{1}{\sh(\mathcal{X}_{123,2}(\boldsymbol{y})-\mathcal{X}_{123,1}(\boldsymbol{1}))\,\mathcal{J}\big(\mathcal{X}_{123,2}(\boldsymbol{y})\big|\pi_{123,1}\big)}.
\end{align}

\paragraph{Recursion relation.} Adding a single box to a Young diagram changes the $\mathcal{J}$-factor in a controlled way. Specifically, let $\mathbf{Y}'_{\mathcal{A}}=\mathbf{Y}_{\mathcal{A}}\cup\{\boldsymbol{w}\}$ be obtained by adding one box $\boldsymbol{w}$ to $\mathbf{Y}_{\mathcal{A}}$. Then:
\begin{multline}
\label{recursion}
    \frac{\prod_{\boldsymbol{x}\in\mathbf{Y}_{\mathcal{A}}\cup\{\boldsymbol{w}\}}\mathcal{J}\big(\mathcal{X}_{\mathcal{A}}(\boldsymbol{x})\big|\mathbf{Y}_{\mathcal{A}}\cup\{\boldsymbol{w}\}\big)}{\prod_{\boldsymbol{x}\in\mathbf{Y}_{\mathcal{A}}}\mathcal{J}\big(\mathcal{X}_{\mathcal{A}}(\boldsymbol{x})\big|\mathbf{Y}_{\mathcal{A}}\big)}
    \\=\mathcal{J}\big(\mathcal{X}_{\mathcal{A}}(\boldsymbol{w})\big|\mathbf{Y}_{\mathcal{A}}\cup\{\boldsymbol{w}\}\big)\left(\sh(\mathcal{X}_{\mathcal{A}}(\boldsymbol{w})-\mathcal{X}_{\mathcal{A}}(\boldsymbol{0}))\,\mathcal{J}\big(\mathcal{X}_{\mathcal{A}}(\boldsymbol{w}+\boldsymbol{1})\big|\mathbf{Y}_{\mathcal{A}}\big)\right)^{(-1)^d},
\end{multline}
where the derivation uses the swapping property~\eqref{swapping prop}. The key point is that the entire ratio reduces to a local contribution at the new box $\boldsymbol{w}$ alone: one factor from the $\mathcal{J}$-factor of the enlarged diagram evaluated at $\boldsymbol{w}$, and one from the original diagram evaluated at the shifted point $\boldsymbol{w}+\boldsymbol{1}$.

The intermediate steps of the derivation are:
\begin{align}
    &\frac{\prod_{\boldsymbol{x}\in\mathbf{Y}_{\mathcal{A}}\cup\{\boldsymbol{w}\}}\mathcal{J}\big(\mathcal{X}_{\mathcal{A}}(\boldsymbol{x})\big|\mathbf{Y}_{\mathcal{A}}\cup\{\boldsymbol{w}\}\big)}{\prod_{\boldsymbol{x}\in\mathbf{Y}_{\mathcal{A}}}\mathcal{J}\big(\mathcal{X}_{\mathcal{A}}(\boldsymbol{x})\big|\mathbf{Y}_{\mathcal{A}}\big)}\cr
    =\,&\mathcal{J}\big(\mathcal{X}_{\mathcal{A}}(\boldsymbol{w})\big|\mathbf{Y}_{\mathcal{A}}\cup\{\boldsymbol{w}\}\big)\frac{\prod_{\boldsymbol{x}\in\mathbf{Y}_{\mathcal{A}}}\mathcal{J}\big(\mathcal{X}_{\mathcal{A}}(\boldsymbol{x})\big|\mathbf{Y}_{\mathcal{A}}\cup\{\boldsymbol{w}\}\big)}{\prod_{\boldsymbol{x}\in\mathbf{Y}_{\mathcal{A}}}\mathcal{J}\big(\mathcal{X}_{\mathcal{A}}(\boldsymbol{x})\big|\mathbf{Y}_{\mathcal{A}}\big)}\cr
    =\,&\mathcal{J}\big(\mathcal{X}_{\mathcal{A}}(\boldsymbol{w})\big|\mathbf{Y}_{\mathcal{A}}\cup\{\boldsymbol{w}\}\big)\frac{\prod_{\boldsymbol{x}\in\mathbf{Y}_{\mathcal{A}}\cup\{\boldsymbol{w}\}}\left(\sh(\mathcal{X}_{\mathcal{A}}(\boldsymbol{x}+\boldsymbol{1})-\mathcal{X}_{\mathcal{A}}(\boldsymbol{1}))\,\mathcal{J}\big(\mathcal{X}_{\mathcal{A}}(\boldsymbol{x}+\boldsymbol{1})\big|\mathbf{Y}_{\mathcal{A}}\big)\right)^{(-1)^d}}{\prod_{\boldsymbol{x}\in\mathbf{Y}_{\mathcal{A}}}\mathcal{J}\big(\mathcal{X}_{\mathcal{A}}(\boldsymbol{x})\big|\mathbf{Y}_{\mathcal{A}}\big)\sh(\mathcal{X}_{\mathcal{A}}(\boldsymbol{x})-\mathcal{X}_{\mathcal{A}}(\boldsymbol{1}))}\cr
    =\,&\mathcal{J}\big(\mathcal{X}_{\mathcal{A}}(\boldsymbol{w})\big|\mathbf{Y}_{\mathcal{A}}\cup\{\boldsymbol{w}\}\big)\left(\sh(\mathcal{X}_{\mathcal{A}}(\boldsymbol{w})-\mathcal{X}_{\mathcal{A}}(\boldsymbol{0}))\,\mathcal{J}\big(\mathcal{X}_{\mathcal{A}}(\boldsymbol{w}+\boldsymbol{1})\big|\mathbf{Y}_{\mathcal{A}}\big)\right)^{(-1)^d}\cr
    &\quad\times\frac{\prod_{\boldsymbol{x}\in\mathbf{Y}_{\mathcal{A}}}\left(\sh(\mathcal{X}_{\mathcal{A}}(\boldsymbol{x})-\mathcal{X}_{\mathcal{A}}(\boldsymbol{0}))\,\mathcal{J}\big(\mathcal{X}_{\mathcal{A}}(\boldsymbol{x}+\boldsymbol{1})\big|\mathbf{Y}_{\mathcal{A}}\big)\right)^{(-1)^d}}{\prod_{\boldsymbol{x}\in\mathbf{Y}_{\mathcal{A}}}\mathcal{J}\big(\mathcal{X}_{\mathcal{A}}(\boldsymbol{x})\big|\mathbf{Y}_{\mathcal{A}}\big)\sh(\mathcal{X}_{\mathcal{A}}(\boldsymbol{x})-\mathcal{X}_{\mathcal{A}}(\boldsymbol{1}))}\cr
    =\,&\mathcal{J}\big(\mathcal{X}_{\mathcal{A}}(\boldsymbol{w})\big|\mathbf{Y}_{\mathcal{A}}\cup\{\boldsymbol{w}\}\big)\left(\sh(\mathcal{X}_{\mathcal{A}}(\boldsymbol{w})-\mathcal{X}_{\mathcal{A}}(\boldsymbol{0}))\,\mathcal{J}\big(\mathcal{X}_{\mathcal{A}}(\boldsymbol{w}+\boldsymbol{1})\big|\mathbf{Y}_{\mathcal{A}}\big)\right)^{(-1)^d},
\end{align}
where the swapping property~\eqref{swapping prop} is used in the penultimate step, and the last step follows because the remaining product over $\mathbf{Y}_{\mathcal{A}}$ telescopes to $1$.

As a concrete example, for 3d Young diagrams the recursion reads:
\begin{align}
    \frac{\prod_{\boldsymbol{x}\in\pi_{123,1}\cup\{\boldsymbol{w}\}}\mathcal{J}\big(\mathcal{X}_{123,1}(\boldsymbol{x})\big|\pi_{123,1}\cup\{\boldsymbol{w}\}\big)}{\prod_{\boldsymbol{x}\in\pi_{123,1}}\mathcal{J}\big(\mathcal{X}_{123,1}(\boldsymbol{x})\big|\pi_{123,1}\big)}
    =\frac{\mathcal{J}\big(\mathcal{X}_{123,1}(\boldsymbol{w})\big|\pi_{123,1}\cup\{\boldsymbol{w}\}\big)}{\sh(\mathcal{X}_{123,1}(\boldsymbol{w})-\mathcal{X}_{123,1}(\boldsymbol{0}))\,\mathcal{J}\big(\mathcal{X}_{123,1}(\boldsymbol{w}+\boldsymbol{1})\big|\pi_{123,1}\big)}.
\end{align}

\paragraph{Splitting property.} Suppose a $d$-dimensional Young diagram $\mathbf{Y}_{\mathcal{A}}$ decomposes as a disjoint union:
\begin{align}
    \mathbf{Y}_{\mathcal{A}}=\mathbf{Y}_{\mathcal{A}}'\cup\mathbf{Y}_{\mathcal{A}}'',
\end{align}
where $\mathbf{Y}_{\mathcal{A}}'$ begins at the same origin as $\mathbf{Y}_{\mathcal{A}}$, and $\mathbf{Y}_{\mathcal{A}}''$ begins at some box $\boldsymbol{y}=(y_1,\ldots,y_d)\in\mathbf{Y}_{\mathcal{A}}$. Then the $\mathcal{J}$-factor factorizes as:
\begin{align}\label{splitting}
    \mathcal{J}\big(x\big|\mathbf{Y}_\mathcal{A}\big)=\mathcal{J}\big(x\big|\mathbf{Y}_{\mathcal{A}}'\big)\,\mathcal{J}\big(x\big|\mathbf{Y}_{\mathcal{A}}''\big)\,\sh(x-\mathcal{X}_{\mathcal{A}}(\boldsymbol{y})).
\end{align}
The extra $\sh$ factor accounts for the interface between $\mathbf{Y}'$ and $\mathbf{Y}''$ at their junction. As a simple example, taking $\{(1,1),(1,2)\}=\{(1,1)\}\cup\{(1,2)\}$ with $\{(1,2)\}$ a single box starting at $(1,2)$:
\begin{align}
    \mathcal{J}\big(x\big|\{(1,1),(1,2)\}_{12,1}\big)=\mathcal{J}\big(x\big|\{(1,1)\}_{12,1}\big)\,\mathcal{J}\big(x\big|\{(1,2)\}_{12,1}\big)\,\sh(x-\mathcal{X}_{12,1}(1,2)).
\end{align}

\section{Instanton of 5d pure SYM}\label{5d pure SYM}
In this section, we employ the shell formula defined above to rewrite the instanton partition function for 5d $\cN=1$ pure SYM with classical gauge group \cite{Nekrasov:2002qd, Shadchin:2005mx, Nekrasov:2004vw}. Although these instanton partition functions can be expressed in terms of the Nekrasov factor, the shell formula representation makes the formulas more intuitive for visualizing the interactions between instantons and various D-branes. Note that in this section, since all space directions lie in $\C_1 \times \C_2=(x^0,x^1,x^2,x^3)$, all the Young diagrams are oriented in the $1,2$-direction, so we will temporarily omit the basis specification in the subsequent discussion.

\subsection{5d pure \texorpdfstring{$\U(N)$}{U(N)} SYM}\label{5dSYM U}

\begin{figure}[ht]
    \centering
    \includegraphics[width=0.2\linewidth]{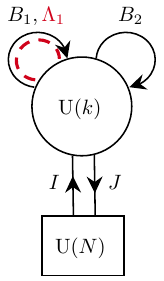}
    \caption{ADHM quiver diagram of D0-D4 system in the infinite adjoint mass
    limit as $\cN=2$ SUSY QM. The black solid lines represent chiral multiplets,
    and the red dashed lines represent fermi multiplets. The circular nodes denote
    gauge groups, while the square nodes denote flavor groups. In this quiver, we
    have a $\U(k)$ gauge group, a $\U(N)$ flavor group, a fundamental chiral $I$,
    an anti-fundamental chiral $J$, two adjoint chirals $B_{1,2}$ respectively,
    and an adjoint fermi $\Lambda_{1}$.}
    \label{fig:U(k)quiver}
\end{figure}

First, we consider the celebrated Nekrasov partition function. For $\cN=1$ SYM
theory with eight supercharges on the spacetime $\C_1\times\C_2\times S^1$, we
begin with the well-known case of pure $\U(N)$ gauge theory for definiteness.
After topological twisting, the partition function localizes to the moduli space
of instantons, $\mathcal{M}^{\mathrm{inst}}_{U(N),k}$, i.e., the space of
solutions to the self-duality equation $F=*F$. This moduli space is described by
the ADHM construction \cite{Atiyah:1978ri}, which for pure $\U(N)$ theory can
be expressed as a quiver diagram in Fig.~\ref{fig:U(k)quiver}.

Then the moduli space can be expressed as:
\begin{align}\label{ADHMeq}
    \mathcal{M}^{\text{inst}}_{\U(N),k}
    =\{(B_1,B_2,I,J)|\mu_{\R}=\mu_\C=0\}/\U(k)
\end{align}
where the ADHM equations are:
\begin{align*}
    \mu_\R=\sum_{i=1}^2[B_i,B_i^\dagger]+II^\dagger-JJ^\dagger,\qquad
    \mu_\C=[B_1,B_2]+IJ.
\end{align*}
$\mathcal{M}^{\text{inst}}_{\U(N),k}$ defined in this way is neither compact
nor smooth. Therefore, we need to slightly modify the conditions of the ADHM
construction~\eqref{ADHMeq} by changing $\mu_\R = 0$ to $\mu_\R =
\zeta\cdot\mathbf{1}_k$, thereby avoiding singularities in the moduli space.
Furthermore, we introduce the $\Omega$-background to render the entire integral
finite. The role of the $\Omega$-background is as follows:
\begin{align}
    (B_1,B_2,I,J,\Lambda_1)
    \xrightarrow{\U(1)_{\e_1}\times\U(1)_{\e_2}}
    (q_1^{-1}B_1,q_2^{-1}B_2,I,q^{-1}_{12}J,q_{12}^{-1}\Lambda_1)
\end{align}
where $q_i\equiv e^{-\e_i}$ and $q_{ij}=q_iq_j$.

The $\Omega$-background effectively localizes 2 complex planes $\C_1\times\C_2$
into a point. Thus the 5d SYM theory can now be viewed as a supersymmetric
quantum mechanics (SUSY QM) on $S^1$. Using the
Losev--Moore--Nekrasov--Shatashvili (LMNS) formalism \cite{Lossev:1997bz}, the
Nekrasov partition function is:
\begin{align}
    \mathcal{Z}_{\text{inst}}^{\U(N)}(v_1,\ldots,v_N)
    =\sum_{k=0}^\infty\mathfrak{q}^k\mathcal{Z}^{\U}_{N,k}(v_1,\ldots,v_N)
\end{align}
where $k$ is the instanton number, and $\mathcal{Z}_k$ is:
\begin{align}
    \mathcal{Z}^{\U}_{N,k}(v_1,\ldots,v_N)
    &=\oint_{\JK}\prod_{i=1}^k\frac{d\phi_i}{2\pi i}\mathcal{I}^{\U}_{N,k},\cr
    \mathcal{I}^{\U}_{N,k}
    &=\frac{1}{k!}\prod_{i\neq j}^k \sh(\phi_i-\phi_j)
    \prod_{i,j=1}^k\frac{\sh(\phi_i-\phi_j-\e_{12})}{\sh(\phi_i-\phi_j-\e_{1,2})}
    \prod_{i=1}^k\prod_{\alpha=1}^N
    \frac{1}{\sh(\phi_i-v_{\alpha})\sh(v_{\alpha}-\e_{12}-\phi_i)}\nonumber
\end{align}
where $\sh(x-\e_{1,2})=\sh(x-\e_1)\sh(x-\e_2)$. The Coulomb branch parameters
$v_{\alpha}$ effectively indicate the location of the $\alpha$-th D4-brane along
the complex planes $\C^2=\C_1\times\C_2$. After applying the JK-residue, the
poles can be classified by a set of $N$ 2d Young diagrams $\vec\lambda =
(\lambda_1, \dots, \lambda_N)$. For the $\alpha$-th Young diagram, the box at
position $\boldsymbol{x}=(i,j)$ contributes a pole at:
\begin{align}\label{2d poles}
    \mathcal{X}_{\alpha}(\boldsymbol{x})
    =v_{\alpha}+(\boldsymbol{x}-\boldsymbol{1})\cdot\boldsymbol{\e}_{12}
    =v_{\alpha}+i\e_1+j\e_2-\e_{12}.
\end{align}
To obtain the closed-form expression for the $k$-instanton partition function,
we introduce the \emph{Nekrasov factor} \cite{Nekrasov:2002qd}:
\begin{align}
    \mathrm{N}^{\vec{\lambda}}_{\alpha,\beta}(\boldsymbol{x})
    \equiv v_{\alpha}-v_{\beta}+L_{\lambda_\alpha}(\boldsymbol{x})\e_1
    -A_{\lambda_\beta}(\boldsymbol{x})\e_2-\e_2
\end{align}
where $L_{\lambda_\alpha}(\boldsymbol{x})$ and $A_{\lambda_\alpha}(\boldsymbol{x})$
are the leg and arm of the box $\boldsymbol{x}$ in $\lambda_\alpha$ respectively
(see Appendix~\ref{charges of shellbox} for definitions). The instanton partition
function is then:
\begin{align}\label{ZU Nekrasov}
    \mathcal{Z}^{\U}_{N,k}(v_1,\ldots,v_N)
    =\sum_{||\vec\lambda||=k}\mathcal{Z}^{\U}(\vec\lambda)
    =\sum_{||\vec\lambda||=k}\prod_{\alpha,\beta=1}^N\prod_{\boldsymbol{x}\in\lambda_\alpha}
    \frac{1}{\sh(-\mathrm{N}^{\vec{\lambda}}_{\alpha,\beta}(\boldsymbol{x}))
              \sh(\mathrm{N}^{\vec{\lambda}}_{\alpha,\beta}(\boldsymbol{x})+\e_{12})}
\end{align}
where $||\vec{\lambda}||\equiv\sum_\alpha|\lambda_\alpha|$ is the total number
of boxes.

While the Nekrasov factor provides a compact encoding of Young diagram data and
yields a concise expression for the instanton partition function, it relies on
arm and leg lengths that are intrinsic to 2d Young diagrams and do not extend
naturally to higher dimensions, and it makes the derivation of algebraic
relations---such as recursion formulas---rather cumbersome. For this reason, we
introduce the $\mathcal{J}$-factor defined in~\eqref{J-def} and rewrite the
instanton partition function as:
\begin{align}\label{ZU shell}
    \mathcal{Z}^{\U}_{N,k}(v_1,\ldots,v_N)
    =\sum_{||\vec{\lambda}||=k}\prod_{\alpha,\beta=1}^N\prod_{\boldsymbol{x}\in\lambda_{\alpha}}
    \frac{\mathcal{J}\big(\mathcal{X}_{\alpha}(\boldsymbol{x})\big|\lambda_{\beta}\big)}
         {\sh(-\mathcal{X}_{\alpha}(\boldsymbol{x})+\mathcal{X}_\beta(\boldsymbol{0}))}.
\end{align}
That~\eqref{ZU shell} is equal to~\eqref{ZU Nekrasov} follows from the identity:
\begin{align}\label{J and nek}
    \prod_{\boldsymbol{x}\in\lambda_{\alpha}}
    \frac{\mathcal{J}\left(\mathcal{X}_{\alpha}(\boldsymbol{x})\big|\lambda_\beta\right)}
         {\sh(\mathcal{X}_{\alpha}(\boldsymbol{x})-\mathcal{X}_\beta(\boldsymbol{0}))}
    =\prod_{\boldsymbol{x}\in\lambda_{\alpha}}
    \frac{1}{\sh(\mathrm{N}^{\vec{\lambda}}_{\alpha,\beta}(\boldsymbol{x}))
              \sh(\mathrm{N}^{\vec{\lambda}}_{\alpha,\beta}(\boldsymbol{x})+\e_{12})},
\end{align}
whose proof is given in Appendix~\ref{app. J and nek}. Thus~\eqref{ZU shell},
\eqref{J and nek}, and~\eqref{ZU Nekrasov} form a commutative triangle: the
shell formula on the left and the Nekrasov factor on the right are two
equivalent representations of the same quantity, related by the identity~\eqref{J and nek}.

\begin{figure}[ht]
    \centering
    \includegraphics[width=0.5\linewidth]{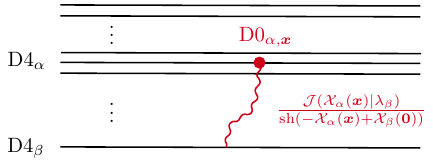}
    \caption{The instanton configuration of 5d pure $\U(N)$ SYM can be
    constructed from the type IIA D0-D4 brane system after integrating out the
    adjoint hypermultiplets. The D4-branes extend along two complex directions
    $\C_1$, $\C_2$, and the time direction $x^0$. D4$_{\alpha}$ refers to the
    $\alpha$-th D4-brane, while D0$_{\alpha,\boldsymbol{x}}$ denotes the
    D0-brane within D4$_{\alpha}$ corresponding to the instanton at coordinate
    $\boldsymbol{x}$ in the Young diagram. The red wavy line represents strings
    connecting D0$_{\alpha,\boldsymbol{x}}$ and D4$_{\beta}$; their contribution
    to the index is
    $\mathcal{J}\big(\mathcal{X}_\alpha\big|\lambda_\beta\big)/
    \sh(-\mathcal{X}_\alpha+\mathcal{X}_\beta(\boldsymbol{0}))$,
    as given in~\eqref{D0-D4 string}.}
    \label{fig:5d SYM brane}
\end{figure}

The shell formula representation makes the D0-D4 interaction structure manifest.
Physically, $v_\alpha$ is the position of the $\alpha$-th D4-brane and
$\mathcal{X}_\alpha(\boldsymbol{x})$ is the position of the D0-brane
(instanton) within D4$_\alpha$ at coordinate $\boldsymbol{x}$. The open strings
stretching from D0$_{\alpha,\boldsymbol{x}}$ to D4$_\beta$ probe the vacuum
configuration encoded in $\lambda_\beta$; the expansion property~\eqref{J-expand}
shows that the $\mathcal{J}$-factor is precisely their Witten index contribution.
Schematically, as illustrated in Fig.~\ref{fig:5d SYM brane}, in the vacuum
corresponding to Young diagram $\lambda_\beta$, the contribution from the
D0-D4 strings is:
\begin{align}\label{D0-D4 string}
    \frac{\mathcal{J}\big(\mathcal{X}_{\alpha}(\boldsymbol{x})\big|\lambda_{\beta}\big)}
         {\sh(-\mathcal{X}_{\alpha}(\boldsymbol{x})+\mathcal{X}_\beta(\boldsymbol{0}))}.
\end{align}

The instanton partition function $\mathcal{Z}_{N,k}^{\SU}$ of $\SU(N)$ SYM is
obtained by imposing the traceless condition $\sum_{\alpha=1}^Nv_\alpha=0$.

Property~\eqref{recursion} also yields recursion relations for the Nekrasov
partition function. When adding the contribution of an instanton within a
D4-brane, the ratio of the partition function contribution from the Young
diagram $\lambda_\alpha \cup \Box$ to that from $\lambda_\alpha$ is:
\begin{align}\label{UN recursion}
    \frac{\mathcal{Z}^{\U}(\lambda_\alpha\cup\Box)}{\mathcal{Z}^{\U}(\lambda_\alpha)}
    &=\left(\prod_{\boldsymbol{x}\in\lambda_\alpha\cup\Box}
      \frac{\mathcal{J}\big(\mathcal{X}_{\alpha}(\boldsymbol{x})\big|
            \lambda_{\alpha}\cup\Box\big)}
           {\sh(-\mathcal{X}_{\alpha}(\boldsymbol{x})+\mathcal{X}_\alpha(\boldsymbol{0}))}\right)
      \Bigg/
      \left(\prod_{\boldsymbol{y}\in\lambda_\alpha}
      \frac{\mathcal{J}\big(\mathcal{X}_{\alpha}(\boldsymbol{y})\big|\lambda_{\alpha}\big)}
           {\sh(-\mathcal{X}_{\alpha}(\boldsymbol{y})+\mathcal{X}_\alpha(\boldsymbol{0}))}\right)
    \cr
    &=-\mathcal{J}\big(\mathcal{X}_\alpha(\Box)\big|\lambda_\alpha\cup\Box\big)
      \times\mathcal{J}\big(\mathcal{X}_\alpha(\Box+\boldsymbol{1})\big|\lambda_\alpha\big)
\end{align}
where $\Box$ denotes the coordinate of the added box in the Young diagram. This
recursion relation is directly connected to the quantum toroidal algebra and
$qq$-characters
\cite{Nekrasov:2015wsu,Nekrasov:2016ydq,Nawata:2023wnk,Bourgine:2017jsi,Kimura:2023bxy,Kimura:2015rgi,Gaiotto-conjecture},
as we now make explicit.

Consider the Gaiotto state, defined as a linear combination of 2d Young diagram
basis states corresponding to the Fock representation of the quantum toroidal
algebra:
\begin{align}
    \ket{\mathfrak{G}}\equiv\sum_{\vec\lambda}
    \left(\mathfrak{q}^{||\vec\lambda||}\mathcal{Z}^{\U}(\vec\lambda)\right)^{1/2}
    \ket{\vec\lambda}.
\end{align}
By the following identities:
\begin{align}
    \mathcal{J}\big(\mathcal{X}_\alpha(\Box)\big|&\lambda_\alpha\cup\Box\big)
    \mathcal{J}\big(\mathcal{X}_\alpha(\Box+\boldsymbol{1})\big|\lambda_\alpha\big)\cr
    =\,&\mathcal{J}\big(\mathcal{X}_\alpha(\Box)\big|\lambda_\alpha\big)
    \mathcal{J}\big(\mathcal{X}_\alpha(\Box)\big|\Box_\alpha\big)
    \sh(\mathcal{X}_\alpha(\Box)-\mathcal{X}_\alpha(\Box))
    \mathcal{J}\big(\mathcal{X}_\alpha(\Box)+\e_{12}\big|\lambda_\alpha\big)\cr
    =\,&-\mathcal{J}\big(\mathcal{X}_\alpha(\Box)\big|\lambda_\alpha\big)
    \mathcal{J}\big(\mathcal{X}_\alpha(\Box)+\e_{12}\big|\Box_\alpha\big)
    \sh(\mathcal{X}_\alpha(\Box)+\e_{12}-\mathcal{X}_\alpha(\Box))
    \mathcal{J}\big(\mathcal{X}_\alpha(\Box)+\e_{12}\big|\lambda_\alpha\big)\cr
    =\,&-\mathcal{J}\big(\mathcal{X}_\alpha(\Box)+\e_{12}\big|\lambda_\alpha\cup\Box\big)
    \mathcal{J}\big(\mathcal{X}_\alpha(\Box)\big|\lambda_\alpha\big),
\end{align}
the recursion relation~\eqref{UN recursion} can be rearranged to:
\begin{align}
    \frac{\mathcal{Z}^{\U}(\lambda_\alpha\cup\Box)}
         {\mathcal{J}\big(\mathcal{X}_\alpha(\Box)+\e_{12}\big|\lambda_\alpha\cup\Box\big)}
    -\mathcal{Z}^{\U}(\lambda_\alpha)\,
    \mathcal{J}\big(\mathcal{X}_\alpha(\Box)\big|\lambda_\alpha\big)=0.
\end{align}
This is precisely the vanishing condition for the $\mathrm{A}_N$ $qq$-character
\cite{Nekrasov:2015wsu}:
\begin{align}\label{D0D4qq}
    \chi(x)\equiv\frac{1}{\mathcal{J}(x+\e_{12})}-\mathfrak{q}\,\mathcal{J}(x),
\end{align}
where the operator $\mathcal{J}(x)$ is defined by:
\begin{align}
    \mathcal{J}(x)\ket{\vec\lambda}\equiv\prod_{\alpha=1}^N
    \mathcal{J}\big(x\big|\lambda_\alpha\big)\ket{\vec\lambda}.
\end{align}
One can verify directly that $\braket{\chi(x)}=\bra{\mathfrak{G}}\chi(x)\ket{\mathfrak{G}}$
is a well-defined polynomial in $x$, which is the defining property of a
$qq$-character. The operator $1/\mathcal{J}(x)$ corresponds precisely to the
$\mathcal{Y}(x)$ observable in the conventions of
\cite{Nekrasov:2015wsu,Kimura:2023bxy},
and the $\mathcal{J}$-factor therefore provides a natural generating function of the $qq$-character.

\subsection{5d pure \texorpdfstring{$\SO(N)$}{SO(N)} SYM}\label{5dSYM SO}
Next, we turn our attention to pure SYM theories for Lie groups of types B and D \cite{Nekrasov:2004vw}. We can express the instanton moduli space via the ADHM construction. First, since $\SO(N)$ involves a symmetric bilinear form, to ensure the moment maps $\mu_{\R}$ and $\mu_{\C}$ are invariant under the gauge group $\SO(N)$, the quotient group must incorporate an antisymmetric bilinear form. That is, the quotient group is of type C: $\Sp(2k) \subset \U(2k)$ \cite{Shadchin:2005mx}. 
The resulting integral form of the instanton partition function is then given by:
\begin{align}
    \mathcal{I}^{\SO}_{N,k}=&\frac{1}{k!\,2^k}\prod_{i\neq j}^k \sh(\phi_i-\phi_j) \prod_{i,j=1}^k\frac{\sh(\phi_i-\phi_j-\e_{12}) }{\sh(\phi_i-\phi_j-\e_{1,2})}\frac{\prod_{i\leq j}^k\sh(\pm(\phi_i+\phi_j))\sh(\pm(\phi_i+\phi_j)-\e_{12})}{\prod_{i< j}^k\sh(\pm(\phi_i+\phi_j)-\e_{1,2})}\cr&\times\prod_{i=1}^k\frac{1}{\prod_{\alpha=1}^{n}\sh(\pm\phi_i\pm v_{ \alpha}-\frac12\e_{12})}\left(\frac{1}{\sh(\pm\phi_i-\frac12\e_{12})}\right)^\chi
\end{align}
where $n=\lfloor\frac{N}{2}\rfloor$ is the rank of $\SO(N)$, and $\chi=N \mod 2$ labels the B and D type Lie group.

Classifying the poles of this integral has been a challenging problem. However, the \emph{unrefined limit} $\epsilon_2 = -\epsilon_1$ simplifies matters considerably. In this limit, as in~\eqref{2d poles}, the non-trivial poles simplify and are classified by 2d Young diagrams \cite{Nawata:2021dlk}. In the unrefined limit, the factor $\mathcal{J}\big(\pm\mathcal{X}_\alpha \big|\lambda_\beta\big)/{\sh(\pm\mathcal{X}_\alpha+\mathcal{X}_\beta)}$ produces singular terms of the form $\sh(a\,\e_{12})/\sh(b\,\e_{12})$ when $\alpha = \beta$ for diagonal boxes (i.e., boxes $\boldsymbol{x}=(i,i)$) in $\lambda_{\alpha}$; the limit $\lim_{\e_2\to-\e_1}$ extracts the coefficient $a/b$, which plays a critical role in the subsequent analysis of $\Sp(2N)$ SYM theory in Sec.~\ref{5dSYM Sp}. The partition function then takes the concise shell formula form:

\begin{align}
    \mathcal{Z}^{\SO}_{N=2n+\chi,k}(v_1,\ldots,v_n)=\lim_{\e_2\to-\e_1}\sum_{||\vec\lambda||=k}\prod_{\alpha=1}^n\prod_{\boldsymbol x\in\lambda_{ \alpha}}&\frac{\sh(2\mathcal{X}_{ \alpha}(\boldsymbol x)+\e_{1,2})\sh^2(2\mathcal{X}_{ \alpha}(\boldsymbol{x}))}{\sh^{\chi}(\pm\mathcal{X}_{ \alpha}(\boldsymbol{x}))}\cr&\times\prod_{\beta=1}^n\frac{\mathcal{J}\big(\pm\mathcal{X}_{ \alpha}(\boldsymbol{x}) \big|\lambda_{ \beta} \big)}{\sh(\pm\mathcal{X}_{ \alpha}(\boldsymbol{x})+\mathcal{X}_\beta(\boldsymbol{0}))}
\end{align}

In order to endow this shell formula with physical meaning, similar to that in Fig.~\ref{fig:5d SYM brane}, we first engineer this system using a 5-brane web in IIB theory \cite{AH,AHK,Zafrir:2015ftn}. The construction requires $\lfloor\frac N2\rfloor
$ D5-branes, 2 NS5-branes, and an O5-plane, with their orientations given in Tab.~\ref{tab:5braneweb}.
\begin{table}[ht]
\centering
\begin{tabular}{|c|c|c|c|c|c|c|c|c|c|c|}
    \hline & \multicolumn{2}{|c|}{$\mathbb{C}_{1}$} & \multicolumn{2}{|c|}{$\mathbb{C}_{2}$} & $x^5$&$x^6$ & $x^7$ & $x^8$ & \multicolumn{2}{|l|}{$\mathbb{R} \times \mathbb{S}^1$}  \\
\cline { 2 - 11 }& 1 & 2 & 3 & 4 & 5 & 6 & 7 & 8 & 9 & 0 \\
\hline D1& $\bullet$ & $\bullet$ & $\bullet$ & $\bullet$ & - & $\bullet$ & $\bullet$ & $\bullet$ & $\bullet$ & - \\
\hline D5& - & - & - & - & - & $\bullet$ & $\bullet$ & $\bullet$ & $\bullet$ & - \\ \hline O5$^\pm$,$\widetilde{\text{O5}}^\pm$& - & - & - & - &-& $\bullet$& $\bullet$& $\bullet$& $\bullet$ & - \\ \hline NS5& - & - & - & - & $\bullet$& -& $\bullet$& $\bullet$& $\bullet$ & - \\ \hline
\end{tabular}
\caption{Brane configuration of 5d $\SO$ or $\Sp$ pure SYM. Consider D5-branes, NS5-branes, and O5-planes in type IIB string theory. The symbol $-$ denotes an extended direction of the D-branes, whereas $\bullet$ denotes a point-like direction. The D1-branes correspond to instantons extend along $x^5$ and $x^0$. The D5-branes and O5-planes extend along the directions $\C_1$, $\C_2$, $x^5$, and $x^0$, while the NS5-branes extend along $\C_1$, $\C_2$, $x^6$, and $x^0$. The non-Abelian gauge group is constructed from the D5-branes. The O5$^-$-plane projects out the symmetric vector states to form $\SO$ gauge groups, while the O5$^+$-plane projects out the antisymmetric vector states to form $\Sp$ gauge groups.}
\label{tab:5braneweb}
\end{table}

\begin{figure}[ht]
    \centering
    \includegraphics[width=0.95\linewidth]{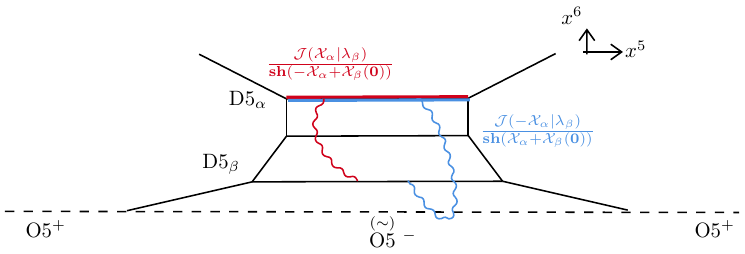}
    \caption{On the $x^5$-$x^6$ plane, the horizontally extending lines represent D5-branes, while the vertical or slanted lines represent NS5-branes. For the $\SO$ gauge group, we need to use an O5$^-$ or $\widetilde{\text{O5}}^-$ orientifold at the bottom, where, when it crosses an NS5-brane, it becomes an O5$^+$. For the $\SO(2n)$ gauge group, we require $n$ D5-branes and an O5$^-$. For $\SO(2n+1)$, we need $n$ D5-branes and an $\widetilde{\text{O5}}^-$ \cite{Zafrir:2015ftn}. The red wavy lines represent the effective contribution of strings connecting D1$_{\alpha,\boldsymbol{x}}$ and D5$_\beta$, which is identical to that of the $\U(N)$ gauge group given in~\eqref{D0-D4 string}. The blue wavy lines correspond to strings connecting D1$_{\alpha,\boldsymbol{x}}$ and D5$_\beta$ whose orientation is reversed by the O-plane; their effective contribution matches the original one, subject to the replacement $\mathcal{X}_\alpha \to -\mathcal{X}_\alpha$.
}
    \label{fig:SON brane}
\end{figure}
The presence of the O5-plane causes strings winding around it to undergo an orientation reversal. Hence, such strings contribute an additional negative sign: $\mathcal{X}_\alpha\to-\mathcal{X_\alpha}$ to the partition function, as shown in Fig.~\ref{fig:SON brane}.

\subsection{5d pure \texorpdfstring{$\Sp(2N)$}{Sp(2N)} SYM}\label{5dSYM Sp}
The shell formula is especially powerful in analyzing the \emph{BPS jumping phenomenon} \cite{Kim:2024vci, Nawata:2021dlk}. Let us consider the case of $\Sp (2N)$. According to the ADHM construction, the quotient group for the $\Sp (2N)$ instanton moduli space for $k$ instantons is the orthogonal group $\mathrm{O}(k)$. Furthermore, since $\pi_4(\Sp(2N)) = \Z_2$ for 5d SYM, the $5$d $\Sp$ SYM partition function naturally includes a discrete topological angle $\theta \in \{0,\pi\}$, corresponding precisely to the two distinct components of the $\mathrm{O}$ group. Based on the ADHM data, the two instanton partition functions for the $\Sp$ group are:
\begin{align}
    \mathcal{Z}^{\Sp,\theta=0}_{2N,k}=\mathcal{Z}^{\Sp,+}_{2N,k}+\mathcal{Z}^{\Sp,-}_{2N,k}\cr  \mathcal{Z}^{\Sp,\theta=\pi}_{2N,k}=\mathcal{Z}^{\Sp,+}_{2N,k}-\mathcal{Z}^{\Sp,-}_{2N,k}
\end{align}
where the integrand of $\mathcal{Z}^{\Sp,\pm}$ are as follow \cite{Shadchin:2005mx,Hwang:2014uwa,Kim:2012gu}:
\begin{footnotesize}
    \begin{align}
    \mathcal{I}^{\Sp,+}_{2N,k}=&\frac{1}{l!\,2^{l+\chi}}\prod_{i\neq j}^l \sh(\phi_i-\phi_j) \prod_{i,j=1}^l\frac{\sh(\phi_i-\phi_j-\e_{12}) }{
\sh(\phi_i-\phi_j-\e_{1,2})}\frac{\prod_{i< j}^l\sh(\pm(\phi_i+\phi_j))\sh(\pm(\phi_i+\phi_j)-\e_{12})}{\prod_{i\leq j}^l\sh(\pm(\phi_i+\phi_j)-\e_{1,2})}\cr&\times\prod_{i=1}^l\frac{1}{\prod_{\alpha=1}^{N}\sh(\pm\phi_i\pm v_{ \alpha}-\frac12\e_{12})}\left(\frac{1}{\sh(\e_{1,2})}\frac{1}{\prod_{\alpha=1}^N\sh(\pm v_{ \alpha}-\frac12\e_{12})}\prod_{i=1}^l\frac{\sh(\pm\phi_i)\sh(\pm \phi_i-\e_{12})}{\sh(\pm\phi_i-\e_{1,2})}\right)^\chi\\ \mathcal{I}^{\Sp,-}_{2N,k}=&\frac{1}{(l-1+\chi)!\,2^{l+\chi}}\prod_{i\neq j}^{l-1+\chi} \sh(\phi_i-\phi_j) \prod_{i,j=1}^{l-1+\chi}\frac{\sh(\phi_i-\phi_j-\e_{12}) }{\sh(\phi_i-\phi_j-\e_{1,2})}\frac{\prod_{i< j}^{l-1+\chi}\sh(\pm(\phi_i+\phi_j))\sh(\pm(\phi_i+\phi_j)-\e_{12})}{\prod_{i\leq j}^{l-1+\chi}\sh(\pm(\phi_i+\phi_j)-\e_{1,2})}\cr&\times\prod_{i=1}^{l-1+\chi}\frac{1}{\prod_{\alpha=1}^{N}\sh(\pm\phi_i\pm v_{ \alpha}-\frac12\e_{12})}\left(\frac{1}{\sh(\e_{1,2})}\frac{1}{\prod_{\alpha=1}^N\ch(\pm v_{ \alpha}-\frac12\e_{12})}\prod_{i=1}^{l-1+\chi}\frac{\ch(\pm\phi_i)\ch(\pm \phi_i-\e_{12})}{\ch(\pm\phi_i-\e_{1,2})}\right)\cr&\times\left(\frac{\ch(\e_{12})}{\sh(2\e_{1,2})}\frac{1}{\prod_{\alpha=1}^N\sh(\pm v_{ \alpha}-\frac12\e_{12})}\prod_{i=1}^{l-1+\chi}\frac{\sh(\pm\phi_i)\sh(\pm \phi_i-\e_{12})}{\sh(\pm\phi_i-\e_{1,2})}\right)^{1-\chi}
\end{align}
\end{footnotesize}
where $l=\lfloor\frac{k}{2}\rfloor$ and $\chi\equiv k \mod 2$. Note that in our simplified notation, the expression written as $\mathcal{I}^{\Sp,-}_{2N,k}$ is not correct after applying the JK-residue when $k$ is even. Fortunately, however, it is valid in the unrefined limit. Since the classification of poles remains unknown for the refined case, we exclusively focus on the unrefined limit.

In the unrefined limit, the poles are classified by $N+4$ Young diagrams \cite{Nawata:2021dlk}. Among these, the first $N$ Young diagrams are labeled by the $N$ coulomb branch parameters $v_{ 1},\ldots v_{ N}$ in the partition function as~\eqref{2d poles}. The additional four poles require distinct labeling according to the following Tab.~\ref{tab:SP poles}.

\begin{table}[ht]
    \centering
    \begin{tabular}{c|c c c c}
         &$v_{ N+1}$&$v_{ N+2}$&$v_{ N+3}$&$v_{ N+4}$ \\
         \hline
        $+,k=$even & $\e_1/2$ & $\e_1/2+\pi i$& $\e_{12}/2$ & $\e_{12}/2+\pi i$\\ $+,k=$odd &$\e_1/2$ & $\e_1/2+\pi i$& $\e_1$ & $\e_{12}/2+\pi i$\\$-,k=$even &$\e_1/2$ & $\e_1/2+\pi i$& $\e_1$ & $\e_1+\pi i$\\ $-,k=$odd &$\e_1/2$ & $\e_1/2+\pi i$& $\e_{12}/2$ & $\e_1+\pi i$
    \end{tabular}
    \caption{The four additional poles required for the $\Sp$ instanton partition function.
}
    \label{tab:SP poles}
\end{table}

The closed-form expression for the $\Sp(2N)$ plus sector is:
\begin{footnotesize}
    \begin{align}\label{ZSp plus}
    \mathcal{Z}^{\Sp,+}_{2N,k=2l+\chi}(v_1,\ldots,v_N)=\lim_{\e_2\to-\e_1}\sum_{||\vec\lambda||=l}&\left(\prod_{\alpha=1}^{N+4}\prod_{\boldsymbol x\in\lambda_{ \alpha}}\frac{\prod_{\beta=1}^{N+4}\mathcal{J}\big(\pm\mathcal{X}_{ \alpha}(\boldsymbol{x}) \big|\lambda_{ \beta} \big)}{\prod_{\beta=1}^{N}\sh(\pm\mathcal{X}_{ \alpha}(\boldsymbol{x})+\mathcal{X}_\beta(\boldsymbol{0}))}\right)\left(\frac{\prod_{\alpha=1}^{N+4}\mathcal{J}\big(0 \big|\lambda_{ \alpha} \big)}{\prod_{\alpha=1}^{N}\sh(0+\mathcal{X}_\alpha(\boldsymbol{0}))}\right)^{\chi}
\end{align}
\end{footnotesize}
where, as Tab.~\ref{tab:SP poles} shows, we need to identify $v_{ N+1}=\frac12\e_1$, $v_{ N+2}=\frac12\e_1+\pi i$, $v_{ N+3}=\chi\,\e_1+\frac12(1-\chi)\,\e_{12}$ and $v_{ N+4}=\frac12\e_{12}+\pi i$.

We remark that without using the $\lim_{\e_2 \to -\e_1}$, the BPS jumping coefficients $C^{\Sp}_{\vec{\lambda},\boldsymbol{v}}$ must be manually included in each term of the summation:
\begin{align}\label{BPS jumping coeff}
    \mathcal{Z}^{\Sp,\pm}_{2N,k}=\sum_{||\vec{\lambda}||=l}C^{\Sp}_{\vec{\lambda},\boldsymbol{v}}\,\mathcal{Z}^{\Sp,\pm}(\vec{\lambda}),\qquad C^{\Sp}_{\vec{\lambda},\boldsymbol{v}}=\prod_{\alpha=N+1}^{N+4}C^{\Sp}_{\lambda_{\alpha},v_\alpha}
\end{align}

These coefficients $C^{\Sp}_{\vec{\lambda},\boldsymbol{v}}$ depend on the specific shapes of the Young diagrams $\vec\lambda$ and the corresponding Coulomb branch parameters $\boldsymbol{v}$: $C^{\Sp}_{\emptyset,v_\alpha}=1$, and
\begin{align}  
    C^{\Sp}_{\lambda_\alpha,v_\alpha=0,\pi i,\frac{\e_1}{2},\frac{\e_1}{2}+\pi i}=\frac{2^{2j-1}}{\binom{2j-1}{j-1}}, \quad& \text{where $j$ is number of diagonal boxes $\boldsymbol{x}=(i,i)$ in $\lambda_\alpha$}, \cr C^{\Sp}_{\lambda_\alpha,v_\alpha=\e_1,\e_1+\pi i}=\frac{2^{2j}}{\binom{2j+1}{j}},\quad& \text{where $j$ is number of superdiagonal boxes $\boldsymbol{x}=(i,i+1)$ in $\lambda_\alpha$}.\cr
\end{align}
Consequently, the introduction of these extra coefficients obstructs the computation of the topological vertex for the O$^+$-plane \cite{Kim:2024ufq} and the derivation of the algebraic properties of the partition functions. Fortunately, by employing the shell formula with the unrefined limit $\lim_{\e_2 \to -\e_1}$, these coefficients are fully absorbed into the limiting procedure. Detailed calculations in Appendix~\ref{Sp2 example} demonstrate how these coefficients arise.

Similarly, the minus sector for $\Sp(2N)$ can be expressed through analogous formulas:
\begin{align}\label{ZSp minus}
    \mathcal{Z}^{\Sp,-}_{2N,k=2l+\chi}(v_1,\ldots,v_N)=\lim_{\e_2\to-\e_1}&\sum_{||\vec\lambda||=l-1+\chi}\left(\prod_{\alpha=1}^{N+4}\prod_{\boldsymbol x\in\lambda_{ \alpha}}\frac{\prod_{\beta=1}^{N+4}\mathcal{J}\big(\pm\mathcal{X}_{ \alpha}(\boldsymbol{x}) \big|\lambda_{ \beta} \big)}{\prod_{\beta=1}^{N}\sh(\pm\mathcal{X}_{ \alpha}(\boldsymbol{x})+\mathcal{X}_\beta(\boldsymbol{0}))}\right)\cr&\qquad\times\left(\frac{\prod_{\alpha=1}^{N+4}\mathcal{J}\big(\pi i \big|\lambda_{ \alpha} \big)}{\prod_{\alpha=1}^{N}\sh(-\pi i+\mathcal{X}_\alpha(\boldsymbol{0}))}\right)\left(\frac{\prod_{\alpha=1}^{N+4}\mathcal{J}\big(0 \big|\lambda_{ \alpha} \big)}{\prod_{\alpha=1}^{N}\sh(0+\mathcal{X}_\alpha(\boldsymbol{0}))}\right)^{1-\chi}\cr
\end{align}
where we need to impose the extra poles conditions as Tab.~\ref{tab:SP poles}, $v_{ N+1}=\frac12\e_1$, $v_{ N+2}=\frac12\e_1+\pi i$, $v_{ N+3}=(1-\chi)\,\e_1+\frac12\chi\,\e_{12}$ and $v_{ N+4}=\e_1+\pi i$.

\begin{figure}[ht]
    \centering
    \includegraphics[width=1\linewidth]{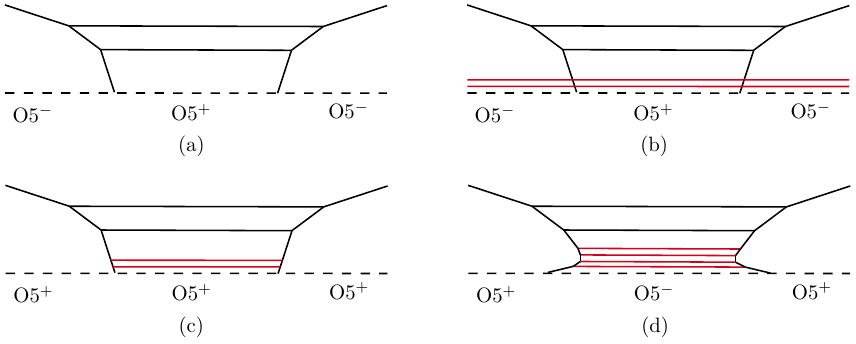}
    \caption{(a) The brane construction for the Sp gauge group is similar to Fig.~\ref{fig:SON brane}. For the $\Sp(2N)$ group, we require $N$ D5-branes, an O5$^+$, and NS5-branes. Furthermore, we can transform this into the brane construction for $\SO(2N+8)$. The transformation process is as follows: (b) We bring two D5-branes from infinity via Higgsing to the vicinity of the O-plane and freeze them. At this point, both the left and right sides at the bottom consist of O5$^-$ plus two D5-branes, while the central part at the bottom consists of O5$^+$ plus two D5-branes. (c) Using the equivalence O5$^- + 2$ D5 $\sim$ O5$^+$, we can replace both sides of the bottom with O5$^+$. (d) Through the equivalence O5$^+$ $\sim$ O5$^- + 2$ D5, the central part can be replaced with O5$^- + 4$ D5. Therefore, $\Sp(2N)$ can be viewed as $\SO(2N+8)$ with 4 Coulomb branch parameters $v_{N+1},\ldots,v_{N+4}$ fixed as Tab.~\ref{tab:SP poles}, corresponding to the frozen D5-branes.}
    \label{fig:Sp to SO brane}
\end{figure}

To provide a physical interpretation for these four additional fixed Coulomb branch parameters $v_{N+1}$ to $v_{N+4}$, we construct the $\Sp$ theory using a five-brane web and compare it with the $\SO$ theory. As illustrated in Tab.~\ref{tab:5braneweb}, the $\Sp$ theory is realized with an O5-plane \cite{Zafrir:2015ftn, Kim:2024vci}. Analysis based on RR charge and monodromy reveals that an Op$^+$-plane is effectively equivalent to an Op$^-$-plane plus $2^{p-4}$ Dp-branes. These Dp-branes are frozen near the orientifold plane and must acquire specific VEVs; hence, an O5$^+$-plane corresponds approximately to an O5$^-$-plane together with two frozen D5-branes. Accordingly, as depicted in Fig.~\ref{fig:Sp to SO brane}, the $\Sp(2N)$ theory is related to the $\SO(2N+8)$ theory by incorporating $4$ D5-branes with specific VEVs.

\begin{figure}[ht]
    \centering
    \includegraphics[width=0.7\linewidth]{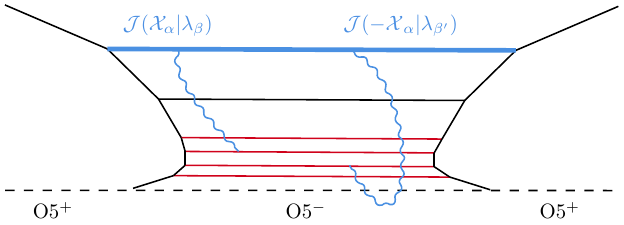}
    \caption{The brane construction of $\Sp$ gauge theory. The blue wavy lines represent the effective contribution of strings connecting the D1$_{\alpha,\boldsymbol x}$ branes and the frozen D5$_{\beta}$ branes, where $\beta = N+1, \dots, N+4$. Unlike the string contributions connecting D1 and ordinary D5 branes mentioned above~\eqref{D0-D4 string}, the contribution of these strings is simply $\mathcal{J}\big(\mathcal{X_{\alpha}}\big|\lambda_{\beta'}\big)$. For the contribution of strings that have been projected by the O-plane and connect D1 and the frozen D5 branes, we also need to replace $\mathcal{X}_{\beta'}$ with $-\mathcal{X}_{\beta'}$.
}
    \label{fig:Sp brane}
\end{figure}

As shown in Fig.~\ref{fig:Sp brane}, the effective contribution of the strings connecting the D1-brane to the four frozen D5-branes is $\mathcal{J}(\pm\mathcal{X}_\alpha|\lambda_\beta)$, which differs from the contribution of ordinary strings.

We briefly summarize the contributions of these various types of strings in the Tab.~\ref{tab:different-strings}.
\begin{table}[ht]
    \centering
    \begin{tabular}{|c|c|}
        \hline
Type & Contribution \\
\hline
D1$_{\alpha,\boldsymbol{x}}$--D5$_{\beta}$ strings 
& $\displaystyle \frac{\mathcal{J}\big(\mathcal{X}_{\alpha}(\boldsymbol{x}) \big|\lambda_{\beta} \big)}
{\operatorname{sh}(-\mathcal{X}_{\alpha}(\boldsymbol{x})+\mathcal{X}_{\beta}(\boldsymbol{0}))}$ \\[8pt]
\hline
O-plane projected D1$_{\alpha,\boldsymbol{x}}$--D5$_{\beta}$ strings 
& $\displaystyle \frac{\mathcal{J}\big(-\mathcal{X}_{\alpha}(\boldsymbol{x}) \big|\lambda_{\beta} \big)}
{\operatorname{sh}(\mathcal{X}_{\alpha}(\boldsymbol{x})+\mathcal{X}_{\beta}(\boldsymbol{0}))}$ \\[8pt]
\hline

D1$_{\alpha,\boldsymbol{x}}$--frozen D5$_{\beta}$ strings 
& $\displaystyle \mathcal{J}\big(\mathcal{X}_{\alpha}(\boldsymbol{x}) \big|\lambda_{\beta} \big)$ \\
\hline

O-plane projected D1$_{\alpha,\boldsymbol{x}}$--frozen D5$_{\beta}$ strings 
& $\displaystyle \mathcal{J}\big(-\mathcal{X}_{\alpha}(\boldsymbol{x}) \big|\lambda_{\beta} \big)$ \\
\hline
    \end{tabular}
    \caption{The four types of string configuration that appear with their contributions}
    \label{tab:different-strings}
\end{table}

We can check the Lie algebra-theoretic relations of instanton partition
functions. The isomorphisms $\Sp(2)\simeq\SU(2)$ and $\Sp(4)\simeq\SO(5)$ of Lie algebras lead to the equality of the partition functions:
\begin{align}
    &\mathcal{Z}^{\SU}_{2,k}(v_1)=\mathcal{Z}^{\Sp,\theta=0}_{2,k}(v_1)\cr&\mathcal{Z}^{\SO}_{5,k}(v_1+v_2,v_1-v_2)=\mathcal{Z}^{\Sp,\theta=0}_{4,k}(v_1,v_2)
\end{align}

\section{Gauge origami}\label{gauge origami}

In this section, we consider a more general setup called gauge origami
\cite{Nekrasov:2016qym}. Although the systems treated below appear at first to
be distinct, they are hierarchically related through tachyon condensation. A
D8-$\overline{\text{D8}}$ pair condenses into a single D6-brane when the
separation between them is tuned to $\epsilon_4$, restricting the relevant 4d
Young diagrams to have at most one layer in the condensed direction and yielding
3d Young diagrams. A further D6-$\overline{\text{D6}}$ condensation produces a
D4-brane and reduces the combinatorics to 2d Young diagrams. Schematically:
\begin{equation}\label{tachyon chain}
  \text{D0-D8}
  \;\xrightarrow{\;\text{D8-}\overline{\text{D8}}\;\text{condensation}\;}
  \text{D0-D6}
  \;\xrightarrow{\;\text{D6-}\overline{\text{D6}}\;\text{condensation}\;}
  \text{D0-D4}.
\end{equation}
The DT3 and DT4 counting problems arise within this hierarchy by placing the
D0-brane system on top of a fixed vacuum configuration---a minimal plane or
solid partition---determined by D2- and D4-brane boundary conditions. The shell
formula provides a unified treatment of all levels of this hierarchy. We will
provide the shell formula for the gauge origami systems on $\C^4 \times \R^1
\times S^1$, including the D0-D8 system known as magnificent four
\cite{Nekrasov:2017cih, Nekrasov:2018xsb, Noshita:2025bzg}, the D0-D6 system
known as tetrahedron instantons \cite{Pomoni:2021hkn,Pomoni:2023nlf}, the D0-D6 system known as tetrahedron instantons \cite{Pomoni:2021hkn, Pomoni:2023nlf}, the D0-D4 system known as spiked instantons \cite{Nekrasov:2016gud, Nekrasov:2015wsu, Nekrasov:2016qym}, the D0-D2-D6 system known as the DT3 counting \cite{Thomas:1998uj,Kimura:2023bxy,Kimura:2025lfo} and the D0-D2-D4-D8 system known as the DT4 counting \cite{Monavari_2022,Kimura:2025lig,Nekrasov:2023nai,Piazzalunga:2023qik}.
These systems are further interconnected by introducing appropriate antibranes \cite{Nekrasov:2017cih, Berkovits:2000hf, Akhmedov:2000zp}, whose tachyon condensation provides a physical mechanism relating different brane configurations within a unified framework.

\subsection{Magnificent four}\label{Magnificent four}
Nekrasov introduced the D0-D8 system on a CY fourfold and dubbed the resulting BPS counting problem the magnificent four \cite{Nekrasov:2017cih}. To investigate the distribution of bound states in SUSY QM on D0-branes, we analyze the energy spectrum of this system. The brane configuration is given in Tab.~\ref{tab:D8 brane}; D8-branes and anti-D8-branes wrap the directions $\C_{1,2,3,4}$ and are compactified on a circle $S^1$ along the $x^0$ direction.

\begin{table}[ht]
\centering
\begin{tabular}{|c|c|c|c|c|c|c|c|c|c|c|}
    \hline & \multicolumn{2}{|c|}{$\mathbb{C}_{1}$} & \multicolumn{2}{|c|}{$\mathbb{C}_{2}$} & \multicolumn{2}{c|}{$\mathbb{C}_{3}$} & \multicolumn{2}{|c|}{$\mathbb{C}_{4}$} & \multicolumn{2}{|l|}{$\mathbb{R} \times \mathbb{S}^1$}  \\
\cline { 2 - 11 }& 1 & 2 & 3 & 4 & 5 & 6 & 7 & 8 & 9 & 0 \\
\hline $k$ D0& $\bullet$ & $\bullet$ & $\bullet$ & $\bullet$ & $\bullet$ & $\bullet$ & $\bullet$ & $\bullet$ & $\bullet$ & - \\ \hline $N$ D8& - & - & - & - & -& -& -& -& $\bullet$ & - \\ \hline $N$ $\overline{\text{D8}}$& - & - & - & - & -& -& -& -& $\bullet$ & - \\ \hline
\end{tabular}
\caption{Brane configuration of the magnificent four. The symbol $-$ denotes an extended direction of the D-branes, whereas $\bullet$ denotes a point-like direction. The D8-branes and anti-D8-branes extend along four complex directions $\C_{1,2,3,4}$ and the time direction $x^0$, while the D0-branes, representing instanton probes, extend only along the time direction.}
\label{tab:D8 brane}
\end{table}

For simplicity, we first consider the case without anti-D8 branes. In the presence of a B-field \cite{Witten:2000mf}, the energy spectrum of the D0-D8 system includes four complex adjoint chiral multiplets $B_{1,2,3,4}$ of 1d $\cN=2$ SUSY QM arising from excitations of strings connecting D0-D0, and a fundamental chiral multiplet $I$ arising from excitations of strings connecting D0-D8. The corresponding 1d $\cN=2$ quiver diagram is given in Fig.~\ref{fig:D0D8quiver}.
\begin{figure}
    \centering
    \includegraphics[width=0.25\linewidth]{figures/D0D8quiver.pdf}
    \caption{Quiver diagram of D0-D8 system as $\cN=2$ SUSY QM. The black solid lines represent chiral multiplets, and the red dashed lines represent fermi multiplets. The circular nodes denote gauge groups, while the square nodes denote flavor groups. In this quiver, we have a $\U(k)$ gauge group, a $\U(N)$ flavor group, a fundamental chiral that contains $I$, four adjoint chirals that contain $B_{1,2,3,4}$ respectively, and three adjoint fermis that contain $\Lambda_{1,2,3}$ respectively.}
    \label{fig:D0D8quiver}
\end{figure}

The moduli space corresponding to this quiver provides the ADHM data for this D0-D8 system as:
\begin{align}
    \mathcal{M}^{\text{D0-D8}}_{N,k}=\{(\boldsymbol{B},I)|\mu_{\R}-\zeta\cdot\mathbf{1}_k=\mu_{ab\in\6}=0\}/\U(k)
\end{align}
where the moment maps are defined as:
\begin{align*}
    &\mu_\R=\sum_{a\in\4}[B_a,B_a^\dagger]+I\cdot I^\dagger\cr&\mu_{ab}=[B_a,B_b]
\end{align*}
Similar to the case of pure SYM, where the D0-D4 system is placed in an $\Omega$-background with $\U(1)_{\e_1}\times\U(1)_{\e_2}$ symmetry, we need to place the entire D0-D8 system in a background with $\SU(4)$ symmetry. This is equivalent to considering the system in a spacetime with $\SU(4)$ holonomy \cite{Szabo:2022zyn}. Consequently,  the charges of the various fields under the $\U(1)^3\subset\SU(4)$ transformation are respectively:
\begin{align}
    \left(\begin{array}{c}
        B_1,B_2,B_3,B_4  \\
        I,\Lambda_1,\Lambda_2,\Lambda_3
    \end{array}\right)\xrightarrow{\U(1)_{\e_1}\times\U(1)_{\e_2}\times\U(1)_{\e_3}} \left(\begin{array}{c}
        q_1^{-1}B_1,q_2^{-1}B_2,q_3^{-1}B_3,q_{123}B_4  \\
        I,q_{23}\Lambda_1,q_{13}\Lambda_2,q_{12}\Lambda_3
    \end{array}\right)
\end{align}

The D0-D8 partition function is:
\begin{align}\label{D0D8partition}
    &\mathcal{Z}^{\text{D0-D8}}_{N,k}(v_{\4,1},\ldots,v_{\4,N})=\oint_{\JK} \prod_{i=1}^k\frac{d\phi_i}{2\pi i}\mathcal{I}^{\text{D0-D8}}_{N,k}\cr&\mathcal{I}^{\text{D0-D8}}_{N,k}=\mathcal{I}^{\text{D0-D0}}_k\times\prod_{i=1}^k\prod_{\alpha=1}^N\frac{1}{\sh(\phi_i-v_{\4,\alpha})}\cr&\mathcal{I}^{\text{D0-D0}}_k=\frac{1}{k!}\prod_{i\neq j}^k\sh(\phi_i-\phi_j)\prod_{i,j}^k\frac{\sh(\phi_i-\phi_j-\e_{1,2,3}-\e_4)}{\sh(\phi_i-\phi_j-\e_{1,2,3,4})}
\end{align}
where we impose the CY four-fold condition  $\e_4=-\e_{123}$ .

The poles of the D0-D8 system are classified by $N$ $4$d Young diagrams, and a closed formula follows from the shell formula. Note, however, that this partition function~\eqref{D0D8partition} differs from the expansion of the $\mathcal{J}$-factor~\eqref{J-expand} under 4d Young diagrams $\rho_{\4}$:
\begin{footnotesize}
    \begin{align}
    \prod_{\boldsymbol{x}\in\rho_{\4}}\mathcal{J}\big(\mathcal{X}_{\4}(\boldsymbol{x})\big|\rho_{\4}\big)=&\prod_{\boldsymbol{x}\in\rho_{\4}}\frac{1}{\sh(\mathcal{X}_{\4}(\boldsymbol{x})-\mathcal{X}_{\4}(\boldsymbol{1}))}\cr&\qquad\times\prod_{\boldsymbol{y}\in\rho_{\4}}\frac{\sh(\mathcal{X}_{\4}(\boldsymbol{x})-\mathcal{X}_{\4}(\boldsymbol{y}))\sh(\mathcal{X}_{\4}(\boldsymbol{x})-\mathcal{X}_{\4}(\boldsymbol{y})-\e_{\4})\prod_{ab\in\6}\sh(\mathcal{X}_{\4}(\boldsymbol{x})-\mathcal{X}_{\4}(\boldsymbol{y})-\e_{ab})}{\prod_{a\in\4}\sh(\mathcal{X}_{\4}(\boldsymbol{x})-\mathcal{X}_{\4}(\boldsymbol{y})-\e_{a})\prod_{A\in\check{\4}}\sh(\mathcal{X}_{\4}(\boldsymbol{x})-\mathcal{X}_{\4}(\boldsymbol{y})-\e_{A})}\cr=&\prod_{\boldsymbol{x}\in\rho_{\4}}\frac{1}{\sh(\mathcal{X}_{\4}(\boldsymbol{x})-\mathcal{X}_{\4}(\boldsymbol{1}))}\prod_{\boldsymbol{y}\in\rho_{\4}}\left(\frac{\sh(\mathcal{X}_{\4}(\boldsymbol{x})-\mathcal{X}_{\4}(\boldsymbol{y}))\sh(\mathcal{X}_{\4}(\boldsymbol{x})-\mathcal{X}_{\4}(\boldsymbol{y})-\e_{1,2,3}-\e_4)}{\sh(\mathcal{X}_{\4}(\boldsymbol{x})-\mathcal{X}_{\4}(\boldsymbol{y})-\e_{1,2,3,4})}\right)^2
\end{align}
\end{footnotesize}

This expression equals the square of the D0-D8 integrand, as can be seen by comparison with~\eqref{D0D8partition}.
Therefore, if we use the original definition~\eqref{J-def} to express the D0-D8 partition function, it is necessary to take the square root of the $\mathcal{J}$-factor, and each term in the summation will exhibit an ambiguous sign. Hence, to obtain a canonical sign choice and a well-defined formula, we define a modified $\mathcal{J}$-factor that selects only those shell boxes whose last coordinate satisfies $x_d \geq y_d$, denoted $\mathcal{J}_{\geq}$:
\begin{align}\label{Jgeq}
    \mathcal{J}_{\geq}\big(\mathcal{X}_{\mathcal{B}}(\boldsymbol{x})\big|\rho_{\mathcal{A}}\big)\equiv \prod_{\substack{\boldsymbol{y}\in\mathcal{S}(\rho_{\mathcal{A}})\\x_{d}\geq y_{d}}}\sh\left(\mathcal{X}_{\mathcal{B}}(\boldsymbol{x})-\mathcal{X}_{\mathcal{A}}(\boldsymbol y)\right)^{\operatorname{Q}_{\rho_{\mathcal{A}}}(\boldsymbol{y})}
\end{align}
where we compare the last coordinates $x_d$ and $y_d$ of two boxes and only select the contribution from boxes where $x_d \geq y_d$. This definition selects precisely the shell boxes needed and yields the correct sign. The D0-D8 partition function is therefore:
\begin{align}\label{D8shell}
    \mathcal{Z}^{\text{D0-D8}}_{N,k}(v_{\4,1},\ldots,v_{\4,N})=\sum_{||\vec{\rho}||=k}\mathcal{Z}^{\text{D0-D8}}(\vec{\rho})=\sum_{||\vec{\rho}||=k}\prod_{\alpha,\beta=1}^N\prod_{\boldsymbol{x}\in\rho_{\4,\alpha}}\mathcal{J}_{\geq}\big(\mathcal{X}_{\4,\alpha}(\boldsymbol x)\big|\rho_{\4,\beta}\big)
\end{align}
where $\vec\rho$ denotes an $N$-tuple of 4d Young diagrams with $||\vec{\rho}||=k$. As in Fig.~\ref{fig:5d SYM brane}, the contribution from strings connecting D0-D8 is:
\begin{align}
    \mathcal{J}_{\geq}\big(\mathcal{X}_{\4,\alpha}(\boldsymbol x)\big|\rho_{\4,\beta}\big)
\end{align}

If we consider an equal number of anti-D8-branes, we need to replace the flavor group with $\U(N|N)$ \cite{Vafa:2001qf}. Effectively, this is equivalent to adding an equal number of Fermi multiplets to the SUSY QM; at the level of the partition function, this amounts to including the corresponding Fermi multiplet contributions:
\begin{align}\label{D0D8D8bar partition}
    &\mathcal{I}^{\text{D0-D8-}\overline{\text{D8}}}_{N,k}=\mathcal{I}^{\text{D0-D8}}_{N,k}\times\prod_{i=1}^k\prod_{\alpha=1}^N\sh(\phi_i-w_{\4,\alpha})\cr&\mathcal{Z}^{\text{D0-D8-}\overline{\text{D8}}}_{N,k}(\{v_{\4,\alpha}\},\{w_{\4,\alpha}\})=\sum_{||\vec{\rho}||=k}\prod_{\alpha,\beta=1}^N\prod_{\boldsymbol{x}\in\rho_{\4,\alpha}}\sh(-\mathcal{X}_{\4,\alpha}(\boldsymbol x)+w_{\4,\beta})\mathcal{J}_{\geq}\big(\mathcal{X}_{\4,\alpha}(\boldsymbol x)\big|\rho_{\4,\beta}\big)
\end{align}
The validity of this formula can be verified by computing the plethystic exponent (PE) expression \cite{Nekrasov:2018xsb}:
\begin{align}\label{D0D8 PE}
    \mathcal{Z}^{\text{D0-D8-}\overline{\text{D8}}}_{N}(\{v_{\4,\alpha}\},\{w_{\4,\alpha}\})=\sum_{k=0}^\infty\mathfrak{q}^k\mathcal{Z}^{\text{D0-D8-}\overline{\text{D8}}}_{N,k}=\PE\left(\frac{\sh(\e_{12,13,23})}{\sh(\e_{1,2,3,4})}\frac{\sh(s)}{\sh(p\pm\frac12s)}\right)
\end{align}
where $s\equiv\sum_{i=1}^N(v_{\4,i}-w_{\4,i})$, and $\mathfrak{q}\equiv e^{-p}$. The PE operation is defined as:
\begin{align}
    \PE f(x_1,\ldots,x_r)\equiv\exp\sum_{m=1}^\infty\frac{1}{m}f(m\,x_1,\ldots,m\,x_r)
\end{align}

The recursion relation in the D0-D8 system, analogous to~\eqref{UN recursion}, describes the contribution from adding a 4d box $\hcube$ to a 4d Young diagram $\rho_{\4,\alpha}$:
\begin{align}\label{4d recursion}
    \frac{\mathcal{Z}^{\text{D0-D8}}(\rho_{\4,\alpha}\cup\hcube)}{\mathcal{Z}^{\text{D0-D8}}(\rho_{\4,\alpha})}=\mathcal{J}_{\geq}\big(\mathcal{X}_{\4,\alpha}(\hcube)\big|\rho_{\4,\alpha}\cup\hcube\big)\mathcal{J}_{<}\big(\mathcal{X}_{\4,\alpha}(\hcube)\big|\rho_{\4,\alpha}\big)
\end{align}
Here, $\hcube$ refers to the coordinate of the 4d box being added. The function $\mathcal{J}_<$ is defined analogously to $\mathcal{J}_{\geq}$ in~\eqref{Jgeq}, but with the inequality $x_{4} \geq y_{4}$ replaced by $x_{4} < y_{4}$. And we use the following identities:
\begin{footnotesize}
    \begin{align}\label{J geq recursion}
    &\frac{\mathcal{J}_{\geq}\big(\mathcal{X}_{\beta}(\boldsymbol{x})\big|\rho_{ \alpha}\cup\{\boldsymbol{y}\}_{ \alpha}\big)}{\mathcal{J}_{\geq}\big(\mathcal{X}_{ \beta}(\boldsymbol{x})\big|\rho_{ \alpha}\big)}=\begin{cases}
        \frac{\sh(\mathcal{X}_{ \beta}(\boldsymbol{x})-\mathcal{X}_{ \alpha}(\boldsymbol{y}))\sh(\mathcal{X}_{ \beta}(\boldsymbol{x})-\mathcal{X}_{ \alpha}(\boldsymbol{y})-\e_{1234})\prod_{ab\in\6}\sh(\mathcal{X}_{ \beta}(\boldsymbol{x})-\mathcal{X}_{ \alpha}(\boldsymbol{y})-\e_{ab})}{\sh(\mathcal{X}_{ \beta}(\boldsymbol{x})-\mathcal{X}_{ \alpha}(\boldsymbol{y})-\e_{1,2,3,4})\prod_{A\in\6}\sh(\mathcal{X}_{ \beta}(\boldsymbol{x})-\mathcal{X}_{ \alpha}(\boldsymbol{y})-\e_A)}, &x_4>y_4\\ \frac{\sh(\mathcal{X}_{ \beta}(\boldsymbol{x})-\mathcal{X}_{ \alpha}(\boldsymbol{y}))\sh(\mathcal{X}_{ \beta}(\boldsymbol{x})-\mathcal{X}_{ \alpha}(\boldsymbol{y})-\e_{14,24,34})}{\sh(\mathcal{X}_{ \beta}(\boldsymbol{x})-\mathcal{X}_{ \alpha}(\boldsymbol{y})-\e_{1,2,3,4})},&x_4=y_4\\1,&x_4<y_4
    \end{cases}
    \cr &\mathcal{J}_{<}\big(\mathcal{X}_{\beta}(\boldsymbol{x})\big|\rho_{ \alpha}\big)=\frac{\mathcal{J}\big(\mathcal{X}_{\beta}(\boldsymbol{x})\big|\rho_{ \alpha}\big)}{\mathcal{J}_{\geq}\big(\mathcal{X}_{\beta}(\boldsymbol{x})\big|\rho_{ \alpha}\big)}
\end{align}
\end{footnotesize}

To illustrate its validity, we provide examples in the Appendix~\ref{D0D8 appendix}.

\subsection{Tetrahedron instanton}\label{Tetrahedron instanton}
We now consider a system whose fixed points are classified by 3d Young diagrams: the D0-D6 system, also known as tetrahedron instantons. This system was first studied in detail by Pomoni, Yan, and Zhang \cite{Pomoni:2021hkn, Pomoni:2023nlf} in type IIB string theory using a D1-D7 system. After T-dualizing along $x^9$, the corresponding brane configuration is listed in Tab.~\ref{tab:D6 brane}.
\begin{table}[ht]
\centering
\begin{tabular}{|c|c|c|c|c|c|c|c|c|c|c|}
    \hline & \multicolumn{2}{|c|}{$\mathbb{C}_{1}$} & \multicolumn{2}{|c|}{$\mathbb{C}_{2}$} & \multicolumn{2}{c|}{$\mathbb{C}_{3}$} & \multicolumn{2}{|c|}{$\mathbb{C}_{4}$} & \multicolumn{2}{|l|}{$\mathbb{R} \times \mathbb{S}^1$}  \\
\cline { 2 - 11 }& 1 & 2 & 3 & 4 & 5 & 6 & 7 & 8 & 9 & 0 \\
\hline $k$ D0& $\bullet$ & $\bullet$ & $\bullet$ & $\bullet$ & $\bullet$ & $\bullet$ & $\bullet$ & $\bullet$ & $\bullet$ & - \\ \hline $N_{\overline{4}}$ D6$_{\overline{4}}$& - & - & - & - & -& -& $\bullet$& $\bullet$& $\bullet$ & - \\ \hline $N_{\overline{3}}$ D6$_{\overline{3}}$& - & - & - & - & $\bullet$& $\bullet$& -& -& $\bullet$ & -\\ \hline $N_{\overline{2}}$ D6$_{\overline{2}}$& - & - & $\bullet$ & $\bullet$ & -& -& -& -& $\bullet$ & -\\ \hline $N_{\overline{1}}$ D6$_{\overline{1}}$& $\bullet$ & $\bullet$ & - & - & -& -& -& -& $\bullet$ & -\\ \hline
\end{tabular}
\caption{Brane configuration of tetrahedron instantons. $-$ represents the direction along which the D-branes extend, while $\bullet$ represents the point-like directions of the D-branes. $\overline{a}$ refers to the complement of $a$ in $\{1,2,3,4\}$. For example, the D6$_{\overline{4}}$ brane refers to D6$_{123}$, i.e., the D6-brane extending along $\C_1$, $\C_2$, $\C_3$ and the time direction. The D0-brane extends only along the time direction. Here, we will need $N_{\overline{a}}$ of D6$_{\overline{a}}$ branes respectively.}
\label{tab:D6 brane}
\end{table}
The ADHM data of this system can be represented by an $\cN=2$ SUSY QM quiver diagram. It includes all adjoint fields $B_{1,2,3,4}$ from the magnificent four in Fig.~\ref{fig:D0D8quiver} as well as $\Lambda_{1,2,3}$. It also contains four distinct fundamental chiral $I_{\overline{4},\overline{3},\overline{2},\overline{1}}$ and Fermi multiplets $\Lambda_{\overline{4},\overline{3},\overline{2},\overline{1}}$, corresponding to the four different D6-branes. The transformations of these fields under the $\U(1)^3$ symmetries are listed as 
\begin{align}
    \left(\begin{array}{c}
        I_{\overline{4}},I_{\overline{3}},I_{\overline{2}},I_{\overline{1}}  \\
        \Lambda_{\overline{1}},\Lambda_{\overline{2}},\Lambda_{\overline{3}},\Lambda_{\overline{4}}
    \end{array}\right)\xrightarrow{\U(1)_{\e_1}\times\U(1)_{\e_2}\times\U(1)_{\e_3}} \left(\begin{array}{c}
        I_{\overline{4}},I_{\overline{3}},I_{\overline{2}},I_{\overline{1}}  \\
        q_1^{-1}\Lambda_{\overline{1}},q_2^{-1}\Lambda_{\overline{2}},q_3^{-1}\Lambda_{\overline{3}},q_{123}\Lambda_{\overline{4}}
    \end{array}\right)
\end{align}

\begin{figure}[ht]
    \centering
    \includegraphics[width=0.5\linewidth]{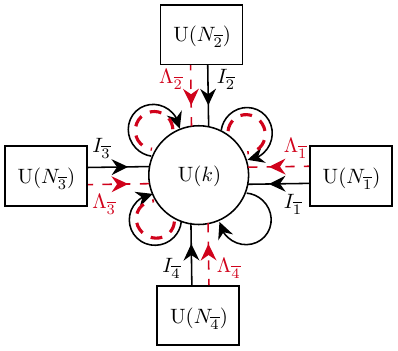}
    \caption{Quiver diagram of D0-D6 system as $\cN=2$ SUSY QM. The black solid lines represent chiral multiplets, and the red dashed lines represent fermi multiplets. The circular nodes denote gauge groups, while the square nodes denote flavor groups. In this quiver, we have the $\U(k)$ gauge group, four flavor groups $\U(N_{\overline1,\overline2,\overline3,\overline4})$, four types of fundamental chiral multiplets $I_{\overline1,\overline2,\overline3,\overline4}$, four types of fundamental fermi multiplets $\Lambda_{\overline1,\overline2,\overline3,\overline4}$, four adjoint chiral multiplets $B_{1,2,3,4}$, and three adjoint fermi multiplets $\Lambda_{1,2,3}$.}
    \label{fig:D0D6quiver}
\end{figure}

The ADHM equations of tetrahedron instantons are:
\begin{align}
    \mathcal{M}^{\text{D0-D6}}_{\boldsymbol{N},k}=\{(\boldsymbol{B},\boldsymbol{I})|\mu_{\R}-\zeta\cdot\mathbf{1}_k=\mu_{ab\in\6}=\sigma_{\overline{a}}=0\}/\U(k)
\end{align}
where the moment maps are defined as:
\begin{align}
    &\mu_\R=\sum_{a\in\4}[B_a,B_a^\dagger]+I_{\overline{a}}I_{\overline{a}}^\dagger\cr&\mu_{ab}=[B_a,B_b]\cr&\sigma_{\overline{a}}=B_a I_{\overline{a}}\ \quad\text{(superpotential F-term coupling the adjoint and fundamental fields)}
\end{align}
Thus, the partition function is expressed as:
\begin{align}\label{D0D6partition}
    \mathcal{Z}_{\boldsymbol{N},k}^{\text{D0-D6}}=\oint_{\JK} \prod_{i=1}^k\frac{d\phi_i}{2\pi i}\mathcal{I}^{\text{D0-D6}}_{\boldsymbol{N},k},\qquad
    \mathcal{I}^{\text{D0-D6}}_{\boldsymbol{N},k}=\mathcal{I}^{\text{D0-D0}}_k\times\prod_{i=1}^k\prod_{\mathcal{A}}\frac{\sh(\phi_i-v_{\mathcal{A}}+\e_{\mathcal{A}})}{\sh(\phi_i-v_{\mathcal{A}})}
\end{align}
where $\mathcal{A}=(A,\alpha)\in\{(\overline{4},1),\ldots,(\overline{4},N_{\overline{4}}),(\overline{3},1),\ldots,(\overline{1},N_{\overline{1}})\}$ label each individual D6-brane, and $\boldsymbol{N}=(N_{\overline{4}},N_{\overline{3}},N_{\overline{2}},N_{\overline{1}})$ denote the numbers of D6-branes in the four distinct orientations. $\mathcal{I}^{\text{D0-D0}}_k$ is the contribution from the D0-D0 strings in \eqref{D0D8partition}. After performing the JK-residue integral, the poles of this system are classified by a set $\{\pi_\mathcal{A}\}$ of 3d Young diagrams in the four different directions. The shell formula then gives:
\begin{align}\label{D0D6shell}
    \mathcal{Z}^{\text{D0-D6}}_{\boldsymbol{N},k}=&\sum_{||\vec{\pi}||=k}\Bigg(\prod_{\mathcal{A},\mathcal{B}}\prod_{\boldsymbol{x}\in\pi_{\mathcal{A}}}\sh(\mathcal{X}_{\mathcal{A}}(\boldsymbol{x})-\mathcal{X}_{\mathcal{B}}(\boldsymbol{0}))\mathcal{J}\big(\mathcal{X}_{\mathcal{A}}(\boldsymbol{x})\big|\pi_\mathcal{B}\big)\Bigg)\cr
    &\times\Bigg(\prod_{\mathcal{A},\mathcal{B}}\prod_{\substack{\boldsymbol{x}\in\pi_{\mathcal{A}}\\\boldsymbol{y}\in\pi_{\mathcal{B}}}}\frac{\sh(\mathcal{X}_{\mathcal{A}}(\boldsymbol{x})-\mathcal{X}_{\mathcal{B}}(\boldsymbol{y})+\e_B)}{\sh(\mathcal{X}_{\mathcal{A}}(\boldsymbol{x})-\mathcal{X}_{\mathcal{B}}(\boldsymbol{y})+\e_A)} \Bigg)\Bigg(\prod_{\substack{\mathcal{A},\mathcal{B}\\A<B}}\prod_{\substack{\boldsymbol{x}\in\pi_{\mathcal{A}}\\\boldsymbol{y}\in\pi_{\mathcal{B}}}}\prod_{\substack{ab\in\6\\\in A\\\notin B}}\frac{\sh(\mathcal{X}_{\mathcal{A}}(\boldsymbol{x})-\mathcal{X}_{\mathcal{B}}(\boldsymbol{y})+\e_{ab} )
}{\sh(\mathcal{X}_{\mathcal{A}}(\boldsymbol{x})-\mathcal{X}_{\mathcal{B}}(\boldsymbol{y})-\e_{ab} )
}\Bigg)\cr
\end{align}
where the order $A < B$ is defined by the canonical ordering $123 < 124 < 134 < 234$. The second factor in~\eqref{D0D6shell} can be interpreted as an additional contribution arising from the interaction between 3d Young diagrams in different directions.

The recursion relation~\eqref{recursion} directly gives the contribution from adding a D0-brane at location $\cube$:

\begin{align}\label{D6recursion}
    \frac{\mathcal{Z}^{\text{D0-D6}}(\pi_\mathcal{A}\cup\cube)}{\mathcal{Z}^{\text{D0-D6}}(\pi_\mathcal{A})}=\frac{\mathcal{J}\big(\mathcal{X}_\mathcal{A}(\cube)\big|\pi_\mathcal{A}\cup\cube\big)}{\mathcal{J}\big(\mathcal{X}_\mathcal{A}(\cube+\boldsymbol{1})\big|\pi_\mathcal{A}\big)}
\end{align}
Detailed calculations for several illustrative examples are collected in Appendix~\ref{D0-D6 appendix}.

The similarity between the D0-D6 system~\eqref{D0D6partition} and the D0-D8-$\overline{\text{D8}}$ system~\eqref{D0D8D8bar partition} is visible at the level of the integrand. Indeed, D6-branes arise from tachyon condensation between D8 and $\overline{\text{D8}}$ branes \cite{Nekrasov:2017cih, Berkovits:2000hf, Akhmedov:2000zp}. For instance, considering a system with only one pair of D8-$\overline{\text{D8}}$-branes, its integrand is given by:
\begin{align}
    \mathcal{I}^{\text{D0-D8-}\overline{\text{D8}}}_{1,k}=\mathcal{I}^{\text{D0-D0}}_k\times\prod_{i=1}^k\frac{\sh(\phi_i-w_{\4,1})}{\sh(\phi_i-v_{\4,1})}
\end{align}

\begin{figure}[ht]
    \centering
    \includegraphics[width=0.8\linewidth]{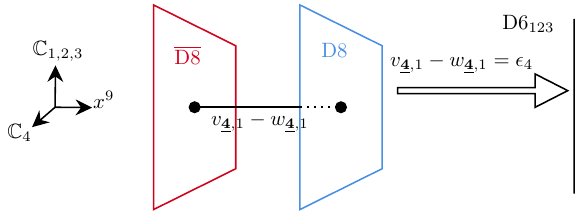}
    \caption{The D8-$\overline{\text{D8}}$ pair annihilates into a D6-brane through tachyon condensation. In the figure, the red brane represents the anti-D8-brane, with $w_{\4,1}$ being its corresponding Coulomb branch parameter. The blue brane represents the D8-brane, with $v_{\4,1}$ being its corresponding Coulomb branch parameter. When we adjust the moduli space coordinate to the place $v_{\4,1} - w_{\4,1} = \e_{4}$, the D8-$\overline{\text{D8}}$ pair becomes equivalent to a single D6-brane oriented along $\C_{\bar{4}}$.}
    \label{fig:D8toD6}
\end{figure}

Once we identify $v_{\4,1}=v_{123,1}$ and $w_{\4,1}=v_{123,1}-\e_{123}$, this integrand is identical to that of a single D6-brane. Setting $w_{\4,1}=v_{123,1}-\e_{123}$ effectively brings the D8 and $\overline{\text{D8}}$-branes together. As shown in Fig.~\ref{fig:D8toD6}, the open strings connecting them develop a negative mass (tachyon) state, which renders the system unstable and causes it to decay into a single D6-brane.

From the perspective of the shell formula, if the corresponding 4d Young diagram contains the box located at $(1, 1, 1, 2)$, then the contribution of the $\overline{\text{D8}}$-brane becomes $\sh(\e_{1234})$ under the condition $w_{\4,1}=v_{123,1}-\e_{123}$ and $v_{\4,1}=v_{123,1}$. Under the CY4 condition $\e_{1234}=0$, this term vanishes. This implies that the $\overline{\text{D8}}$-brane contribution selects only those 4d Young diagrams that have exactly one layer in the fourth direction, i.e., 3d Young diagrams. In this case, the condition $x_4 \geq y_4$ in the $\mathcal{J}_{\geq}$-factor~\eqref{Jgeq} becomes trivial.

It follows that the PE expression of the tetrahedron instanton \cite{Nekrasov:2018xsb, Pomoni:2023nlf}, derivable from that of the magnificent four~\eqref{D0D8 PE}, is manifestly independent of all the Coulomb branch parameters $\{v_{\mathcal{A}}\}$:
\begin{align}
    \mathcal{Z}^{\text{D0-D6}}_{\boldsymbol{N}}(\{v_{\mathcal{A}}\})=\sum_{k=0}^\infty\mathfrak{q}^k\mathcal{Z}^{\text{D0-D6}}_{\boldsymbol{N},k}=\PE\left(\frac{\sh(\e_{12,13,23})}{\sh(\e_{1,2,3,4})}\frac{\sh(s)}{\sh(p\pm\frac12s)}\right)
\end{align}
where the parameter $s\equiv\sum_{i=1}^N(v_{\4,i}-w_{\4,i})$ after tachyon condensation becomes:
\begin{align}
    s=&N_{\overline{4}}\e_{123}+N_{\overline{3}}\e_{124}+N_{\overline{2}}\e_{134}+N_{\overline{1}}\e_{234}\cr=&(N_{123}-N_{234})\e_1+(N_{123}-N_{134})\e_2+(N_{123}-N_{124})\e_3
\end{align}

We briefly discuss the generalized tetrahedron instanton, namely the theory with the addition of anti-D6 branes \cite{Kimura:2023bxy, Kimura:2024osv}. Similar to the anti-D8 case, the $\overline{\text{D6}}$-brane contribution to the index is the reciprocal of that of a D6-brane. The partition function is:
\begin{align}\label{D6D6bar partition}
    \mathcal{I}^{\text{D0-D6-}\overline{\text{D6}}}_{\boldsymbol{N},\boldsymbol{M},k}=\mathcal{I}^{\text{D0-D6}}_{\boldsymbol{N},k}\times\prod_{i=1}^k\prod_{\mathcal{B}}\frac{\sh(-\phi_i+w_{\mathcal{B}})}{\sh(-\phi_i+w_{\mathcal{B}}-\e_\mathcal{B})}
\end{align}
where $w_\mathcal{B}$ is the fugacity for the $\overline{\text{D6}}$-brane with $\mathcal{B}=(B,\beta)\in\{(\overline{4},1),\ldots,(\overline{4},M_{\overline{4}}),(\overline{3},1),\ldots,(\overline{1},M_{\overline{1}})\}$ label each $\overline{\text{D6}}$-brane.

\subsection{Spiked instanton}\label{Spiked instanton}
\begin{figure}[ht]
    \centering
    \includegraphics[width=0.45\linewidth]{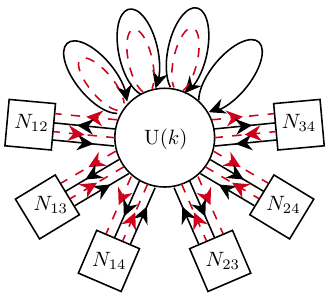}
    \caption{Quiver diagram of D0-D4 system as $\cN=2$ SUSY QM. In this quiver, we have the $\U(k)$ gauge group, six flavor groups $\U(N_{ab\in\6})$, six types of fundamental chiral multiplets that contain $I_{ab\in\6}$, six types of anti-fundamental chirals that contain $J_{ab\in\6}$, twelve types of fermi multiplets $\Lambda_{I_{ab}},\Lambda_{J_{ab}}$ correspond to $I_{ab}$ and $J_{ab}$ respectively, four adjoint chiral multiplets $B_{1,2,3,4}$, and three adjoint fermi multiplets $\Lambda_{1,2,3}$.}
    \label{fig:D0D4quiver}
\end{figure}

Spiked instantons arise from D0-branes bound to D4-branes with multiple orientations \cite{Nekrasov:2016gud, Nekrasov:2015wsu, Nekrasov:2016qym}. They can be obtained from the 5d $\cN=1$ theory of Sec.~\ref{5dSYM U} by incorporating D4-branes with different orientations and turning on adjoint multiplet masses, or alternatively from the tetrahedron instanton of Sec.~\ref{Tetrahedron instanton} by introducing an equal number of anti-D6-branes through tachyon condensation. The corresponding brane configuration involves six types of D4-branes with distinct orientations, as shown in Tab.~\ref{D0D4brane}. In the low-energy regime, D0-instantons attach to any of these D4-branes in the form of 2d Young diagrams.
\begin{table}[ht]\centering
\begin{tabular}{|c|c|c|c|c|c|c|c|c|c|c|}
\hline & \multicolumn{2}{|c|}{$\C_1$} & \multicolumn{2}{|c|}{$\C_2$} & \multicolumn{2}{c|}{$\C_3$} & \multicolumn{2}{|c|}{$\C_4$} & \multicolumn{2}{|l|}{$\mathbb{R} \times \mathbb{S}^1$}  \\
\cline { 2 - 11 } & 1 & 2 & 3 & 4 & 5 & 6 & 7 & 8 & 9 & 0 \\
\hline D0 & $\bullet$ & $\bullet$ & $\bullet$ & $\bullet$ & $\bullet$ & $\bullet$ & $\bullet$ & $\bullet$ & $\bullet$ & - \\
\hline $\mathrm{D} 4_{12}$ & - & - & - & - & $\bullet$ & $\bullet$ & $\bullet$ & $\bullet$ & $\bullet$ & - \\
\hline${\mathrm{D} 4_{13}}$ & - & - & $\bullet$ & $\bullet$ & - & - & $\bullet$ & $\bullet$ & $\bullet$ & - \\
\hline $\mathrm{D} 4_{14}$ & - & - & $\bullet$ & $\bullet$ & $\bullet$ & $\bullet$ & - & - & $\bullet$ & - \\
\hline $\mathrm{D} 4_{23}$ & $\bullet$ & $\bullet$ & - & - & - & - & $\bullet$ & $\bullet$ & $\bullet$ & - \\
\hline $\mathrm{D} 4_{24}$ & $\bullet$ & $\bullet$ & - & - & $\bullet$ & $\bullet$ & - & - & $\bullet$ & - \\
\hline $\mathrm{D} 4_{34}$ & $\bullet$ & $\bullet$ & $\bullet$ & $\bullet$ & - & - & - & - & $\bullet$ & - \\
\hline
\end{tabular}
\caption{Brane configuration of spiked instantons. $-$ represents the direction along which the D-branes extend, while $\bullet$ represents the point-like directions of the D-branes. In the spiked instanton, we consider six types of D4-branes, where the D4$_{ab\in\6}$-branes extend along $\C_{ab}$ and the time direction $x^0$.}
\label{D0D4brane}
\end{table}
With the quiver diagram as $\cN=2$ SUSY QM as in Fig.~\ref{fig:D0D4quiver}, the Instanton moduli space is defined as:
\begin{align}
     \mathcal{M}^{\text{D0-D4}}_{\boldsymbol{N},k}=\{(\boldsymbol{B},\boldsymbol{I},\boldsymbol{J})|\mu_{\R}-\zeta\cdot\mathbf{1}_k=\mu_{ab}=\sigma_{a;bc}=\tilde{\sigma}_{a;bc}=0\}/\U(k)
\end{align}
where $a\in\4$, $ab,bc\in\6$. The moment maps in the spiked instantons cases are modified as:
\begin{align}
    &\mu_\R=\sum_{a\in\4}[B_a,B_a^\dagger]+\sum_{ab\in\6}(I_{ab}I^\dagger_{ab}-J_{ab}^\dagger J_{ab})\cr&\mu_{ab}=[B_a,B_b]+I_{ab}J_{ab}\cr&\sigma_{a;bc}=B_aI_{bc}\cr&\tilde{\sigma}_{a;bc}=J_{bc}B_a
\end{align}

The spiked instanton partition function is:
\begin{align}
    \mathcal{Z}_{\boldsymbol{N},k}^{\text{D0-D4}}=&\oint_{\JK} \prod_{i=1}^k\frac{d\phi_i}{2\pi i}\mathcal{I}^{\text{D0-D4}}_{\boldsymbol{N},k}\cr
    \mathcal{I}^{\text{D0-D4}}_{\boldsymbol{N},k}=&\mathcal{I}^{\text{D0-D0}}_k\times\prod_{i=1}^k\prod_{\mathbf{ab}}\frac{\prod_{c\in\overline{ab}}\sh(\phi_i-v_{\mathbf{ab}}-\e_{c})}{\sh(\phi_i-v_{\mathbf{ab}})\sh(-\phi_i+v_{\mathbf{ab}}-\e_{ab})}
\end{align}
where $\mathbf{ab}=(ab,\alpha)\in\{(12,1),\ldots,(12,N_{12}),(13,1),\ldots,(34,N_{34})\}$ label each D4 brane, $\overline{ab}$ denotes the complement of $ab$ within $\4$, and $\boldsymbol{N} = (N_{12}, \dots, N_{34})$ corresponds to the numbers of D4-branes with different orientations. The poles of this residue integral are classified by a set of 2d Young diagrams, and the shell formula gives:

\begin{align}
    \mathcal{Z}_{\boldsymbol{N},k}^{\textrm{D4},\bC^4}=&\sum_{||\vec{\lambda}||=k}(-1)^{k}\prod_{\mathbf{ab},\mathbf{ab}'}\Bigg(\prod_{\boldsymbol{x}\in\lambda_{\mathbf{ab}}}\frac{\mathcal{J}\big(\mathcal{X}_{\mathbf{ab}}(\boldsymbol{x})\big|\lambda_{\mathbf{ab'}}\big)}{\sh(-\mathcal{X}_{\mathbf{ab}}(\boldsymbol{x})+\mathcal{X}_{\mathbf{ab}'}(\boldsymbol{0}))}\prod_{c\in\overline{ab}}\sh(\mathcal{X}_{\mathbf{ab}}(\boldsymbol{x})-\mathcal{X}_{\mathbf{ab}'}(\boldsymbol{0})+\e_c) \Bigg)\cr&\qquad\times\left(\prod_{\substack{\boldsymbol{x}\in\lambda_{\mathbf{ab}}\\\boldsymbol{y}\in\lambda_{\mathbf{ab}'}}}\frac{\sh(\mathcal{X}_{\mathbf{ab}}(\boldsymbol{x})-\mathcal{X}_{\mathbf{ab}'}(\boldsymbol{y})-\e_{1,2,3}-\e_4)}{\sh(\mathcal{X}_{\mathbf{ab}}(\boldsymbol{x}+\boldsymbol{1})-\mathcal{X}_{\mathbf{ab}'}(\boldsymbol{y}))~\prod_{c\in \overline{ab}}\sh(\mathcal{X}_{\mathbf{ab}}(\boldsymbol{x})-\mathcal{X}_{\mathbf{ab}'}(\boldsymbol{y})+\e_{c})}\right)
\end{align}

From the perspective of tachyon condensation, this spiked instanton system can arise from a D6-anti-D6 system \cite{Kimura:2024osv, Kimura:2023bxy}. For instance, on the integrand level, a D4$_{12}$-brane can be obtained from a D6$_{123}$-$\overline{\text{D6}}_{123}$ system by taking $v_{123,1}= v_{12,1}$, $w_{123,1}=v_{12,1}+\e_3$, or from a D6$_{124}$-$\overline{\text{D6}}_{124}$ system by taking $v_{123,1}= v_{12,1}$, $w_{123,1}=v_{12,1}+\e_4$ as in \eqref{D6D6bar partition}. Furthermore, if we consider only a single type of D4-brane, the resulting theory is precisely the 5d $\U(N)$ SYM theory with an additional adjoint hypermultiplet, where $\e_3$ can be interpreted as the mass fugacity for the adjoint hypermultiplet \cite{Kim:2024vci}.

\subsection{Donaldson-Thomas 3 counting}\label{DT3 section}

Another natural application of the shell formula is DT3 counting \cite{Thomas:1998uj,Kimura:2023bxy,Kimura:2025lfo,Kimura:2024osv,Nekrasov:2014nea,Galakhov:2021xum}. Mathematically, the DT3 invariants measure the virtual Euler characteristics of moduli spaces of ideal sheaves on a CY threefold, or equivalently, the virtual counts of curve and point subschemes. Physically, they enumerate bound states of D0-D2-D6 branes. Since a single D2-brane corresponds to an infinitely long 3d Young diagram composed of individual boxes, the $\mathcal{J}$-factor provides a natural tool for computing the partition function of the D0-D2-D6 system.

We consider the simplest DT3 invariant, corresponding to the system placed on $\C^3$. The brane construction data are summarized in Tab.~\ref{tab:D6-D2 brane}.
\begin{table}[ht]
\centering
\begin{tabular}{|c|c|c|c|c|c|c|c|c|c|c|}
    \hline & \multicolumn{2}{|c|}{$\mathbb{C}_{1}$} & \multicolumn{2}{|c|}{$\mathbb{C}_{2}$} & \multicolumn{2}{c|}{$\mathbb{C}_{3}$} & \multicolumn{2}{|c|}{$\mathbb{C}_{4}$} & \multicolumn{2}{|l|}{$\mathbb{R} \times \mathbb{S}^1$}  \\
\cline { 2 - 11 }& 1 & 2 & 3 & 4 & 5 & 6 & 7 & 8 & 9 & 0 \\
\hline $k$ D0& $\bullet$ & $\bullet$ & $\bullet$ & $\bullet$ & $\bullet$ & $\bullet$ & $\bullet$ & $\bullet$ & $\bullet$ & - \\ \hline $1$ D6$_{123}$& - & - & - & - & -& -& $\bullet$& $\bullet$& $\bullet$ & - \\ \hline $N_{1}$ D2$_{1}$& - & - & $\bullet$ & $\bullet$ & $\bullet$& $\bullet$& $\bullet$& $\bullet$& $\bullet$ & -\\ \hline $N_{2}$ D2$_{2}$& $\bullet$ & $\bullet$ & - & - & $\bullet$& $\bullet$& $\bullet$& $\bullet$& $\bullet$ & -\\ \hline $N_{3}$ D2$_{3}$& $\bullet$ & $\bullet$ & $\bullet$ & $\bullet$ & -& -& $\bullet$& $\bullet$& $\bullet$ & -\\ \hline
\end{tabular}
\caption{Brane configuration for DT3 counting. The symbol $-$ denotes an extended worldvolume direction, while $\bullet$ denotes a transverse (point-like) direction. Each D2-brane extends along one of the complex directions $\C_{1,2,3}$ and the time direction $x^0$. We consider a single D6-brane extending over $\C^{3}_{123}$.}
\label{tab:D6-D2 brane}
\end{table}

\begin{figure}[ht]
    \centering
    \includegraphics[width=0.4\linewidth]{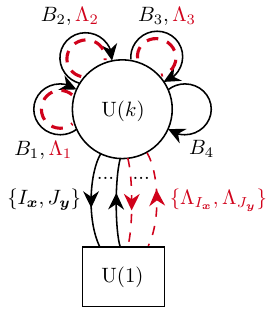}
    \caption{Quiver diagram of the D0-D2-D6 system as $\cN=2$ SUSY quantum mechanics. Solid black lines represent chiral multiplets: the adjoint chirals $B_{1,2,3,4}$, fundamental chirals $\{I_{\boldsymbol{x}}\}$, and anti-fundamental chirals $\{J_{\boldsymbol{y}}\}$. Red dashed lines denote Fermi multiplets: the adjoint fermis $\Lambda_{1,2,3}$, fundamental fermis $\{\Lambda_{I_{\boldsymbol{x}}}\}$, and anti-fundamental fermis $\{\Lambda_{J_{\boldsymbol{y}}}\}$. The multiplicities of $\{I_{\boldsymbol{x}}\}$, $\{J_{\boldsymbol{y}}\}$, $\{\Lambda_{I_{\boldsymbol{x}}}\}$, and $\{\Lambda_{J_{\boldsymbol{y}}}\}$, together with their $\U(1)^3$ charges, are determined by the vacuum configuration of the D2-branes.}
    \label{fig:D0D2D6quiver}
\end{figure}

The D6-brane extends over the full CY threefold $\C^3_{123}$, with D2-branes wrapping curve subschemes of $\C^3_{123}$. Fixing a stable D2-brane configuration (the minimal plane partition, or vacuum) reduces the problem to counting bound states on the worldvolume of the $k$ D0-branes—equivalently, counting placements of $k$ boxes that extend the 3d Young diagram. The corresponding D0-D2-D6 framed quiver, shown in Fig.~\ref{fig:D0D2D6quiver}, is constructed following \cite{Galakhov:2021xum}.

\begin{figure}[ht]
    \centering
    \includegraphics[width=1\linewidth]{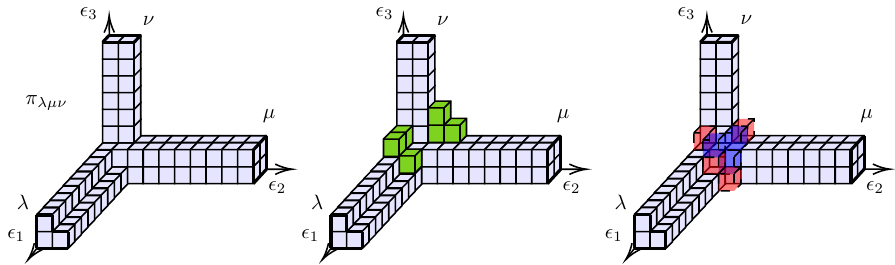}
    \caption{Left panel: the minimal 3d Young diagram $\pi_{\lambda\mu\nu}$ (minimal plane partition) for given asymptotic boundary conditions $(\lambda, \mu, \nu)$. The boundary condition in the 1-direction is the 2d Young diagram $\lambda$ on the 23-plane, and so forth. Center panel: a representative DT3 configuration above the vacuum $\pi_{\lambda\mu\nu}$; green boxes correspond to D0-branes, and their total number equals the instanton number $k$. The placement rule requires that the full assembly—vacuum plus green boxes—forms a valid 3d Young diagram. Right panel: the shell boxes with non-zero charge for the vacuum $\pi_{\lambda\mu\nu}$; red boxes carry charge $-1$ and blue boxes carry charge $+1$.}
    \label{fig:D0D2D6 3d YD}
\end{figure}

The vacuum configuration of the D2-branes is identified with the minimal infinite 3d Young diagram $\pi_{\lambda\mu\nu}$ (minimal plane partition) admitting three prescribed asymptotic boundary conditions $\lambda$, $\mu$, $\nu$ satisfying $|\lambda|=N_1$, $|\mu|=N_2$, and $|\nu|=N_3$. As illustrated in Fig.~\ref{fig:D0D2D6 3d YD}, the three asymptotic planes at infinity are precisely the 2d Young diagrams $\lambda$, $\mu$, and $\nu$. We refer to the configuration as the 3-leg case when all three boundaries are non-empty, the 2-leg case when exactly two are non-empty, and the 1-leg case when only one is non-empty.

Given a vacuum configuration, the multiplicities and $\U(1)^3$ charges of the fields $\{I_{\boldsymbol{x}}\}$, $\{J_{\boldsymbol{y}}\}$, $\{\Lambda_{I_{\boldsymbol{x}}}\}$, and $\{\Lambda_{J_{\boldsymbol{y}}}\}$ are determined via the framed quiver and its superpotential \cite{Galakhov:2021xum,Kimura:2025lfo} (the construction of the framed quiver and superpotential is omitted here). This in turn yields the integral expression for the partition function.

By applying the shell formula together with the recursion relation~\eqref{D6recursion} for 3d Young diagrams, the partition function is obtained directly from the minimal plane partition $\pi_{\lambda\mu\nu}$:
\begin{align}\label{D0D2D6 partition}
    \mathcal{I}^{\text{D0-D2-D6}}_{\lambda,\mu,\nu;k}=\mathcal{I}^{\text{D0-D0}}_k\times\prod_{i=1}^k\frac{\mathcal{J}\big(\phi_i\big|\pi_{\lambda\mu\nu}\big)}{\mathcal{J}_-\big(\phi_i+\e_{123}\big|\pi_{\lambda\mu\nu}\big)},
\end{align}
where the D6-brane label $(123,1)$ and the Coulomb branch parameter $v$ are suppressed for brevity. To reproduce the correct $\U(1)^3$ charges and multiplet content, the signs of the terms in the denominator $\mathcal{J}$-factor must be flipped via $\sh(x) \to -\sh(-x)$. Accordingly, $\mathcal{J}_-$ is defined by:
\begin{align}
    \mathcal{J}_-\big(x\big|\mathbf{Y}_{\mathcal{A}}\big)&\equiv \prod_{\boldsymbol{y}\in\mathcal{S}(\mathbf{Y}_{\mathcal{A}})}-\sh\left(\mathcal{X}_{\mathcal{A}}(\boldsymbol y)-x\right)^{\operatorname{Q}_{\mathbf{Y}_{\mathcal{A}}}(\boldsymbol{y})},\cr\mathcal{J}_-\big(x\big|\emptyset_{\mathcal{A}}\big)&\equiv \frac{-1}{\sh(\mathcal{X}_\mathcal{A}(\boldsymbol{1})-x)}.
\end{align}

\begin{figure}[ht]
    \centering
    \includegraphics[width=1\linewidth]{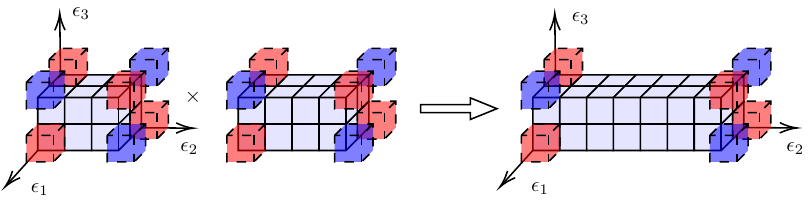}
    \caption{When two long legs with identical 2d Young diagram cross-sections are glued together, non-trivial shell boxes appear only at the two ends of the combined leg. In the figure, the cross-section is $\mu=\{(1,1),(1,2),(2,1),(2,2)\}$. Accordingly, for the infinitely long leg $\pi_{\emptyset\mu\emptyset}$, the non-trivial shell boxes are likewise confined to the finite end and the end at infinity. The contribution of the shell boxes at the infinite end to the $\mathcal{J}$-factor is exactly $1$ and may therefore be discarded.}
    \label{fig:1leg gluing}
\end{figure}

The $\mathcal{J}$-factor for $\pi_{\lambda\mu\nu}$ is computed via the following inclusion-exclusion relation:
\begin{align}\label{J 3leg}
    \mathcal{J}\big(x\big|\pi_{\lambda\mu\nu}\big)=\frac{\mathcal{J}\big(x\big|\pi_{\lambda\emptyset\emptyset}\big)\,\mathcal{J}\big(x\big|\pi_{\emptyset\mu\emptyset}\big)\,\mathcal{J}\big(x\big|\pi_{\emptyset\emptyset\nu}\big)\,\mathcal{J}\big(x\big|\pi_{\lambda\emptyset\emptyset}\cap\pi_{\emptyset\mu\emptyset}\cap\pi_{\emptyset\emptyset\nu}\big)}{\mathcal{J}\big(x\big|\pi_{\lambda\emptyset\emptyset}\cap\pi_{\emptyset\mu\emptyset}\big)\,\mathcal{J}\big(x\big|\pi_{\lambda\emptyset\emptyset}\cap\pi_{\emptyset\emptyset\nu}\big)\,\mathcal{J}\big(x\big|\pi_{\emptyset\mu\emptyset}\cap\pi_{\emptyset\emptyset\nu}\big)},
\end{align}
where the pairwise and triple intersections $\pi_{\lambda\emptyset\emptyset}\cap\pi_{\emptyset\mu\emptyset}$, $\pi_{\lambda\emptyset\emptyset}\cap\pi_{\emptyset\emptyset\nu}$, $\pi_{\emptyset\mu\emptyset}\cap\pi_{\emptyset\emptyset\nu}$, and $\pi_{\lambda\emptyset\emptyset}\cap\pi_{\emptyset\mu\emptyset}\cap\pi_{\emptyset\emptyset\nu}$ are all finite 3d Young diagrams, so their $\mathcal{J}$-factors are computed directly from the charges of their shell boxes. For the 1-leg factors $\mathcal{J}(x|\pi_{\lambda\emptyset\emptyset})$, $\mathcal{J}(x|\pi_{\emptyset\mu\emptyset})$, and $\mathcal{J}(x|\pi_{\emptyset\emptyset\nu})$, a recursive argument via~\eqref{splitting}—illustrated in Fig.~\ref{fig:1leg gluing}—shows that shell boxes with non-zero charge are confined to the two ends of each leg. Moreover, the numbers of $+1$ and $-1$ charged shell boxes at the infinite end are equal, so their net contribution to the $\mathcal{J}$-factor is:
\begin{align}\label{D2 infinite end}
    \frac{\sh(\infty \e_2+\cdots)}{\sh(\infty \e_2+\cdots)}\,\frac{\sh(\infty \e_2+\cdots)}{\sh(\infty \e_2+\cdots)}\,\frac{\sh(\infty \e_2+\cdots)}{\sh(\infty \e_2+\cdots)}\times\cdots=1.
\end{align}
Since the contribution from the infinite end is exactly $1$, the $\mathcal{J}$-factor for each 1-leg diagram reduces to a contribution from the 2d Young diagram defining its asymptotic boundary condition:
\begin{align}\label{1leg J}
    \mathcal{J}\big(x\big|\pi_{\lambda\emptyset\emptyset}\big)&=\mathcal{J}\big(x+v_{23,1}-v\big|\lambda_{23,1}\big),\cr
    \mathcal{J}\big(x\big|\pi_{\emptyset\mu\emptyset}\big)&=\mathcal{J}\big(x+v_{13,1}-v\big|\mu_{13,1}\big),\cr
    \mathcal{J}\big(x\big|\pi_{\emptyset\emptyset\nu}\big)&=\mathcal{J}\big(x+v_{12,1}-v\big|\nu_{12,1}\big).
\end{align}

Applying the JK residue, the refined DT3 invariant is:
\begin{align}\label{DT partition}
    \mathcal{Z}^{\text{D0-D2-D6}}_{\lambda,\mu,\nu;k}=&\oint_{\JK} \prod_{i=1}^k\frac{d\phi_i}{2\pi i}\,\mathcal{I}^{\text{D0-D2-D6}}_{\lambda,\mu,\nu;k}\cr=&\sum_{|\tilde{\pi}_{\lambda\mu\nu}|=k}\,\prod_{\boldsymbol{x}\in\tilde{\pi}_{\lambda\mu\nu}}\frac{\mathcal{J}\big(\mathcal{X}(\boldsymbol{x})\big|\pi_{\lambda\mu\nu}\big)}{\mathcal{J}\big(\mathcal{X}(\boldsymbol{x})+\e_{123}\big|\pi_{\lambda\mu\nu}\big)}\prod_{\boldsymbol{y}\in\tilde{\pi}_{\lambda\mu\nu}}\sh(\mathcal{X}(\boldsymbol{x})-\mathcal{X}(\boldsymbol{y}))\,\mathcal{J}\big(\mathcal{X}(\boldsymbol{x})\big|\{\boldsymbol{y}\}\big),\cr
\end{align}
where, for any infinite 3d Young diagram $\pi\supset \pi_{\lambda\mu\nu}$ containing the vacuum, we set $\widetilde{\pi}_{\lambda\mu\nu}\equiv \pi\backslash\pi_{\lambda\mu\nu}$. Here $\{\boldsymbol{x}\}$ denotes the 3d Young diagram consisting of the single box at $\boldsymbol{x}$ (rather than at $\boldsymbol{1}$), whose $\mathcal{J}$-factor is computed via~\eqref{J-translation}. Note that although $\sh(0)$ appears when $\boldsymbol{x}=\boldsymbol{y}$, the overall expression remains finite.

In the special case $\lambda=\mu=\nu=\emptyset$, i.e., $N_1=N_2=N_3=0$ with no D2-branes present, the system reduces to the simplest sector of the tetrahedron instanton of Sec.~\ref{Tetrahedron instanton}—namely, the D0-D6 system with a single D6-brane. The DT3 invariant~\eqref{DT partition} for $\pi_{\emptyset\emptyset\emptyset}=\emptyset$ then reads:
\begin{align}
    \mathcal{Z}^{\text{D0-D2-D6}}_{\emptyset,\emptyset,\emptyset;k}=&\sum_{|\pi|=k}\prod_{\boldsymbol{x}\in\pi}\frac{\sh(\mathcal{X}(\boldsymbol{x})-\mathcal{X}(\boldsymbol{0}))}{\sh(\mathcal{X}(\boldsymbol{x})-\mathcal{X}(\boldsymbol{1}))}\prod_{\boldsymbol{y}\in\pi}\sh(\mathcal{X}(\boldsymbol{x})-\mathcal{X}(\boldsymbol{y}))\,\mathcal{J}\big(\mathcal{X}(\boldsymbol{x})\big|\{\boldsymbol{y}\}\big)\cr=&\sum_{|\pi|=k}\prod_{\boldsymbol{x}\in\pi}\frac{\sh(\mathcal{X}(\boldsymbol{x})-\mathcal{X}(\boldsymbol{0}))}{\sh(\mathcal{X}(\boldsymbol{x})-\mathcal{X}(\boldsymbol{1}))}\prod_{\boldsymbol{y}\in\pi}\frac{\sh(\mathcal{X}(\boldsymbol{x})-\mathcal{X}(\boldsymbol{y}))\,\sh(\mathcal{X}(\boldsymbol{x})-\mathcal{X}(\boldsymbol{y})-\e_{12,13,23})}{\sh(\mathcal{X}(\boldsymbol{x})-\mathcal{X}(\boldsymbol{y})-\e_{1,2,3})\,\sh(\mathcal{X}(\boldsymbol{x})-\mathcal{X}(\boldsymbol{y})-\e_{123})}\cr=&\sum_{|\pi|=k}\prod_{\boldsymbol{x}\in\pi}\sh(\mathcal{X}(\boldsymbol{x})-\mathcal{X}(\boldsymbol{0}))\,\mathcal{J}\big(\mathcal{X}(\boldsymbol{x})\big|\pi\big),
\end{align}
where the first equality uses $\mathcal{J}(x|\emptyset)=1/\sh(x-\mathcal{X}(\boldsymbol{1}))$, and the second and third equalities follow from the expansion formula~\eqref{J-expand}. Since $\widetilde{\pi}_{\emptyset\emptyset\emptyset}=\pi$, the final line coincides precisely with the partition function~\eqref{D0D6shell} of the tetrahedron instanton with a single D6-brane, giving:
\begin{align}
    \mathcal{Z}^{\text{D0-D2-D6}}_{\emptyset,\emptyset,\emptyset;k}=\mathcal{Z}^{\text{D0-D6}}_{(1,0,0,0),k}.
\end{align}
Detailed calculations for simple 1-leg and 3-leg examples, together with the derivation of~\eqref{DT partition}, are collected in Appendix~\ref{appendix DT}.

\subsection{Donaldson-Thomas 4 counting}\label{DT4 section}

Analogous to DT3 counting, the shell formula extends naturally to a 4d generalization, governing the enumeration of bound states in D8-D2-D0 systems on a CY fourfold. This setup is referred to as DT4 counting with leg boundary conditions. In contrast to DT3 counting, one may also impose surface boundary conditions on the CY fourfold, corresponding to bound states of D8-D4-D0 systems \cite{Monavari_2022,Kimura:2025lig,Nekrasov:2023nai,Piazzalunga:2023qik}.

We consider DT4 counting on $\C^4_{1234}$. The setup consists of $k$ D0-branes wrapping $S^1$, a single D8-brane extending over $\C^4_{1234} \times S^1$, $N_a$ D2$_a$-branes on $\C_a \times S^1$ for $a \in \4$, and $N_{ab}$ D4$_{ab}$-branes on $\C_{ab} \times S^1$ for $ab \in \6$. For brevity, throughout this chapter we suppress the ${\4,1}$ labels on Young diagrams.

DT4 counting is formulated by taking the minimal 4d Young diagram $\rho_{\{\pi_a\},\{\lambda_{ab}\}}$—characterized by four leg boundaries $\pi_a$ and six surface boundaries $\lambda_{ab}$—as the vacuum configuration, and enumerating BPS bound states of D0-branes with $|\pi_a| = N_a$ and $|\lambda_{ab}| = N_{ab}$. The worldvolume theory on the D0-branes is described by an $\cN=2$ supersymmetric quantum mechanics quiver analogous to that in Fig.~\ref{fig:D0D2D6quiver}, with differences arising in the specific content of fundamental and anti-fundamental multiplets. For a given minimal 4d Young diagram $\rho_{\{\pi_a\},\{\lambda_{ab}\}}$, the integrand associated with the D0-branes is:
\begin{align}\label{DT4 integrand}
    \mathcal{I}^{\text{D0-D2-D4-D8}}_{\{\pi_a\},\{\lambda_{ab}\};k}=\mathcal{I}^{\text{D0-D0}}_k\times\prod_{i=1}^k\mathcal{J}_{-\mathfrak{A}}\big(\phi_i\big|\rho_{\{\pi_a\},\{\lambda_{ab}\}}\big).
\end{align}
To obtain the correct DT4 character within the JK-residue formalism, one must perform the sign reversal $\sh(x)\to-\sh(-x-\e_{1234})$ on certain terms in the $\mathcal{J}$-factor, as in \eqref{D0D2D6 partition}; specifically, the contributions from certain fundamental multiplets are replaced by those from anti-fundamental ones. Accordingly, $\mathcal{J}_{-\mathfrak{A}}$ is defined by:
\begin{align}
    \mathcal{J}_{-\mathfrak{A}}\big(x\big|\mathbf{Y}_{\mathcal{A}}\big)&\equiv \left(\prod_{\boldsymbol{y}\in\mathfrak{A}(\mathbf{Y}_{\mathcal{A}})}\sh\left(x-\mathcal{X}_{\mathcal{A}}(\boldsymbol y)\right)^{\operatorname{Q}_{\mathbf{Y}_{\mathcal{A}}}(\boldsymbol{y})}\right)\left(\prod_{\boldsymbol{y}\in\mathcal{S}(\mathbf{Y}_{\mathcal{A}})\backslash\mathfrak{A}(\mathbf{Y}_{\mathcal{A}})}-\sh\left(\mathcal{X}_{\mathcal{A}}(\boldsymbol y)-x-\e_{1234}\right)^{\operatorname{Q}_{\mathbf{Y}_{\mathcal{A}}}(\boldsymbol{y})}\right),\cr\mathcal{J}_{-\mathfrak{A}}\big(x\big|\emptyset_{\mathcal{A}}\big)&\equiv \frac{1}{\sh(x-\mathcal{X}_\mathcal{A}(\boldsymbol{1}))}.
\end{align}
Here $\mathfrak{A}(\mathbf{Y})$ denotes the set of all \emph{addable boxes} of $\mathbf{Y}$, so that $\mathcal{J}_{-\mathfrak{A}}$ is obtained from $\mathcal{J}$ by flipping the sign of every term except those corresponding to addable boxes.

After performing the JK residue, the partition function takes the form:
\begin{footnotesize}
    \begin{align}\label{DT4 partition function}
     \mathcal{Z}^{\text{D0-D2-D4-D8}}_{\{\pi_a\},\{\lambda_{ab}\};k}=&\oint_{\JK}\,\prod_{i=1}^k\frac{d\phi_i}{2\pi i}\,\mathcal{I}^{\text{D0-D2-D4-D8}}_{\{\pi_a\},\{\lambda_{ab}\};k}\cr=&\sum_{|\tilde{\rho}_{\{\pi_a\},\{\lambda_{ab}\}}|=k}\,\prod_{\boldsymbol{x}\in\tilde{\rho}_{\{\pi_a\},\{\lambda_{ab}\}}}\mathcal{J}\big(\mathcal{X}(\boldsymbol{x})\big|\rho_{\{\pi_a\},\{\lambda_{ab}\}}\big)\prod_{\boldsymbol{y}\in\tilde{\rho}_{\{\pi_a\},\{\lambda_{ab}\}}}\sh(\mathcal{X}(\boldsymbol{x})-\mathcal{X}(\boldsymbol{y}))\,\mathcal{J}_{\geq}\big(\mathcal{X}(\boldsymbol{x})\big|\{\boldsymbol{y}\}\big),\cr
\end{align}
\end{footnotesize}
where the notation parallels that of DT3 counting introduced earlier, and for a 4d Young diagram $\rho \supset \rho_{\{\pi_a\},\{\lambda_{ab}\}}$ we write $\widetilde{\rho}_{\{\pi_a\},\{\lambda_{ab}\}} = \rho \backslash \rho_{\{\pi_a\},\{\lambda_{ab}\}}$.

\begin{figure}[ht]
    \centering
    \includegraphics[width=0.5\linewidth]{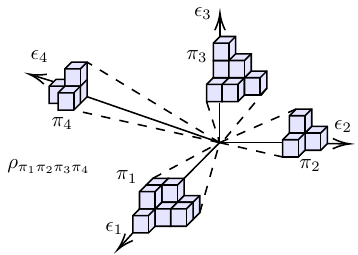}
    \caption{A schematic illustration of the minimal 4d Young diagram (minimal solid partition) $\rho_{\pi_1\pi_2\pi_3\pi_4}$. Its basis is spanned by $\epsilon_{1,2,3,4}$, with four distinct 3d Young diagrams $\pi_1,\pi_2,\pi_3,\pi_4$ serving as asymptotic leg boundary conditions.}
    \label{fig:DT4_4leg}
\end{figure}

Specializing to the D8-D2-D0 system—i.e., DT4 counting with leg boundaries $\pi_{a\in\4}$—the minimal solid partition $\rho_{\{\pi_a\}}=\rho_{\pi_1\pi_2\pi_3\pi_4}$ is schematically illustrated in Fig.~\ref{fig:DT4_4leg}. The associated $\mathcal{J}$-factor is computed by an inclusion-exclusion analogous to \eqref{J 3leg}:
\begin{align}
    \mathcal{J}\big(x\big|\rho_{\pi_1\pi_2\pi_3\pi_4}\big)=\frac{\left(\prod_{a\in\4}\mathcal{J}\big(x\big|\rho_{\pi_a}\big)\right)\left(\prod_{abc\in\check{\4}}\mathcal{J}\big(x\big|\rho_{\pi_a}\cap\rho_{\pi_b}\cap\rho_{\pi_c}\big)\right)}{\left(\prod_{ab\in\6}\mathcal{J}\big(x\big|\rho_{\pi_a}\cap\rho_{\pi_b}\big)\right)\mathcal{J}\big(x\big|\rho_{\pi_1}\cap\rho_{\pi_2}\cap\rho_{\pi_3}\cap\rho_{\pi_4}\big)},
\end{align}
where $\rho_{\pi_1}$ is shorthand for $\rho_{\pi_1\emptyset\emptyset\emptyset}$, and similarly for $\rho_{\pi_{2,3,4}}$. The pairwise, triple, and quadruple intersections $\rho_{\pi_a} \cap \rho_{\pi_b}$, $\rho_{\pi_a} \cap \rho_{\pi_b}\cap \rho_{\pi_c}$, and $\rho_{\pi_1} \cap \rho_{\pi_2}\cap \rho_{\pi_3}\cap \rho_{\pi_4}$ are all finite 4d Young diagrams, so the charges of their boxes are computed in the standard way. For the infinite Young diagram $\rho_{\pi_a}$, as in the DT3 case, the $\mathcal{J}$-factor is determined entirely by the boundary condition $\pi_a$:
\begin{align}
    \mathcal{J}\big(x\big|\rho_{\pi_a}\big)=\mathcal{J}\big(x+v_{\overline{a},1}-v\big|(\pi_a)_{\overline{a},1}\big).
\end{align}

The computation of the $\mathcal{J}$-factor for the minimal 4d Young diagram $\rho_{\{\lambda_{ab}\}}$ with surface boundaries $\{\lambda_{ab}\}_{ab\in\6}$ is more involved. Proceeding by the same inclusion-exclusion method yields:
\begin{align}
    \mathcal{J}\big(x\big|\rho_{\lambda_{12}\lambda_{13}\lambda_{14}\lambda_{23}\lambda_{24}\lambda_{34}}\big)=&\frac{\left(\prod_{ab\in\6}\mathcal{J}\big(x\big|\rho_{\lambda_{ab}}\big)\right)\left(\prod_{ab_1< ab_2<ab_3}\mathcal{J}\big(x\big|\cap_{i=1}^3\rho_{\lambda_{ab_i}}\big)\right)}{\left(\prod_{ab_1< ab_2}\mathcal{J}\big(x\big|\cap_{i=1}^2\rho_{\lambda_{ab_i}}\big)\right)\left(\prod_{ab_1< ab_2<ab_3<ab_4}\mathcal{J}\big(x\big|\cap_{i=1}^4\rho_{\lambda_{ab_i}}\big)\right)}\cr&\times\frac{\left(\prod_{ab_1< ab_2<ab_3<ab_4<ab_5}\mathcal{J}\big(x\big|\cap_{i=1}^5\rho_{\lambda_{ab_i}}\big)\right)}{\mathcal{J}\big(x\big|\cap_{i=1}^6\rho_{\lambda_{ab_i}}\big)},
\end{align}
where for each surface boundary:
\begin{align}
    \mathcal{J}\big(x\big|\rho_{\lambda_{ab}}\big)=\mathcal{J}\big(x+v_{\overline{ab},1}-v\big|(\lambda_{ab})_{\overline{ab},1}\big).
\end{align}

In practice, a more convenient computational strategy is available. As established in \eqref{D2 infinite end}, all contributions from asymptotic boundary conditions cancel pairwise. One may therefore impose a cutoff on $\rho_{\{\pi_a\},\{\lambda_{ab}\}}$ at a sufficiently large distance from the origin, compute the $\mathcal{J}$-factor directly on the truncated diagram, and then discard all terms that depend on the cutoff. Concretely, for a sufficiently large integer $m$, we define:
\begin{align}
    \rho_{\{\pi_a\},\{\lambda_{ab}\};m}=\{(x_1,x_2,x_3,x_4)\in\rho_{\{\pi_a\},\{\lambda_{ab}\}}\mid x_i\leq m\};
\end{align}
then $\mathcal{J}\big(x\big|\rho_{\{\pi_a\},\{\lambda_{ab}\}}\big)$ is recovered from $\mathcal{J}\big(x\big|\rho_{\{\pi_a\},\{\lambda_{ab}\};m}\big)$ by discarding all terms of the form $\sh(x+v-m\,\e_a+\cdots)$. Detailed computational examples are collected in Appendix~\ref{appendix DT4}.

Finally, the inclusion of D6-branes modifies the D0–D2–D4–D8 system by introducing a nontrivial background on which the 4d Young diagram is supported. Combinatorially, this is equivalent to placing the original configuration on top of a semi-infinite bulk, or, equivalently, shifting the origin of the minimal 4d Young diagram $\rho_{\{\pi_a\},\{\lambda_{ab}\}}$, so that all boxes are measured relative to a displaced reference corner.

This shift does not alter the local growth rules of the diagram, but changes the vacuum structure and the asymptotic data entering the shell formula, effectively reweighting the contributions of boxes. In this way, the shell formula naturally extends to the D0–D2–D4–D6–D8 system, i.e., the 4G network system \cite{Nekrasov:2023nai}. While we leave a detailed analysis of this system to future work, its partition function is expected to be directly obtained from that of the DT4 system through this shift of the reference configuration. For example, for a DT4 system with $\boldsymbol{N}=(N_{\overline{4}},N_{\overline{3}},N_{\overline{2}},N_{\overline{1}})$ D6-branes (of types D6$_{\overline{4}}$, D6$_{\overline{3}}$, D6$_{\overline{2}}$, D6$_{\overline{1}}$), the origin of the corresponding 4d Young diagram $\rho_{\{\pi_a\},\{\lambda_{ab}\}}$ is shifted from $(1,1,1,1)$ to $\boldsymbol{N}+\boldsymbol{1}$. Consequently, the integrand is given by:
\begin{align}
    \mathcal{I}^{\text{D0-D2-D4-D6-D8}}_{\{\pi_a\},\{\lambda_{ab}\},\boldsymbol{N};k}=\mathcal{I}^{\text{D0-D0}}_k\times\prod_{i=1}^k\mathcal{J}_{-\mathfrak{A}}\big(\phi_i-\boldsymbol{N}\cdot\boldsymbol{\e}\big|\rho_{\{\pi_a\},\{\lambda_{ab}\}}\big).
\end{align}

\section{Discussion}
The physical systems analyzed in this work share two key features.

\medskip\noindent\textbf{Feature 1: Universal D0-D0 sector structure.} The partition functions of all systems exhibit a universal structure in the D0-D0 sector; specifically, the contributions from this sector all take the form of the expansion of the $\mathcal{J}$-factor~\eqref{J-expand}:
\begin{itemize}
    \item Spiked instanton and 5d SYM with classical gauge groups (for instance, on $\C_1\times\C_2$):
    \begin{align}
        \mathcal{Z}^{\text{D0-D4},\U,\SO,\Sp}_k\supset\prod_{i\neq j}^k\sh(\phi_i-\phi_j)\prod_{i,j}^k\frac{\sh(\phi_i-\phi_j-\e_{12})}{\sh(\phi_i-\phi_j-\e_{1,2})}
    \end{align}
    \item Tetrahedron instanton and DT3 counting (for instance, on $\C_1\times\C_2\times \C_3$):
    \begin{align}
        \mathcal{Z}^{\text{D0-D6},\text{D0-D2-D6}}_k\supset\prod_{i\neq j}^k\sh(\phi_i-\phi_j)\prod_{i,j}^k\frac{\sh(\phi_i-\phi_j-\e_{12,13,23})}{\sh(\phi_i-\phi_j-\e_{1,2,3})\sh(\phi_i-\phi_j-\e_{123})}
    \end{align}
    \item Magnificent four and DT4 counting:
    \begin{footnotesize}
        \begin{align}
        \mathcal{Z}^{\text{D0-D8},\text{D0-D2-D4-D8}}_k\supset\left(\prod_{i\neq j}^k\sh(\phi_i-\phi_j)\prod_{i,j}^k\frac{\sh(\phi_i-\phi_j-\e_{1234})\prod_{ab\in\6}\sh(\phi_i-\phi_j-\e_{ab})}{\sh(\phi_i-\phi_j-\e_{1,2,3,4})\prod_{A\in\check{\4}}\sh(\phi_i-\phi_j-\e_{A})}\right)^{1/2}
    \end{align}
    \end{footnotesize}
    
\end{itemize}

\medskip\noindent\textbf{Feature 2: Classification of BPS bound states.} The BPS bound states can be classified by Young diagrams of different dimensions:
\begin{itemize}
    \item For 5d SYM with classical gauge groups and spiked instantons, the poles are classified by tuples of 2d Young diagrams.
    \item For tetrahedron instantons and DT3 counting, the poles are classified by tuples of 3d Young diagrams.
    \item  For magnificent four, the poles are classified by tuples of 4d Young diagrams.
\end{itemize}
Any physical system whose partition function satisfies the above two criteria can be described using the shell formula:
\begin{itemize}
    \item Spiked instanton and 5d SYM with classical gauge groups:
    \begin{align}
        \mathcal{Z}^{\text{D0-D4},\U,\SO,\Sp}_k\supset\prod_{\boldsymbol{x}\in\lambda}\frac{\mathcal{J}\big(\mathcal{X}(\boldsymbol{x})\big|\lambda\big)}{\sh(-\mathcal{X}(\boldsymbol{x})+\mathcal{X}(\boldsymbol{0}))}
    \end{align}
    \item Tetrahedron instanton:
    \begin{align}
        \mathcal{Z}^{\text{D0-D6}}_k\supset\prod_{\boldsymbol{x}\in\pi}\sh(\mathcal{X}(\boldsymbol{x})-\mathcal{X}(\boldsymbol{0}))\mathcal{J}\big(\mathcal{X}(\boldsymbol{x})\big|\pi\big)
    \end{align}
    \item DT3 counting:
    \begin{align}
        \mathcal{Z}^{\text{D0-D2-D6}}_k\supset\prod_{\boldsymbol{x}\in\widetilde{\pi}_{\lambda\mu\nu}}\frac{\mathcal{J}\big(\mathcal{X}(\boldsymbol{x})\big|\pi_{\lambda\mu\nu}\big)}{\mathcal{J}\big(\mathcal{X}(\boldsymbol{x})+\e_{123}\big|\pi_{\lambda\mu\nu}\big)}
    \end{align}
    \item Magnificent four:
    \begin{align}
        \mathcal{Z}^{\text{D0-D8}}_k\supset\prod_{\boldsymbol{x}\in\rho}\mathcal{J}_{\geq}\big(\mathcal{X}(\boldsymbol{x})\big|\rho\big)
    \end{align}
    \item DT4 counting:
    \begin{align}
        \mathcal{Z}^{\text{D0-D2-D4-D8}}_k\supset\prod_{\boldsymbol{x}\in\widetilde{\rho}_{\{\pi_a\},\{\lambda_{ab}\}}}\mathcal{J}\big(\mathcal{X}(\boldsymbol{x})\big|\rho_{\{\pi_a\},\{\lambda_{ab}\}}\big)
    \end{align}
\end{itemize}
The availability of these precise partition function expressions enables the systematic computation of additional properties, such as algebraic identities and recurrence relations.

A promising future direction is to employ the shell formula in any system meeting the above conditions, as well as to generalize its relationship to other known algebraic frameworks:
\begin{itemize}
    \item We aim to clarify the precise relationship between the shell formula and the topological vertex. In particular, it would be valuable to derive an explicit expression for the O$^+$-plane \cite{Hayashi:2020hhb,Nawata:2021dlk,Kim:2024ufq} vertex and understand how orientifold projections modify the combinatorial and representation-theoretic structure of the vertex.
    
    \item Another promising direction is to generalize the shell formula to SYM theories with matter fields in various representations. This includes both classical and exceptional Lie groups, where the structure of instanton moduli spaces becomes more general \cite{Shadchin:2005mx,Kim:2024vci,Chen:2023smd,Kim:2023qwh}. Such generalizations may reveal how the shell formula encodes representation-dependent contributions and could shed new light on exceptional gauge symmetries in non-perturbative string theory.

    \item Using the exact closed form of the magnificent four partition function, one can investigate its interplay with $qq$-characters and the representation theory of quantum algebras. This includes examining how the partition function furnishes generating functions for protected operators and how it realizes the action of quantum toroidal (or DIM-type) symmetries \cite{Nekrasov:2015wsu, Nekrasov:2016ydq, Nawata:2023wnk, Bourgine:2017jsi, Kimura:2023bxy, Kimura:2015rgi}.

    \item The configurations analyzed in this work are formulated for D-branes extended along flat complex planes. A natural direction is to generalize the shell formula to more intricate geometric and physical settings. On the geometric side, one may consider orbifolds or more general Calabi–Yau manifolds \cite{Ooguri:2009ijd,Galakhov:2021xum}, where the combinatorics of Young diagrams is replaced by colored or fractional configurations, and the shell formula is expected to incorporate discrete data associated with the orbifold action. On the physical side, the formalism should extend to more general gauge origami setups, including the full 4G system of D0–D2–D4–D6–D8 branes \cite{Nekrasov:2023nai}, where additional defect sectors and couplings arise.

    \item From the viewpoint of enumerative geometry, the shell formula already captures DT3 and DT4 invariants on $\C^3$ and $\C^4$, and it is natural to expect its extension to Donaldson–Thomas theory on more general Calabi–Yau threefolds and fourfolds, where new features such as self-dual obstruction theories appear. It would also be interesting to explore its relation to other curve-counting theories, such as Pandharipande–Thomas stable pair invariants \cite{Pandharipande:2007kc, Cao:2019tnw, Kimura:2025lig}. More broadly, these generalizations suggest that the shell formula may provide a universal framework for organizing BPS counting across different dimensions, geometries, and brane configurations.
\end{itemize}

\acknowledgments
The author is grateful to Satoshi Nawata for his generous guidance and constant support throughout the development of this work. The author also warmly thanks Taro Kimura, Go Noshita, and Jiahao Zheng for many insightful conversations on gauge origami, DT counting, and for their thoughtful suggestions that greatly improved the presentation of this paper. This work is supported by the Shanghai Municipal Science and Technology Major Project (No.24ZR1403900).

\appendix
\section{Examples of charges of shellboxes}\label{charges of shellbox}
In this appendix, we summarize notations necessary for the paper and present various examples of the definitions in Sec.~\ref{YD and Shell}.
\subsection{Labels of Young diagram and notations}
To concisely indicate the required $\epsilon_i$ parameter and the basis of Young diagrams, we adopt the following simplified notation:
\begin{itemize}
    \item $\4=\{1,2,3,4\}$, $a,b\in\4$. Elements of $\4$ index the four complex directions of $\C^4$; they label individual D-brane worldvolume directions and appear as subscripts on the $\Omega$-background parameters $\e_a$.
    \item $\check{\4}=\{123,124,134,234\}=\{\overline{4},\overline{3},\overline{2},\overline{1}\}$, $A,B\in\check{\4}$. Elements of $\check{\4}$ index the four coordinate hyperplanes (complements of each direction in $\4$); $A\in\check{\4}$ specifies the basis directions of a 3d or 4d Young diagram, and appears in the labels of tetrahedron instanton and D0-D8 systems (Sec.~\ref{Tetrahedron instanton}--\ref{DT4 section}).
    \item $\6=\{12,13,23,14,24,34\}$, $ab\in\6$. Elements of $\6$ index the six coordinate 2-planes; $ab\in\6$ labels D4-brane orientations in the spiked instanton and DT3 systems (Sec.~\ref{Spiked instanton}--\ref{DT3 section}), and specifies the basis of a 2d Young diagram.
\end{itemize}

The $\Omega$-background parameters $\e_{1,2,3,4}$ are also denoted as:
\begin{align}
    q_{i}\equiv e^{-\e_i},\quad q_{i_1\ldots i_s}=q_{i_1}\ldots q_{i_s}=e^{-(\e_{i_1}+\ldots+\e_{i_s})}
\end{align}
In this paper, we adopt the CY fourfold condition $\epsilon_4 = -\epsilon_{123}$ as a default assumption.

Each label $\mathcal{A}=(A,\alpha)$ consists of a basis specification $A\in\check{\4}\cup\6\cup\4$ and a color index $\alpha$ counting Young diagrams with the same basis. We employ $\mathcal{A}$, $\mathcal{B}$, $\mathbf{ab}$ and $\mathbf{ab}'$ to denote such labels; for instance, $\mathcal{A}=(234, 7)$ means $\mathbf{Y}_{\mathcal{A}}$ is the 7th Young diagram in the basis $\e_2,\e_3,\e_4$.

The coordinate function converting box positions to integration variables is:
\begin{align}
    \mathcal{X}_{\mathcal{A}}(\boldsymbol{x})\equiv v_{\mathcal{A}}+(\boldsymbol{x}-\boldsymbol{1})\cdot \boldsymbol{\e}_A=v_{\mathcal{A}}+\sum_{i=1}^d(x_i-1)\,\e_{a_i}
\end{align}
This is the function appearing in the pole classification of Sec.~\ref{YD and Shell}: a JK-selected pole at $\boldsymbol{\phi}_*$ corresponds to setting $\phi_{i*}=\mathcal{X}_{\mathcal{A}}(\boldsymbol{x})$ for each box $\boldsymbol{x}$ in the corresponding Young diagram.

\begin{figure}[ht]
    \centering
    \includegraphics[width=0.3\linewidth]{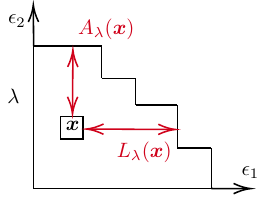}
    \caption{Leg length $L_\lambda(\boldsymbol{x})$ and arm length $A_\lambda(\boldsymbol{x})$ of a box $\boldsymbol{x}$ in a Young diagram $\lambda$.}
    \label{fig:legarm}
\end{figure}

Given a 2d Young diagram $\lambda$, we can define, for each box $\boldsymbol{x}$ in $\lambda$, its leg $L_\lambda(\boldsymbol{x})$ and arm $A_\lambda(\boldsymbol{x})$ as illustrated in Fig.~\ref{fig:legarm}. The leg is the number of boxes in direction 1 from $x$ within $\lambda$, while the arm is the number of boxes in direction 2.

Given a tuple of Young diagrams $\vec{\mathbf{Y}} = (\mathbf{Y}_\mathcal{A},\ldots)$, the total number of boxes is defined as follows. For a single Young diagram $\mathbf{Y}_\mathcal{A}$, we denote its number of boxes by $|\mathbf{Y}_\mathcal{A}|$. Then, the total number of boxes for the tuple is defined as $||\vec{\mathbf{Y}}||\equiv\sum_{\mathcal{A}} |\mathbf{Y}_\mathcal{A}|$.

Throughout this paper, the functions $\sh$ and $\ch$ that appear in all the partition functions are defined by:
\begin{align}
    \sh(x)\equiv e^{x/2}-e^{-x/2},\qquad \ch(x)\equiv e^{x/2}+e^{-x/2}
\end{align}
We also use the following shorthand for products over multiplicative parameters:
\begin{align}
    &\sh(\pm x\pm y)=\sh(x+y)\sh(x-y)\sh(-x+y)\sh(-x-y)\cr&\sh(x+\e_{12,13,\ldots})=\sh(x+\e_{12})\sh(x+\e_{13})\times\ldots
\end{align}
The sets $\4$, $\check{\4}$, and $\6$ together cover all the index types needed in this paper: $\check{\4}$ appears in the tetrahedron instanton and D0-D8 system (Sec.~\ref{Tetrahedron instanton}--\ref{D0D8 appendix}), $\6$ in the spiked instanton and DT3 (Sec.~\ref{Spiked instanton}--\ref{appendix DT}), and $\4$ in the magnificent four and DT4 (Sec.~\ref{Magnificent four}--\ref{appendix DT4}).

\subsection{Shell and \texorpdfstring{$\mathcal{J}$}{J}-factor}\label{shell and J}
This subsection presents three progressively higher-dimensional examples of shell and $\mathcal{J}$-factor computations: a single-box 2d diagram worked out fully from Definition~\ref{shell}, a general 2d diagram showing the charge-to-box correspondence, and a single-box 3d diagram that is the fundamental building block of all 3d partition functions in this paper.
\begin{itemize}
    \item \textbf{2d single box: $\lambda_{12,1}=\{(1,1)\}$.} With $\mathbf{B}_2=\{(0,0),(0,1),(1,0),(1,1)\}$ from Definition~\ref{shell}, the shell is:
    \begin{align}
        \mathcal{S}(\lambda_{12,1})=&(\lambda_{12,1}+\mathbf{B}_2)\backslash\lambda_{12,1}\cr=&\{(1,1),(1,2),(2,1),(2,2)\}\backslash\{(1,1)\}\cr=&\{(1,2),(2,1),(2,2)\}
    \end{align}

    Therefore, the charge of each shellbox is defined as~\eqref{charge}:
    \begin{align}
        \operatorname{Q}_{\lambda_{12,1}}(1,2)=&(-1)^{|(0,1)|}=-1\cr\operatorname{Q}_{\lambda_{12,1}}(2,1)=&(-1)^{|(1,0)|}=-1\cr\operatorname{Q}_{\lambda_{12,1}}(2,2)=&(-1)^{|(1,1)|}=1
    \end{align}
    The $\mathcal{J}$-factor~\eqref{J-def} is therefore:
    \begin{align}\label{J 2dYD 1box}
        \mathcal{J}\big(x\big|\lambda_{12,1}\big)=&\frac{\sh(x-\mathcal{X}_{12,1}(2,2))}{\sh(x-\mathcal{X}_{12,1}(1,2))\sh(x-\mathcal{X}_{12,1}(2,1))}\cr=&\frac{\sh(x-v_{12,1}-\e_{12})}{\sh(x-v_{12,1}-\e_{1})\sh(x-v_{12,1}-\e_{2})}
    \end{align}
    The Young diagram, its shell, and the corresponding charges are illustrated in Fig.~\ref{fig:2dYD 1box}.
    \begin{figure}[ht]
        \centering
        \includegraphics[width=0.8\linewidth]{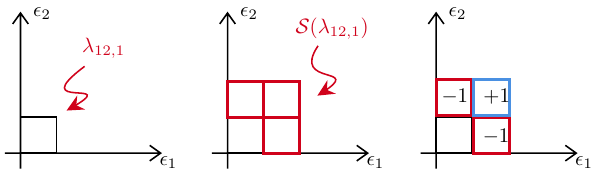}
        \caption{The leftmost diagram shows a 2d Young diagram $\lambda_{12,1}$ with only one box, labeled $\{12,1\}$ indicating it is the first Young diagram in the 12-plane. In the middle diagram, the red boxes represent the shell $\mathcal{S}(\lambda_{12,1})$ of $\lambda_{12,1}$; for a 2d Young diagram with only one box, its shell $\mathcal{S}(\lambda_{12,1})$ consists of only 3 shellboxes. In the rightmost diagram, the charge of each shellbox is shown, where the $+1$ shellboxes are marked in blue and the $-1$ shellboxes are marked in red.}
        \label{fig:2dYD 1box}
    \end{figure}
    \item For a more general 2d Young diagram $\lambda$, Fig.~\ref{fig:2dYD shell and charge} shows the shell and the charges of each shellbox. The charge pattern reveals a precise combinatorial correspondence:

\begin{figure}[ht]
    \centering
    \includegraphics[width=0.9\linewidth]{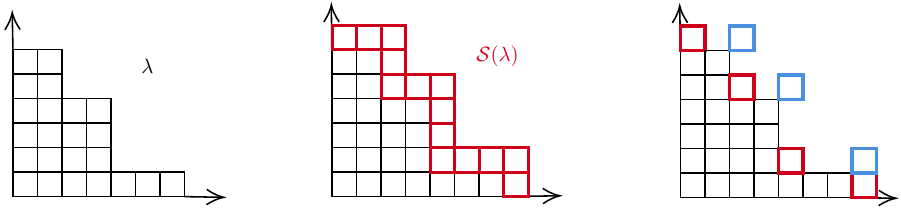}
    \caption{For a general 2d Young diagram $\lambda$ (leftmost), the shellboxes $\mathcal{S}(\lambda)$ are marked in the middle figure. In the rightmost figure, red boxes have $-1$ charge, blue boxes have $+1$ charge, and unmarked shellboxes have $0$ charge.}
    \label{fig:2dYD shell and charge}
\end{figure}

\noindent\textbf{Property (2d charge-box correspondence).} \emph{For any 2d Young diagram $\lambda$, the shellboxes with charge $-1$ are exactly the addable boxes $\mathfrak{A}(\lambda)$ (positions where a new box may be placed while preserving the Young diagram property), and each shellbox with charge $+1$ at position $(i,j)$ corresponds to a removable box $\mathfrak{R}(\lambda)$ at position $(i-1,j-1)$.}

This correspondence is a consequence of the inclusion-exclusion definition~\eqref{charge}: for a 2d diagram, each shell box has at most one binary neighbor inside $\lambda$, so only charges $\pm 1$ and $0$ appear, and the geometric roles of $\pm 1$ boxes are exactly as stated. As a result, the $\mathcal{J}$-factor for a 2d Young diagram can be expressed directly in terms of addable and removable boxes:
\begin{align}\label{addable removable}
    \mathcal{J}\big(x\big|\lambda_{\mathbf{ab}}\big)=\frac{\prod_{\boldsymbol{y}\in\mathfrak{R}}\sh(x-\mathcal{X}_{\mathbf{ab}}(\boldsymbol{y}+\boldsymbol{1}))}{\prod_{\boldsymbol{y}\in\mathfrak{A}}\sh(x-\mathcal{X}_{\mathbf{ab}}(\boldsymbol{y}))}
\end{align}
    \item For 3d Young diagrams, the charge-to-box correspondence of the previous item breaks down. As illustrated in Fig.~\ref{fig:3dcharges}, a 3d shellbox may carry charges $0$, $\pm 1$, or even $\pm 2$—values that do not correspond to single addable or removable boxes. Consequently, the $\mathcal{J}$-factor for a 3d Young diagram cannot be expressed in the same form as~\eqref{addable removable}.
    \begin{figure}[ht]
    \centering
    \includegraphics[width=0.4\linewidth]{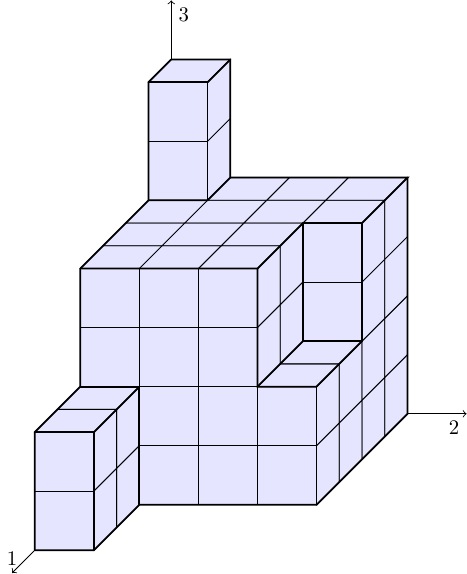}
    \includegraphics[width=0.45\linewidth]{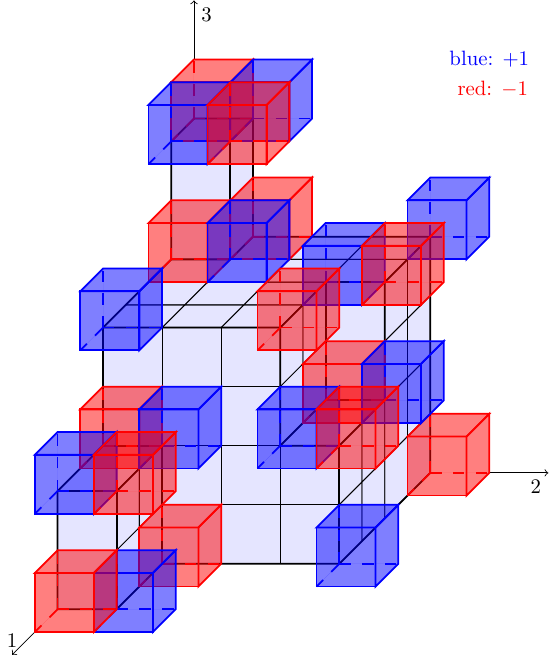}
    \caption{For an arbitrary 3d Young diagram (left), the shellbox charges are shown on the right: red boxes carry charge $-1$ and blue boxes carry charge $+1$.}
    \label{fig:3dcharges}
\end{figure}

    \medskip\noindent\textbf{Example (3d single box).} The 3d Young diagram $\pi_{\bar{4},1}=\{(1,1,1)\}$ has binary set $\mathbf{B}_3$ of size $2^3=8$. Applying~\eqref{shell} and~\eqref{charge} gives:
    \begin{align*}
        \operatorname{Q}=+1:\quad (1,2,2),\,(2,1,2),\,(2,2,1)\qquad\operatorname{Q}=-1:\quad (1,1,2),\,(1,2,1),\,(2,1,1),\,(2,2,2)
    \end{align*}
    The $\mathcal{J}$-factor~\eqref{J-def} is therefore:
    \begin{align}\label{J 3dYD 1box}
        \mathcal{J}\big(x\big|\{(1,1,1)\}_{\bar{4},1}\big)=\frac{\sh(x-v_{\bar{4},1}-\e_{12})\sh(x-v_{\bar{4},1}-\e_{13})\sh(x-v_{\bar{4},1}-\e_{23})}{\sh(x-v_{\bar{4},1}-\e_{1,2,3})\sh(x-v_{\bar{4},1}-\e_{123})}
    \end{align}
    This is the building block of the tetrahedron instanton and DT3 partition functions in Sec.~\ref{Tetrahedron instanton}--\ref{appendix DT}.
    \item For a $d$-dimensional Young diagram, the charge of a shellbox can take any integer value from $-d$ to $d$. The extreme values $\pm d$ arise only at the corner of the Young diagram, where the box $(1,1,\ldots,1)$ is adjacent to all $2^d$ binary neighbors simultaneously: $Q=-d$ occurs when none of those neighbors is in $\mathbf{Y}$ (the box is addable in all $d$ directions at once), and $Q=+d$ when all are in $\mathbf{Y}$. In practice, charges beyond $\pm 1$ first appear in 3d diagrams, and for 4d diagrams the charges $Q=+2$ and $Q=-3$ visible in Tab.~\ref{tab:4leg charges} reflect the geometry of four infinite legs meeting at a common origin.
\end{itemize}

\section{Witten index and JK-residue}\label{witten inedx and JK}
\subsection{1d \texorpdfstring{$\cN=2$}{N=2} quiver and Witten index}
Given a SUSY QM, the definition of the Witten index is:
\begin{align}
    \mathcal{I}=\Tr_{\mathcal{H}}(-1)^Fe^{-\beta H-\sum_iT_iu_i}
\end{align}
where $F$ is the fermion number operator, $\{u_i\}$ are the chemical potentials of the flavor symmetries, $\{T_i\}$ are the Cartan generators of the flavor symmetries, and $\beta$ denotes the size of the time circle $\mathbb{S}^1$.

Given a supersymmetric field theory, a quiver encodes the gauge symmetry, flavor symmetry, and all fields that transform non-trivially under these symmetries. For a given brane system, the ADHM data determine the quiver; the Witten index is the product of contributions from each quiver element evaluated at the complexified gauge variables $\boldsymbol{\phi}$, giving precisely the integrand $\mathcal{I}(\boldsymbol{\phi})$ appearing in all formulas of Sec.~\ref{5d pure SYM} and Sec.~\ref{gauge origami}. For 1d $\cN=2$ SUSY QM, the quiver elements and their contributions are as follows:
\begin{itemize}
    \item A circular node represents a gauge group, and each gauge group carries the contribution of the vector multiplet. In our cases, we only focus on the $\U(k)$ gauge group:
    
    \begin{minipage}{0.1\textwidth}
        $\,$
    \end{minipage}
    \begin{minipage}{0.3\textwidth}
    \includegraphics[width=0.4\linewidth]{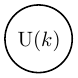}
    \end{minipage}
    \begin{minipage}{0.1\textwidth}
        $\Longrightarrow$
    \end{minipage}
    \begin{minipage}{0.4\textwidth}
\begin{align*}
    \left(\prod_{i=1}^k\frac{d\phi_i}{2\pi i }\right)\prod_{i\neq j}^k\sh(\phi_i-\phi_j)
\end{align*} 
    \end{minipage}
    \item A square node represents a flavor (global symmetry) group. A black solid line with arrows connecting a circular (gauge) node and a square (flavor) node represents a chiral multiplet $\Phi_{\alpha}^\beta$: it transforms in the fundamental representation under the node at the arrow's tip, and in the anti-fundamental under the node at the tail. For such a chiral multiplet with additional $\U(1)$ flavor charges $q(\Phi_{\alpha}^\beta)$, its contribution to the index is:
    
    \begin{minipage}{0.4\textwidth}
    \includegraphics[width=0.9\linewidth]{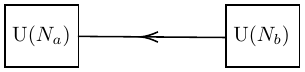}
    \end{minipage}
    \begin{minipage}{0.1\textwidth}
        $\Longrightarrow$
    \end{minipage}
    \begin{minipage}{0.4\textwidth}
\begin{align*}
   \prod_{\alpha=1}^{N_a}\prod_{\beta=1}^{N_b}\frac{1}{\sh(a_\alpha-b_\beta+q(\Phi_{\alpha}^\beta))}
\end{align*} 
    \end{minipage}

    \noindent Here $a_\alpha$ and $b_\beta$ are the eigenvalues of the groups at each end of the arrow: $a_\alpha=\phi_i$ when the source is the gauge group $\U(k)$, and $a_\alpha=v_\mathcal{A}$ (a Coulomb branch parameter) when it is a flavor node.

    \item The red dashed lines represent Fermi multiplets $\Psi_\alpha^\beta$, connecting gauge and flavor nodes with the same orientation convention as above. Their contribution with additional $\U(1)$ flavor charges $q(\Psi_\alpha^\beta)$ to the index is:

    \begin{minipage}{0.4\textwidth}
    \includegraphics[width=0.9\linewidth]{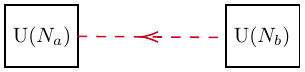}
    \end{minipage}
    \begin{minipage}{0.1\textwidth}
        $\Longrightarrow$
    \end{minipage}
    \begin{minipage}{0.4\textwidth}
\begin{align*}
   \prod_{\alpha=1}^{N_a}\prod_{\beta=1}^{N_b}\sh(a_\alpha-b_\beta+q(\Psi_\alpha^\beta))
\end{align*} 
    \end{minipage}
\end{itemize}
In our context, all $\U(1)$ flavor symmetries are contained in the CY4 holonomy $\U(1)^3=\U(1)_{\e_1}\times\U(1)_{\e_2}\times\U(1)_{\e_3}$. The four adjoint chirals $B_{1,2,3,4}$ correspond to motion in the four complex directions $\C_{1,2,3,4}$: $B_a$ carries $\U(1)_{\e_a}$ charge $-1$ and all other $\U(1)$ charges zero. For example, for $B_1$ with charges $(-1,0,0)$, its contribution to the index is:
\begin{align}
    \mathcal{I}(B_1)=\prod_{i,j}^k\frac{1}{\sh(\phi_i-\phi_j-\e_1)}
\end{align}
\subsection{Jeffrey-Kirwan residue}\label{JK}
The computation of the Witten index requires explicit evaluation of contour integrals. The JK-residue prescription \cite{Jeffrey:1993cun,Szenes_2004,Benini:2013nda,Benini:2013xpa,Nawata:2023aoq} provides the correct method for performing these integrals. It is applied to the integrands of Sec.~\ref{5d pure SYM}--\ref{gauge origami}; the resulting poles are classified by the Young diagrams introduced in Sec.~\ref{YD and Shell}. Here we review the JK-residue procedure.

We consider a gauge theory with rank-$k$ gauge group, which is $\U(k)$ in our context. The Witten index is expressed as an integral of a meromorphic $k$-form over a specific cycle:
\begin{align}
\mathcal{Z} = \oint_{\text{JK}} \prod_{I=1}^k \frac{d\phi_I}{2\pi i}  \mathcal{I}(\boldsymbol{\phi}) = \oint_{\text{specific cycles}} \mathcal{I}(\boldsymbol{\phi})  d\phi_1 \wedge \cdots \wedge d\phi_k,
\end{align}
where $\boldsymbol{\phi} = (\phi_1, \ldots, \phi_k)$ denotes the complexified gauge variables. The integrand $\mathcal{Z}(\boldsymbol{\phi})$ is periodic in each $\phi_I$:
\begin{align}
\mathcal{I}(\ldots, \phi_I, \ldots) = \mathcal{I}(\ldots, \phi_I + 2\pi i, \ldots).
\end{align}

The poles of the integrand arise from the denominator, which takes the schematic form:
\begin{align}
\mathcal{I}(\boldsymbol{\phi}) \propto \prod_a \frac{1}{\sh\left(\sum_{I=1}^k Q_a^I \phi_I + m_a\right)^{N_a}},
\end{align}
where $Q_a^I \in \mathbb{Q}$ are charge vectors and $m_a$ are masses or equivariant parameters. The poles of $\mathcal{I}$ are thus located at solutions of the equations
\begin{align}\label{poles-eq}
\sum_{I=1}^k Q_a^I \phi_I + m_a = 2\pi i n_a, \quad n_a \in \mathbb{Z}, \quad a = 1, \ldots, k.
\end{align}
The periodicity leads to multiple copies of poles shifted by $2\pi i$, and the allowed values of $n_a$ can be parametrized by an invertible matrix $Q_{ai}$ as
\begin{align}
\begin{pmatrix}
n_1 \\
\vdots \\
n_k
\end{pmatrix}
=\frac{Q}{|\det Q|} \cdot
\begin{pmatrix}
l_1 \\
\vdots \\
l_k
\end{pmatrix}, \quad
l_I \in {0, 1, \ldots, |\det Q|-1}.
\end{align}

Given a specific pole $\boldsymbol{\phi}_*$ satisfying~\eqref{poles-eq}, the JK-residue is evaluated through the following steps 
\begin{itemize}
    \item Given a pole $\boldsymbol{\phi}_*$, we identify an associated set of charge vectors $Q_*=\{Q_1,\ldots,Q_r\}$ with $r\geq k$, such that $Q_\ell\cdot\boldsymbol{\phi}_*+m_{\ell}=n_\ell2\pi i$ for any $Q_\ell\in Q_*$. We can construct a flag $F$ from any $k$-sequence of linearly independent charge vectors $\{Q_{a_1},\ldots,Q_{a_k}\}\subset Q_*$ that satisfies:
    \begin{align}
        \{0\}\subset F_1\subset\ldots\subset F_k=\bR^k,\quad F_\ell=\operatorname{span}\{Q_{a_1},\ldots,Q_{a_\ell}\}
    \end{align}
    The sequence $\{Q_{a_1},\ldots,Q_{a_k}\}$ is called a basis $\mathcal{B}(F,Q_*)$ of $F$ in $Q_*$.
    \item From each flag $F$ and its basis $\mathcal{B}(F,Q_*)$, a sequence of vectors is constructed: \begin{align} \kappa(F,Q_*)\equiv(\kappa_1,\ldots,\kappa_k),\quad\text{where } \kappa_\ell=\sum_{\substack{Q\in Q_*\\Q\in F_\ell}}Q
    \end{align}
    Intuitively, $\kappa_\ell$ is the sum of all charge vectors in $Q_*$ that belong to the $\ell$-th subspace $F_\ell$ of the flag; the condition~\eqref{in cone} then checks whether $\eta$ lies in the cone spanned by $\kappa_1,\ldots,\kappa_k$, which determines whether this flag contributes to the residue. If different flags $F$, $F'$ yield the same $\kappa(F,Q_*)=\kappa(F',Q_*)$, either choice may be taken.
    \item We need to choose a reference vector $\eta=(\eta_1,\ldots,\eta_k)\in(\bR^k)^*$. And we only pick the sequence of vectors $\kappa(F,Q_*)$ that satisfies:
    \begin{align}\label{in cone}
        \kappa(F,Q_*)^T\cdot\boldsymbol\lambda=\eta,\quad\text{where}\quad\boldsymbol{\lambda}=(\lambda_1,\ldots,\lambda_k)\in\bR^k_+
    \end{align}
    For this purpose, one can define a delta function:
    \begin{align}
        \delta(F,\eta)=\left\{\begin{aligned}
            1,\qquad&\kappa(F,Q_*) \text{ satisfies~\eqref{in cone}}\\0,\qquad&\text{else}
        \end{aligned}\right.
    \end{align}
\end{itemize}
With these objects defined, the JK-residue of the given pole $\boldsymbol{\phi}_*$ is:
\begin{align}
    \underset{\boldsymbol{\phi}=\boldsymbol{\phi}_*}{\operatorname{JK-Res}}(\eta)\mathcal{I}=\sum_F\delta(F,\eta)\frac{\sgn\det\kappa(F,Q_*)}{\det \mathcal{B}(F,Q_*)}\underset{\varepsilon_k=0}{\operatorname{Res}}\ldots\underset{\varepsilon_1=0}{\operatorname{Res}}\mathcal{I}\Bigg|_{\substack{Q_{a_1}\boldsymbol\phi_*+m_{a_1}+\varepsilon_1=n_{a_1}2\pi i\\\scriptscriptstyle{\vdots}\\Q_{a_k}\boldsymbol\phi_*+m_{a_k}+\varepsilon_k=n_{a_k}2\pi i}}
\end{align}
where the sum is over all flags constructed from $Q_*$ associated to $\boldsymbol\phi_*$. The $\varepsilon_1,\ldots,\varepsilon_k$ represent the order of integral induced by the chosen flag $F$.

Finally, given a generic $\eta$, the JK-residue can be computed as follows:
\begin{align}
    \oint_{\text{JK}}\prod^k_{I=1}\frac{d\phi_I}{2\pi i}\mathcal{I}(\boldsymbol{\phi})=\sum_{\boldsymbol{\phi}_*}\underset{\boldsymbol{\phi}=\boldsymbol{\phi}_*}{\operatorname{JK-Res}}(\eta)\mathcal{I}(\boldsymbol{\phi})
\end{align}
Note that in most cases, the results of the JK-residue are independent of the choice of the reference vector $\eta$. However, in our problems, the results sometimes depend on the choice of $\eta$; this occurs when poles lie on the boundary of a cone, a situation arising in the $\Sp$ and $\SO$ theories at special values of the Coulomb parameters. The standard choice is $\eta=(1,\ldots,1)$, corresponding to a positive FI parameter, which agrees with the standard ADHM prescription \cite{Shadchin:2005mx,Hwang:2014uwa}.

In all physical systems of Sec.~\ref{5d pure SYM}--\ref{gauge origami}, the JK-selected poles take the form $\phi_{i*}=\mathcal{X}_{\mathcal{A}}(\boldsymbol{x})$ for boxes $\boldsymbol{x}$ in a tuple of Young diagrams $\vec{\mathbf{Y}}=(\mathbf{Y}_{\mathcal{A}},\mathbf{Y}_{\mathcal{B}},\ldots)$, as described in Sec.~\ref{YD and Shell}. This Young diagram structure is not an assumption but a consequence of the charge matrix $Q$ being built from the ADHM data; the JK-prescription then selects the signs and multiplicities that reproduce the correct instanton counting.

\section{Detail computations for various cases}\label{examples}

This appendix provides explicit low-instanton calculations that confirm the main-text formulas and illustrate how the shell formula operates in practice. The subsections are organized in order of increasing Young diagram dimension, and each is self-contained.

\begin{itemize}
\item  Appendix~\ref{app. J and nek}  ($\U(N)$ SYM, 2d Young diagrams) provides a proof of 
the equivalence \eqref{J and nek} between the shell formula and the Nekrasov
factor.
    \item Appendix~\ref{Sp2 example} ($\Sp(2)$ SYM, 2d Young diagrams) verifies the closed-form expressions~\eqref{ZSp plus}--\eqref{ZSp minus} at $k=1,2$ and demonstrates how the BPS jumping coefficients~\eqref{BPS jumping coeff} are absorbed by the unrefined limit.
    \item Appendix~\ref{D0-D6 appendix} (D0-D6, 3d Young diagrams) confirms the $k=1,2$ contributions match the MacMahon function under the CY3 condition, and verifies the recursion relation~\eqref{D6recursion}.
    \item Appendix~\ref{appendix DT} (DT3, 3d Young diagrams with boundary) presents explicit 1-leg residue computations, derives the general DT3 integrand~\eqref{DT partition} via the recursion relation, and computes the simplest 3-leg example.
    \item Appendix~\ref{D0D8 appendix} (D0-D8, 4d Young diagrams) demonstrates the $\mathcal{J}_{\geq}$-factor computation, verifies the 4d recursion relation~\eqref{4d recursion}, and explains the sign discrepancy between the partition function convention of \cite{Nekrasov:2018xsb} and ours.
    \item Appendix~\ref{appendix DT4} (DT4, 4d Young diagrams with boundary) illustrates the cutoff method for two boundary configurations and derives the DT4 integrand~\eqref{DT4 integrand}.
\end{itemize}

\subsection{Shell formula and Nekrasov factor}\label{app. J and nek}

We want to show that the shell formula \eqref{J and nek} is equivalent to the Nekrasov
factor. The idea is simple: we first handle a small rectangular piece of a Young diagram,
then tile the whole diagram with such pieces, and finally check that leftover terms
cancel. We treat the single-diagram case $\alpha=\beta$ first, then explain how the
argument carries over to $\alpha\neq\beta$.

\subsubsection*{Step 1: A rectangle case}

\begin{figure}[ht]
    \centering
    \includegraphics[width=0.8\linewidth]{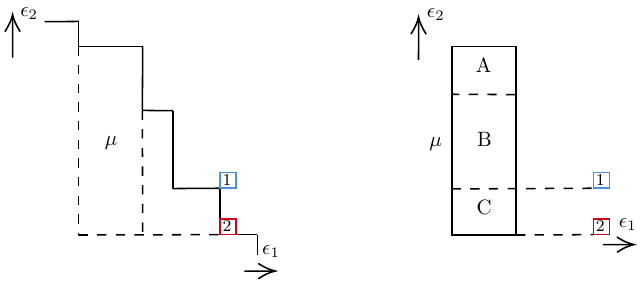}
    \caption{Left: the Young diagram $\lambda$ with a rectangular subdiagram $\mu$, an
    addable box $\boldsymbol{x}_2$ flush with the bottom of $\mu$, and an addable box
    $\boldsymbol{x}_1$ directly above $\boldsymbol{x}_2$. Right: $\mu$ split into
    regions A, B, C (with $\mathrm{A}\simeq\mathrm{C}$), where C is the strip between
    $\boldsymbol{x}_1$ and $\boldsymbol{x}_2$.}
    \label{fig:rectangle and pair box}
\end{figure}

Fix a rectangular subdiagram $\mu\subset\lambda_\alpha$ and two shell boxes
$\boldsymbol{x}_1$ (addable) and $\boldsymbol{x}_2$ (removable) as in
Fig.~\ref{fig:rectangle and pair box}: $\boldsymbol{x}_2$ sits flush with the bottom
of $\mu$, and $\boldsymbol{x}_1$ sits directly above it. Split $\mu$ into three
regions A, B, C (with $\mathrm{A}\simeq\mathrm{C}$, and C being the strip between
$\boldsymbol{x}_1$ and $\boldsymbol{x}_2$). Two things can be checked directly:
\begin{align}
    \prod_{\boldsymbol{y}\in\mathrm{A}}
    \frac{1}{\sh\!\bigl(\mathcal{X}_\alpha(\boldsymbol{y})-\mathcal{X}_\alpha(\boldsymbol{x}_{2})\bigr)}
    &= \prod_{\boldsymbol{y}\in\mathrm{C}}
    \frac{1}{\sh\!\bigl((L_\lambda(\boldsymbol{y})+1)\e_1
              -A_\lambda(\boldsymbol{y})\e_2\bigr)},
    \label{eq:rect-id1}\\[4pt]
    \frac{\displaystyle\prod_{\boldsymbol{y}\in\mathrm{A}\cup\mathrm{B}}
          \sh\!\bigl(\mathcal{X}_\alpha(\boldsymbol{y})-\mathcal{X}_\alpha(\boldsymbol{x}_{1})\bigr)}
         {\displaystyle\prod_{\boldsymbol{y}\in\mathrm{B}\cup\mathrm{C}}
          \sh\!\bigl(\mathcal{X}_\alpha(\boldsymbol{y})-\mathcal{X}_\alpha(\boldsymbol{x}_{2})\bigr)}
    &= 1.
    \label{eq:rect-id2}
\end{align}
The first says that the contribution of A can be rewritten using arm/leg lengths over C.
The second says that B drops out. Putting them together, the ratio over all of $\mu$
reduces to just a product over C:
\begin{align}\label{rectangle relation}
    \prod_{\boldsymbol{y}\in\mu}
    \frac{\sh\!\bigl(\mathcal{X}_\alpha(\boldsymbol{y})-\mathcal{X}_\alpha(\boldsymbol{x}_{1})\bigr)}
         {\sh\!\bigl(\mathcal{X}_\alpha(\boldsymbol{y})-\mathcal{X}_\alpha(\boldsymbol{x}_{2})\bigr)}
    =\prod_{\boldsymbol{y}\in\mathrm{C}}
    \frac{\sh\!\bigl(\mathcal{X}_\alpha(\boldsymbol{y})-\mathcal{X}_\alpha(\boldsymbol{x}_{1})\bigr)}
         {\sh\!\bigl((L_\lambda(\boldsymbol{y})+1)\e_1
              -A_\lambda(\boldsymbol{y})\e_2\bigr)}.
\end{align}
Note that even if A and C intersect, this equality still holds. We will use this repeatedly in the next step.

\subsubsection*{Step 2: Tiling the full Young diagram}

\begin{figure}[ht]
    \centering
    \includegraphics[width=0.4\linewidth]{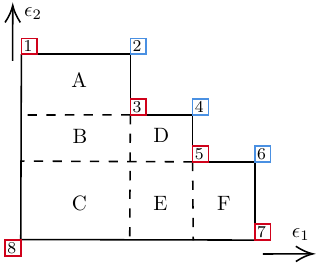}
    \caption{The 2d Young diagram $\lambda_\alpha$ split into regions A--F. Red boxes
    $\boldsymbol{x}_{1,3,5,7}$: addable; blue boxes $\boldsymbol{x}_{2,4,6}$:
    removable; $\boldsymbol{x}_8$: accounts for $1/\sh(x-\mathcal{X}(\boldsymbol{0}))$
    in~\eqref{J and nek}.}
    \label{fig:shell and nek}
\end{figure}

Now take an arbitrary Young diagram $\lambda_\alpha$ as in Fig.~\ref{fig:shell and nek}.
Label its addable boxes $\boldsymbol{x}_{1,3,5,7}$ (red) and removable boxes
$\boldsymbol{x}_{2,4,6}$ (blue), with $\boldsymbol{x}_8$ keeping track of the factor
$1/\sh(x-\mathcal{X}(\boldsymbol{0}))$ in \eqref{J and nek}. Partition $\lambda_\alpha$
into regions A,\ldots,F using the addable boxes as dividers.

Apply \eqref{rectangle relation} to each rectangle in $\lambda_\alpha$---for instance,
to $\mathrm{A}\cup\mathrm{B}\cup\mathrm{C}$ paired with $\boldsymbol{x}_{6,7}$, and
to $\mathrm{E}\cup\mathrm{F}$ paired with $\boldsymbol{x}_{3,4}$, and so on. Writing
\begin{align}
    (\mathrm{A}-\boldsymbol{x}_{2})
    &\;\equiv\;\prod_{\boldsymbol{y}\in\mathrm{A}}
    \sh\!\bigl(\mathcal{X}_{\alpha}(\boldsymbol{y})-\mathcal{X}_\alpha(\boldsymbol{x}_2)\bigr),
    \\
    (\mathrm{B}-\boldsymbol{x}_{2,4})
    &\;\equiv\;\prod_{\boldsymbol{y}\in\mathrm{B}}
    \sh\!\bigl(\mathcal{X}_{\alpha}(\boldsymbol{y})-\mathcal{X}_\alpha(\boldsymbol{x}_2)\bigr)
    \sh\!\bigl(\mathcal{X}_{\alpha}(\boldsymbol{y})-\mathcal{X}_\alpha(\boldsymbol{x}_4)\bigr),
\end{align}
and so on for the other regions, multiplying all the rectangle contributions together
gives:
\begin{align}\label{rectangle contribution}
    \frac{(\mathrm{A}-\boldsymbol{x}_2)(\mathrm{D}-\boldsymbol{x}_4)
          (\mathrm{F}-\boldsymbol{x}_6)(\mathrm{B}-\boldsymbol{x}_{2,4})
          (\mathrm{C}-\boldsymbol{x}_{2,6})(\mathrm{E}-\boldsymbol{x}_{4,6})}
         {\displaystyle\prod_{\boldsymbol{y}\in\lambda_{\alpha}}
          \sh\!\bigl((L_{\lambda_\alpha}(\boldsymbol{y})+1)\e_1
              -A_{\lambda_\alpha}(\boldsymbol{y})\e_2\bigr)
          \sh\!\bigl((A_{\lambda_\alpha}(\boldsymbol{y})+1)\e_2
              -L_{\lambda_\alpha}(\boldsymbol{y})\e_1\bigr)}.
\end{align}
The denominator is exactly the Nekrasov factor in \eqref{J and nek}.

\subsubsection*{Step 3: The remaining terms cancel}

Comparing \eqref{rectangle contribution} with the full right-hand side of \eqref{J and
nek}, there are still terms not yet accounted for:
\begin{align}\label{remain}
    \frac{(\mathrm{C}-\boldsymbol{x}_{4})}
         {(\mathrm{A}-\boldsymbol{x}_{8})(\mathrm{B}-\boldsymbol{x}_{3,8})
          (\mathrm{C}-\boldsymbol{x}_{3,5,8})(\mathrm{D}-\boldsymbol{x}_{8})
          (\mathrm{E}-\boldsymbol{x}_{5,8})(\mathrm{F}-\boldsymbol{x}_{8})}.
\end{align}
One can check directly that when all these terms are put together, they cancel:
\begin{align}
    \frac{(\mathrm{A}-\boldsymbol{x}_2)(\mathrm{B}-\boldsymbol{x}_{2,4})
          (\mathrm{C}-\boldsymbol{x}_{2,4,6})(\mathrm{D}-\boldsymbol{x}_4)
          (\mathrm{E}-\boldsymbol{x}_{4,6})(\mathrm{F}-\boldsymbol{x}_6)}
         {(\mathrm{A}-\boldsymbol{x}_{8})(\mathrm{B}-\boldsymbol{x}_{3,8})
          (\mathrm{C}-\boldsymbol{x}_{3,5,8})(\mathrm{D}-\boldsymbol{x}_{8})
          (\mathrm{E}-\boldsymbol{x}_{5,8})(\mathrm{F}-\boldsymbol{x}_{8})}=1,
\end{align}
which follows from five identities, each of which is just \eqref{rectangle relation}
applied to a different sub-region:
\begin{footnotesize}
\begin{align}
    \frac{(\mathrm{A}\cup\mathrm{B}\cup\mathrm{C}-\boldsymbol{x}_2)}
         {(\mathrm{A}\cup\mathrm{B}\cup\mathrm{C}-\boldsymbol{x}_8)}
    =\frac{(\mathrm{B}\cup\mathrm{C}-\boldsymbol{x}_8)}
          {(\mathrm{B}\cup\mathrm{C}-\boldsymbol{x}_3)}
    =\frac{(\mathrm{C}\cup\mathrm{E}\cup\mathrm{F}-\boldsymbol{x}_6)}
          {(\mathrm{C}\cup\mathrm{E}\cup\mathrm{F}-\boldsymbol{x}_8)}
    =\frac{(\mathrm{C}\cup\mathrm{E}-\boldsymbol{x}_8)}
          {(\mathrm{C}\cup\mathrm{E}-\boldsymbol{x}_5)}
    =\frac{(\mathrm{B}\cup\mathrm{C}\cup\mathrm{D}\cup\mathrm{E}-\boldsymbol{x}_4)}
          {(\mathrm{B}\cup\mathrm{C}\cup\mathrm{D}\cup\mathrm{E}-\boldsymbol{x}_8)}
    =1.
\end{align}
\end{footnotesize}
This finishes the proof for $\alpha=\beta$. The same argument works for Young diagrams
of any shape.

\subsubsection*{The case of two distinct Young diagrams ($\alpha\neq\beta$)}

\begin{figure}[ht]
    \centering
    \includegraphics[width=0.8\linewidth]{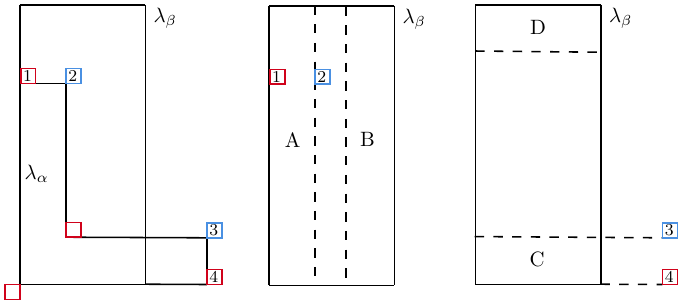}
    \caption{For two distinct Young diagrams $\lambda_\alpha$ and $\lambda_\beta$, we
    apply the same rectangle-tiling strategy to both diagrams at once.}
    \label{fig:2 lambda}
\end{figure}

When $\lambda_\alpha\neq\lambda_\beta$, we use the same rectangle-tiling idea but now
applied to both diagrams simultaneously, as shown in Fig.~\ref{fig:2 lambda}. The
relevant identities become:
\begin{align}
    \prod_{\boldsymbol{y}\in\mathrm{B}}
    \frac{1}{\sh\!\bigl(\mathcal{X}_\beta(\boldsymbol{y})-\mathcal{X}_\alpha(\boldsymbol{x}_{1})\bigr)}
    &=\prod_{\boldsymbol{y}\in\mathrm{A}}
    \frac{1}{\sh\!\bigl((A_{\lambda_\alpha}(\boldsymbol{y})+1)\e_2
              -L_{\lambda_\beta}(\boldsymbol{y})\e_1\bigr)},\\[4pt]
    \prod_{\boldsymbol{y}\in\mathrm{D}}
    \frac{1}{\sh\!\bigl(\mathcal{X}_\beta(\boldsymbol{y})-\mathcal{X}_\alpha(\boldsymbol{x}_{4})\bigr)}
    &=\prod_{\boldsymbol{y}\in\mathrm{C}}
    \frac{1}{\sh\!\bigl((L_{\lambda_\alpha}(\boldsymbol{y})+1)\e_1
              -A_{\lambda_\beta}(\boldsymbol{y})\e_2\bigr)}.
\end{align}
The remaining terms cancel by the same kind of check as in Step~3. Since the calculation
is entirely parallel and adds nothing new, we omit it.

\subsection{5d \texorpdfstring{$\Sp(2)$}{Sp(2)} SYM}\label{Sp2 example}

We verify the shell formula expressions~\eqref{ZSp plus}--\eqref{ZSp minus} for $\Sp(2)$ at $k=1$ and $k=2$, and demonstrate how the BPS jumping coefficient~\eqref{BPS jumping coeff} arises and is absorbed by the limiting procedure $\lim_{\e_2\to-\e_1}$. As a consistency check, the $k=1$ result should reproduce the $\SU(2)$ partition function by the Lie algebra isomorphism $\Sp(2)\simeq\SU(2)$.

\paragraph{$k=1$.} Since the corresponding auxiliary gauge group has rank $0$ at level $k=1$, this reduces to a single term with no integration. The partition function follows directly from~\eqref{ZSp plus} and~\eqref{ZSp minus}:
    \begin{align}
    \mathcal{Z}^{\Sp,+}_{2,k=1}=&\lim_{\e_2\to-\e_1}\frac{\prod_{\alpha=1}^{5}\mathcal{J}\big(0 \big|\emptyset_{ \alpha} \big)}{\sh(0+v_{1}-\e_{12})}\cr=&\lim_{\e_2\to-\e_1}\frac{1}{\sh(v_1-\e_{12})\prod_{\alpha=1}^5\sh(-v_\alpha)}\cr=&\lim_{\e_2\to-\e_1}\frac{-1}{\sh(\frac{\e_1}{2})\sh(\frac{\e_1}{2}+\pi i)\sh(\e_1)\sh(\frac{\e_{12}}{2}+\pi i)\sh(v_1)\sh(v_1-\e_{12})}\cr=&\frac{1}{2\sh(v_1)^2\sh(\e_1)^2}\cr\mathcal{Z}^{\Sp,-}_{2,k=1}=&\lim_{\e_2\to-\e_1}\frac{\prod_{\alpha=1}^{5}\mathcal{J}\big(\pi i\big|\emptyset_{ \alpha} \big)}{\sh(-\pi i+v_{1}-\e_{12})}\cr=&\lim_{\e_2\to-\e_1}\frac{-1}{\sh(\frac{\e_1}{2})\sh(\frac{\e_1}{2}+\pi i)\sh(\e_1)\sh(\frac{\e_{12}}{2}+\pi i)\sh(v_1+\pi i)\sh(v_1-\e_{12}+\pi i)}\cr=&-\frac{\sh(v_1)^2}{2\sh(2v_1)^2\sh(\e_1)^2}
\end{align}
where we substituted the frozen brane values $v_2,\ldots,v_5$ from Tab.~\ref{tab:SP poles} and used the identity $\sh(x+\pi i)=i\sh(2x)/\sh(x)$. A further identity $\sh(x+\pi i)^2+\sh(x)^2=-4$ then gives the full $k=1$ partition function:
\begin{align}
    \mathcal{Z}^{\Sp}_{2,k=1}=\mathcal{Z}^{\Sp,+}_{2,k=1}+\mathcal{Z}^{\Sp,-}_{2,k=1}=\frac{2}{\sh(2v_1)^2\sh(\e_1)^2}
\end{align}
This agrees exactly with the $\SU(2)$ unrefined instanton partition function, as required by the Lie algebra isomorphism $\Sp(2)\simeq\SU(2)$.

\paragraph{$k=2$.} For the minus sector, $k=2$ admits no non-trivial Young diagrams, so the result is immediate:
\begin{align}
    \mathcal{Z}^{\Sp,-}_{2,k=2}=&\lim_{\e_2\to-\e_1}\frac{\prod_{\alpha=1}^{5}\mathcal{J}\big(0 \big|\emptyset_{ \alpha} \big)}{\sh(0+v_{1}-\e_{12})}\frac{\prod_{\alpha=1}^{5}\mathcal{J}\big(\pi i\big|\emptyset_{ \alpha} \big)}{\sh(-\pi i+v_{1}-\e_{12})}\cr=&\frac{-1}{\sh(2v_1)^2\sh(\e_1)^2\sh(2\e_1)^2}
\end{align}
However, the plus sector at level $k=2$ possesses one degree of freedom corresponding to a single box. This box can be placed on any of the five empty Young diagrams, resulting in a total of five distinct Young diagram configurations $\vec\lambda$:
\begin{align}
    &(\{(1,1)\}_{1},\emptyset_2,\emptyset_3,\emptyset_4,\emptyset_5),\quad(\emptyset_1,\{(1,1)\}_{2},\emptyset_3,\emptyset_4,\emptyset_5),\quad(\emptyset_1,\emptyset_2,\{(1,1)\}_{3},\emptyset_4,\emptyset_5)\cr&(\emptyset_1,\emptyset_2,\emptyset_3,\{(1,1)\}_{4},\emptyset_5),\quad(\emptyset_1,\emptyset_2,\emptyset_3,\emptyset_4,\{(1,1)\}_{5})
\end{align}
We also know the $\mathcal{J}$-factor for a single 2d box from~\eqref{J 2dYD 1box}. The plus sector partition function is then (first equality uses $\mathcal{J}(x|\emptyset_\alpha)=1/\sh(x-v_\alpha)$; second equality substitutes the frozen brane values from Tab.~\ref{tab:SP poles}; the limit $\e_2\to-\e_1$ reduces the five contributions to three distinct types):
\begin{footnotesize}
    \begin{align}
    \mathcal{Z}^{\Sp,+}_{2,k=2}=&\lim_{\e_2\to-\e_1}\sum_{i=1}^5\frac{\mathcal{J}\big(\pm v_i\big|\{(1,1)\}_i\big)}{\sh(v_i\pm v_i-\e_{12})\prod_{j\neq i}^5\sh(\pm v_i-v_j)}\cr=&\lim_{\e_2\to-\e_1}\Big(-\frac{1}{\sh(2v_1\pm\e_1)\sh(2v_1+\e_{1,2})\sh(\e_{1,2})\sh(2v_1-\e_{12})^2}\cr&\qquad +\frac{1}{2\sh(v_1\pm\frac{\e_1}{2})\sh(v_1-\e_{12}\pm\frac{\e_1}{2})\sh(2\e_1)^2\sh(\e_2)^2}\cr&\qquad+\frac{\sh(v_1\pm\frac{\e_1}{2})\sh(v_1-\e_{12}\pm\frac{\e_1}{2})}{2\sh(2v_1\pm\e_1)\sh(2v_1-2\e_{12}\pm\e_1)\sh(2\e_1)^2\sh(\e_2)^2}\cr&\qquad -\frac{\sh(\e_{12}^2)}{2\sh(v_1\pm\frac{\e_{12}}{2})\sh(v_1-\e_{12}\pm\frac{\e_{12}}{2})\sh(\e_1)\sh(\e_2)^2\sh(2\e_1+\e_2)^2\sh(\e_1+2\e_2)}\cr&\qquad -\frac{\sh(\e_{12})^2\sh(v_1\pm\frac{\e_12}{2})\sh(v_1-\e_{12}\pm\frac{\e_{12}}{2})}{2\sh(2v_1\pm\e_{12})\sh(2v_1-2\e_{12}\pm\e_{12})\sh(\e_1)\sh(\e_2)^2\sh(\e_1+2\e_2)\sh(2\e_1+\e_2)^2}\Big)\cr=&\frac{1}{\sh(2v_1)^2\sh(2v_1\pm\e_1)^2\sh(\e_1)^2}+\frac{1}{2\sh(v_1\pm\frac{\e_1}{2})^2\sh(\e_1)^2\sh(2\e_1)^2}+\frac{\sh(v_1\pm\frac{\e_1}{2})^2}{2\sh(2v_1\pm\e_1)^2\sh(\e_1)^2\sh(2\e_1)^2}\cr
\end{align}
\end{footnotesize}
Here, at the first equality, we have used $\mathcal{J}\big(x\big|\emptyset_\alpha\big)=1/\sh(x-v_\alpha)$. At the second equality, we substituted the specific values of the frozen branes as Tab.~\ref{tab:SP poles}. The complete partition function for the $\Sp(2)$ $k=2$ theory is:
\begin{align}
    \mathcal{Z}^{\Sp}_{2,k=2}=&\frac{1}{\sh(2v_1)^2\sh(2v_1\pm\e_1)^2\sh(\e_1)^2}+\frac{1}{2\sh(v_1\pm\frac{\e_1}{2})^2\sh(\e_1)^2\sh(2\e_1)^2}\cr&+\frac{\sh(v_1\pm\frac{\e_1}{2})^2}{2\sh(2v_1\pm\e_1)^2\sh(\e_1)^2\sh(2\e_1)^2}-\frac{1}{\sh(2v_1)^2\sh(\e_1)^2\sh(2\e_1)^2}
\end{align}

\paragraph{BPS jumping coefficient.} To illustrate the emergence of the BPS jumping coefficients~\eqref{BPS jumping coeff} and their absorption into the shell formula, we consider the configuration $\vec\lambda=(\emptyset_1,\lambda_2,\emptyset_3,\emptyset_4,\emptyset_5)$, where $\lambda_2$ is the first Young diagram shape that produces a non-trivial coefficient. It is the L-shaped diagram:
    \begin{align}
        \lambda_2=\{(1,1),(1,2),(1,3),(2,1),(3,1),(2,2)\}
    \end{align}
    The corresponding $\mathcal{J}$-factor is given by:
    \begin{align}
        \mathcal{J}\big(x\big|\lambda_2\big)=\frac{\sh(x-v_2-\e_1-3\e_2)\sh(x-v_2-2\e_{12})\sh(x-v_2-3\e_1-\e_2)}{\sh(x-v_2-3\e_{1,2})\sh(x-v_2-\e_1-2\e_2)\sh(x-v_2-2\e_1-\e_2)}
    \end{align}
    The corresponding contribution of $\vec\lambda$ is then:
    \begin{align}
        \mathcal{Z}^{\Sp,+}(\vec{\lambda})\propto\lim_{\e_2\to-\e_1} \frac{\sh(\e_{12})^3}{2\sh(2\e_{12})^2\sh(3\e_{12})}=\frac{1}{24}
    \end{align}
    The term in $\mathcal{Z}^{\Sp,+}$ corresponding to $\vec{\lambda}$ thus carries an extra coefficient $\frac{1}{24}$ at the unrefined limit. Indeed, the exact result is:
    \begin{align}\label{Sp result 1/24}
        \mathcal{Z}^{\Sp,+}(\vec{\lambda})=&\frac{1}{24\sh(\e_1)^2\sh(2\e_1)^4\sh(3\e_1)^6\sh(4\e_1)^6\sh(5\e_1)^5\sh(6\e_1)^2}\cr&\times\frac{1}{\sh(v_1\pm\frac52\e_1)^2\sh(v_1\pm\frac32\e_1)^4\sh(v_1\pm\frac12\e_1)^6}
    \end{align}
    By comparing the Eq.(2.12) of \cite{Nawata:2021dlk}, the expression for the $\Sp(2)$ plus sector with the configuration $\vec{\lambda}=(\emptyset_1,\lambda_2,\emptyset_3,\emptyset_4,\emptyset_5)$, is given by:
\begin{align}\label{Sp result 1/64}
    \mathcal{Z}^{\Sp,+}(\vec{\lambda})=&C^{\Sp}_{\vec{\lambda},\boldsymbol{v}}\times\frac{1}{64\sh(\e_1)^2\sh(2\e_1)^4\sh(3\e_1)^6\sh(4\e_1)^6\sh(5\e_1)^5\sh(6\e_1)^2}\cr&\times\frac{1}{\sh(v_1\pm\frac52\e_1)^2\sh(v_1\pm\frac32\e_1)^4\sh(v_1\pm\frac12\e_1)^6}
\end{align}

In this configuration, the number of diagonal boxes in $\lambda_2$ is $j=2$. Hence, by~\eqref{BPS jumping coeff}, the coefficient $C^{\Sp}_{\vec{\lambda},\boldsymbol{v}}$ reads:
\begin{align}
    C^{\Sp}_{\vec{\lambda},\boldsymbol{v}}= C^{\Sp}_{\emptyset,\frac{\e_1}{2}}C^{\Sp}_{\lambda_2,\frac{\e_1}{2}+\pi i}C^{\Sp}_{\emptyset,0}C^{\Sp}_{\emptyset,\pi i}=\frac{2^{2j-1}}{\binom{2j-1}{j-1}}=\frac{8}{3}
\end{align}
Thus, the factor $\frac{1}{64}$ contained in~\eqref{Sp result 1/64}, combined with the prefactor $C^{\Sp}_{\vec{\lambda},\boldsymbol{v}}=\frac{8}{3}$, yields the complete coefficient $\frac{1}{24}$, which matches exactly the result in~\eqref{Sp result 1/24}. This example demonstrates that, through careful implementation of the limiting procedure $\lim_{\e_2\to-\e_1}$, the coefficient $C^{\mathrm{Sp}}_{\vec{\lambda},\boldsymbol{v}}$ can be fully absorbed into the shell formula.

\subsection{D0-D6 partition function}\label{D0-D6 appendix}

We compute the $k=1$ and $k=2$ instanton contributions for a single D6$_{\bar{4}}$-brane using the shell formula~\eqref{D0D6shell}, verify that the results match the MacMahon function coefficients under the CY3 condition $\e_{123}=0$, and confirm the recursion relation~\eqref{D6recursion} explicitly. We focus on the configuration $\{(1,1,1),(1,1,2)\}$ at $k=2$ as a representative; the other two $k=2$ configurations follow by symmetry.

\paragraph{$k=1$.} The partition function~\eqref{D0D6partition} involves only a first-order contour integral, and there is a unique 3d Young diagram configuration $\vec\pi=(\{(1,1,1)\})$. Computing the shellboxes and their charges (Tab.~\ref{tab:3d 1box charge}) gives:
   \begin{table}[ht]
       \centering
       \begin{tabular}{|c|cccc|}
           \hline  $\operatorname{Q}=+1$ & $(1,2,2)$ & $(2,1,2)$ & $(2,2,1)$ & \\ \hline
           $\operatorname{Q}=-1$ & $(1,1,2)$ & $(1,2,1)$ & $(2,1,1)$ & $(2,2,2)$\\ \hline
       \end{tabular}
       \caption{The charges of the shellboxes of the 3d Young diagram $\{(1,1,1)\}$. The second column displays the coordinates of each shellbox.}
       \label{tab:3d 1box charge}
   \end{table}
   
    Therefore the $\mathcal{J}$-factor of $\{(1,1,1)\}$ is:
   \begin{align}
       \mathcal{J}\big(x\big|\{(1,1,1)\}_{\overline{4},1}\big)=\frac{\sh(x-v_{\overline{4},1}-\e_{12})\sh(x-v_{\overline{4},1}-\e_{13})\sh(x-v_{\overline{4},1}-\e_{23})}{\sh(x-v_{\overline{4},1}-\e_{1,2,3})\sh(x-v_{\overline{4},1}-\e_{123})}
   \end{align}
   
   The partition function~\eqref{D0D6shell} becomes:
    \begin{align}
        \mathcal{Z}^{\text{D0-D6}}_{(1,0,0,0),k=1}=&\sh(\mathcal{X}_{\overline{4},1}(1,1,1)-\mathcal{X}_{\overline{4},1}(0,0,0))\mathcal{J}\big(\mathcal{X}_{\overline{4},1}(1,1,1)\big|\{(1,1,1)\}_{\overline{4},1}\big)\cr=&-\frac{\sh(\e_{12})\sh(\e_{13})\sh(\e_{23})}{\sh(\e_1)\sh(\e_2)\sh(\e_3)}
    \end{align}
Under the CY threefold condition $\e_{123} = 0$ \cite{Cirafici_2009}, this reduces to $1$, matching the first-order coefficient of the MacMahon function:
\begin{align}\label{MacMahon}
    \prod_{k=1}^\infty\frac{1}{(1-\mathfrak{q}^k)^k}=1+\mathfrak{q}+3\mathfrak{q}^2+6\mathfrak{q}^3+13\mathfrak{q}^4+\ldots
\end{align}

\paragraph{$k=2$.} There are three 3d Young diagram configurations:
    \begin{center}
\begin{minipage}[t]{0.3\textwidth}
    \centering
    \includegraphics[width=0.25\linewidth]{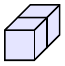}
    \par\vspace{2mm}
    $\{(1,1,1),(2,1,1)\}$
    
\end{minipage}
\hfill
\begin{minipage}[t]{0.3\textwidth}
    \centering
    \includegraphics[width=0.3\linewidth]{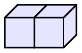}
    \par\vspace{2mm}
    $\{(1,1,1),(1,2,1)\}$
\end{minipage}
\hfill
\begin{minipage}[t]{0.3\textwidth}
    \centering
    \includegraphics[width=0.18\linewidth]{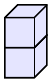}
    \par\vspace{2mm}
    $\{(1,1,1),(1,1,2)\}$
    
\end{minipage}
\end{center}
    Let us focus on the configuration $(\{(1,1,1),(1,1,2)\})$, the charges of the shellboxes are listed in Tab.~\ref{tab:3d 2box charge}.
    \begin{table}[ht]
        \centering
        \begin{tabular}{|c|cccc|}
           \hline  $\operatorname{Q}=+1$ & $(1,2,3)$ & $(2,1,3)$ & $(2,2,1)$ & \\ \hline
           $\operatorname{Q}=-1$ & $(1,1,3)$ & $(1,2,1)$ & $(2,1,1)$ & $(2,2,3)$\\ \hline
       \end{tabular}
         \caption{The charges of the shellboxes of the 3d Young diagram $\{(1,1,1),(1,1,2)\}$.}
       \label{tab:3d 2box charge}
    \end{table}

    Therefore the corresponding $\mathcal{J}$-factor is:
    \begin{align}
        \mathcal{J}\big(x\big|\{(1,1,1),(1,1,2)\}_{\overline{4},1}\big)=\frac{\sh(x-v_{\overline{4},1}-\e_{12})\sh(x-v_{\overline{4},1}-\e_{1}-2\e_3)\sh(x-v_{\overline{4},1}-\e_{2}-2\e_{3})}{\sh(x-v_{\overline{4},1}-\e_{1,2})\sh(x-v_{\overline{4},1}-2\e_3)\sh(x-v_{\overline{4},1}-\e_{12}-2\e_3)}
    \end{align}
    The contribution from $\{(1,1,1),(1,1,2)\}$ is:
    \begin{align}
        \mathcal{Z}^{\text{D0-D6}}_{(1,0,0,0)}(\{(1,1,1),(1,1,2)\}_{\4,1})=\frac{\sh(\e_{12})\sh(\e_{13})\sh(\e_{23})\sh(\e_{12}-\e_3)\sh(\e_1+2\e_3)\sh(\e_2+2\e_3)}{\sh(\e_1)\sh(\e_2)\sh(\e_3)\sh(2\e_3)\sh(\e_1-\e_3)\sh(\e_2-\e_3)}
    \end{align}
    By symmetry among $\e_1$, $\e_2$, $\e_3$, the contributions from $\{(1,1,1),(2,1,1)\}$ and $\{(1,1,1),(1,2,1)\}$ follow immediately. Under the CY3 condition $\e_{123}=0$ each of the three terms reduces to $1$, giving $\mathcal{Z}^{\text{D0-D6}}_{(1,0,0,0),k=2}=3$, which matches the second-order MacMahon coefficient~\eqref{MacMahon}. This confirms that the shell formula reproduces the expected enumerative geometry result at low instanton number.

\paragraph{Recursion relation~\eqref{D6recursion}.} The ratio of consecutive instanton contributions is:
    \begin{align}
        \frac{\mathcal{Z}^{\text{D0-D6}}(\{(1,1,1),(1,1,2)\}_{\overline{4},1})}{\mathcal{Z}^{\text{D0-D6}}(\{(1,1,1)\}_{\overline{4},1})}=-\frac{\sh(\e_{12}-\e_3)\sh(\e_{1}+2\e_3)\sh(\e_2+2\e_3)}{\sh(\e_1-\e_3)\sh(\e_2-\e_3)\sh(2\e_3)}
    \end{align}
    The recursion relation~\eqref{D6recursion} is verified computationally: evaluating the $\mathcal{J}$-factors at the new box location $(1,1,2)$ in the enlarged diagram, and at the shifted point $(2,2,3)$ in the original diagram, gives:
    \begin{align}
        &\mathcal{J}\big(\mathcal{X}_{\overline{4},1}(1,1,2)\big|\{(1,1,1),(1,1,2)\}_{\overline{4},1}\big)=-\frac{\sh(\e_{13})\sh(\e_{23})\sh(\e_{12}-\e_3)}{\sh(\e_3)\sh(\e_{123})\sh(\e_1-\e_3)\sh(\e_2-\e_3)}\cr&\mathcal{J}\big(\mathcal{X}_{\overline{4},1}(2,2,3)\big|\{(1,1,1)\}_{\overline{4},1}\big)=\frac{\sh(2\e_3)\sh(\e_{13})\sh(\e_{23})}{\sh(\e_3)\sh(\e_{123})\sh(\e_1+2\e_3)\sh(\e_2+2\e_3)}
    \end{align}
    Their ratio reproduces the partition function ratio, confirming~\eqref{D6recursion}:
    \begin{align}
       \frac{\mathcal{Z}^{\text{D0-D6}}(\{(1,1,1),(1,1,2)\}_{\overline{4},1})}{\mathcal{Z}^{\text{D0-D6}}(\{(1,1,1)\}_{\overline{4},1})}=\frac{\mathcal{J}\big(\mathcal{X}_{\overline{4},1}(1,1,2)\big|\{(1,1,1),(1,1,2)\}_{\overline{4},1}\big)}{\mathcal{J}\big(\mathcal{X}_{\overline{4},1}(2,2,3)\big|\{(1,1,1)\}_{\overline{4},1}\big)}
    \end{align}
\subsection{DT3 counting}\label{appendix DT}

This subsection presents three calculations: (i) an explicit $k=1$ residue computation for a 1-leg vacuum with boundary $\lambda=\{(1,1),(1,2),(2,1)\}$, demonstrating that the asymptotic contribution cancels and the $\mathcal{J}$-factor reduces to a finite 2d contribution as in~\eqref{1leg J}; (ii) a recursive derivation of the same result using~\eqref{D6recursion}, which generalizes to arbitrary $k$ and yields the DT3 integrand~\eqref{DT partition}; and (iii) the simplest 3-leg case $\pi_{\lambda\lambda\lambda}$ with $\lambda=\{(1,1)\}$. Since there is only a single D6-brane throughout, we omit the Young diagram label $(123,1)$.

\begin{figure}[ht]
    \centering
    \includegraphics[width=0.85\linewidth]{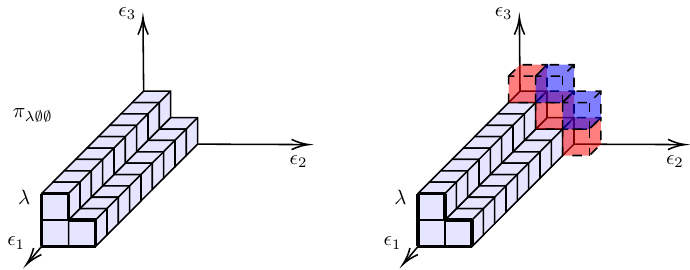}
    \caption{The left figure shows a one-leg 3d Young diagram $\pi_{\lambda\emptyset\emptyset}$ with boundaries, where the boundary in the $\e_1$ direction is a 2d Young diagram $\lambda$ in the 23-plane, and the boundaries in the other two directions are empty sets $\emptyset$. The right figure shows the shellboxes with non-zero charges for $\pi_{\lambda\emptyset\emptyset}$, where red represents $-1$ charge and blue represents $+1$ charge. Note that the contributions from all shellboxes at the infinite boundary are canceled out; therefore, only the finite-distance shellboxes provide non-trivial contributions to the partition function.}
    \label{fig:D6D2 1leg}
    \end{figure}

\paragraph{1-leg case: direct residue computation.} We consider the partition function of three D2$_1$-branes inside a D6$_{123}$-brane. The vacuum is a minimal 3d Young diagram $\pi_{\lambda\emptyset\emptyset}$ extending infinitely in the $\C_1$ direction with boundary condition $\lambda_{23,1}=\{(1,1),(1,2),(2,1)\}$, as shown in Fig.~\ref{fig:D6D2 1leg}. 
    The charges of the shellboxes of $\pi_{\lambda\emptyset\emptyset}$ are listed in Tab.~\ref{tab:3d 1leg charge}.
 \begin{table}[ht]
        \centering
        \begin{tabular}{|c|cccccc|}
           \hline  $\operatorname{Q}=+1$ & $(1,2,3)$&$(1,3,2)$&$(\infty,1,3)$&$(\infty,2,2)$&$(\infty,3,1)$&\\ \hline
           $\operatorname{Q}=-1$ &$(1,1,3)$&$(1,2,2)$&$(1,3,1)$&$(\infty,1,1)$&$(\infty,2,3)$&$(\infty,3,2)$\\ \hline
       \end{tabular}
         \caption{The charges of the shellboxes of the 3d Young diagram $\pi_{\lambda\emptyset\emptyset}$.}
       \label{tab:3d 1leg charge}
 \end{table}
 
As~\eqref{D2 infinite end} mentioned, the contribution to the $\mathcal{J}$-factor from the shellboxes at the infinite end reads:
\begin{align}
    \frac{\sh(x-\infty\e_1-2\e_3)\sh(x-\infty\e_1-\e_{23})\sh(x-\infty\e_1-2\e_2)}{\sh(x-\infty\e_1)\sh(x-\infty\e_1-\e_2-2\e_3)\sh(x-\infty\e_1-2\e_2-\e_3)}=1
\end{align}
Therefore, the complete $\mathcal{J}$-factor as~\eqref{1leg J} is:
\begin{align}
    \mathcal{J}\big(x\big|\pi_{\lambda\emptyset\emptyset}\big)=&\frac{\sh(x-v-\e_2-2\e_3)\sh(x-v-2\e_2-\e_3)}{\sh(x-v-2\e_2)\sh(x-v-2\e_3)\sh(x-v-\e_{23})}\cr=&\mathcal{J}\big(x+v_{23,1}-v\big|\lambda_{23,1}\big)
\end{align}
The partition function before integration reads:
    \begin{align}
    \mathcal{I}^{\text{D0-D2-D6}}_{\lambda,\emptyset,\emptyset;k}=(-1)^k\mathcal{I}^{\text{D0-D0}}_k\times\prod_{i=1}^k\frac{\sh(\phi_i-v-\e_{234})}{\sh(\phi_i-v-\e_{23})}\frac{\sh(\phi_i-v-2\e_{2,3}-\e_4)}{\sh(\phi_i-v-2\e_{2,3})}\frac{\sh(-\phi_i+v+\e_{2,3}-\e_{14})}{\sh(-\phi_i+v+\e_{2,3}-\e_{1})}
\end{align}

For the vacuum configuration $\pi_{\lambda\emptyset\emptyset}$ at $k=1$, there are three poles, which correspond to three distinct box locations:
 \begin{align}
     \widetilde{\pi}_{\lambda\emptyset\emptyset}=\pi\backslash\pi_{\lambda\emptyset\emptyset}=\{(1,2,2)\},\quad\{(1,3,1)\},\text{ and }\{(1,1,3)\}
 \end{align}
 where $|\widetilde{\pi}_{\lambda\emptyset\emptyset}|=k=1$. Using ~\eqref{DT partition}, the partition functions read:
 \begin{align}
      &\mathcal{Z}^{\text{D0-D2-D6}}_{\lambda,\emptyset,\emptyset}(1,2,2)=\frac{\sh(\e_{23})\sh(\e_1+2\e_{2,3})}{\sh(\e_1)\sh(\e_2-\e_3)^2}\cr&\mathcal{Z}^{\text{D0-D2-D6}}_{\lambda,\emptyset,\emptyset}(1,3,1)=\frac{\sh(\e_{13})\sh(\e_{23})\sh(\e_1+2\e_2)\sh(\e_2-2\e_3)\sh(\e_1+3\e_2-\e_3)}{\sh(\e_{1,2})\sh(\e_2-\e_3)\sh(2\e_2-2\e_3)\sh(\e_1+2\e_2-\e_3)}\cr&\mathcal{Z}^{\text{D0-D2-D6}}_{\lambda,\emptyset,\emptyset}(1,1,3)=-\frac{\sh(\e_{12})\sh(\e_{23})\sh(2\e_2-\e_3)\sh(\e_1+2\e_3)\sh(\e_1-\e_2+3\e_3)}{\sh(\e_{1,3})\sh(\e_2-\e_3)\sh(2\e_2-2\e_3)\sh(\e_1-\e_2+2\e_3)}\cr&\mathcal{Z}^{\text{D0-D2-D6}}_{\lambda,\emptyset,\emptyset;k=1}=\mathcal{Z}^{\text{D0-D2-D6}}_{\lambda,\emptyset,\emptyset}(1,2,2)+\mathcal{Z}^{\text{D0-D2-D6}}_{\lambda,\emptyset,\emptyset}(1,3,1)+\mathcal{Z}^{\text{D0-D2-D6}}_{\lambda,\emptyset,\emptyset}(1,1,3)\cr
 \end{align}

\paragraph{Derivation via recursion relation~\eqref{D6recursion}.} An equivalent approach identifies the D2-brane with an infinite row of D0-branes. Interpreted as additional boxes placed on the vacuum 3d Young diagram $\pi_{\lambda\emptyset\emptyset}$ with $\lambda=\{(1,1),(1,2),(2,1)\}$, the DT3 invariant follows from the D0-D6 recursion relation~\eqref{D6recursion}:
 \begin{align}
     \mathcal{Z}^{\text{D0-D2-D6}}_{\lambda,\emptyset,\emptyset}(1,2,2)&=\frac{\mathcal{J}\big(\mathcal{X}(1,2,2)\big|\pi_{\lambda\emptyset\emptyset}\cup\{(1,2,2)\}\big)}{\mathcal{J}\big(\mathcal{X}(1,2,2)+\e_{123}\big|\pi_{\lambda\emptyset\emptyset}\big)}\cr&=\frac{\mathcal{J}\big(\mathcal{X}(1,2,2)\big|\pi_{\lambda\emptyset\emptyset}\big)}{\mathcal{J}\big(\mathcal{X}(1,2,2)+\e_{123}\big|\pi_{\lambda\emptyset\emptyset}\big)}\sh(0)\mathcal{J}\big(\mathcal{X}(1,2,2)\big|\{(1,2,2)\}\big)\cr&=\frac{\sh(\e_{23})\sh(\e_1+2\e_{2,3})}{\sh(\e_1)\sh(\e_2-\e_3)^2}
 \end{align}
 where the second equality is obtained by applying the splitting property~\eqref{splitting}: the $\mathcal{J}$-factor on the enlarged diagram $\pi_{\lambda\emptyset\emptyset}\cup\{(1,2,2)\}$ factorizes into the $\mathcal{J}$-factor on the vacuum times a single-box contribution at the new box location, with the interface $\sh(0)$ factor canceling trivially. The contributions $\mathcal{Z}^{\text{D0-D2-D6}}_{\lambda,\emptyset,\emptyset}(1,1,3)$ and $\mathcal{Z}^{\text{D0-D2-D6}}_{\lambda,\emptyset,\emptyset}(1,3,1)$ from the other two poles follow similarly.

\paragraph{Derivation of the general DT3 integrand~\eqref{DT partition}.} The recursive approach generalizes to arbitrary $k$. For a given pole corresponding to a set $\widetilde{\pi}_{\lambda\mu\nu} = \{\boldsymbol{x}_1, \ldots, \boldsymbol x_k\}$ arranged so that each step $\pi_{\lambda\mu\nu} \cup \{\boldsymbol x_1\},\, \pi_{\lambda\mu\nu} \cup \{\boldsymbol x_1\} \cup \{\boldsymbol x_2\}, \ldots$ remains a valid 3d Young diagram, iterating the recursion relation~\eqref{D6recursion} gives:
        \begin{align}
        \mathcal{Z}^{\text{D0-D2-D6}}_{\lambda,\mu,\nu}(\boldsymbol{x}_1,\boldsymbol{x}_2,\ldots,\boldsymbol{x}_k)=&\frac{\mathcal{J}\big(\mathcal{X}(\boldsymbol{x}_1)\big|\pi_{\lambda\mu\nu}\cup\{\boldsymbol{x}_1\}\big)}{\mathcal{J}\big(\mathcal{X}(\boldsymbol{x}_1)+\e_{123}\big|\pi_{\lambda\mu\nu}\big)}\frac{\mathcal{J}\big(\mathcal{X}(\boldsymbol{x}_2)\big|\pi_{\lambda\mu\nu}\cup\{\boldsymbol{x}_1\}\cup\{\boldsymbol{x}_2\}\big)}{\mathcal{J}\big(\mathcal{X}(\boldsymbol{x}_2)+\e_{123}\big|\pi_{\lambda\mu\nu}\cup\{\boldsymbol{x}_1\}\big)}\times\ldots\cr=&\left(\prod_{i=1}^k\frac{\mathcal{J}\big(\mathcal{X}(\boldsymbol{x}_i)\big|\pi_{\lambda\mu\nu}\big)}{\mathcal{J}\big(\mathcal{X}(\boldsymbol{x}_i)+\e_{123}\big|\pi_{\lambda\mu\nu}\big)}\right)\prod_{i,j}^k\sh(\mathcal{X}(\boldsymbol{x}_i)-\mathcal{X}(\boldsymbol{x}_j))\mathcal{J}\big(\mathcal{X}(\boldsymbol{x}_i)\big|\boldsymbol{x}_j\big)\cr=&\left(\prod_{i=1}^k\frac{\mathcal{J}\big(\mathcal{X}(\boldsymbol{x}_i)\big|\pi_{\lambda\mu\nu}\big)}{\mathcal{J}\big(\mathcal{X}(\boldsymbol{x}_i)+\e_{123}\big|\pi_{\lambda\mu\nu}\big)}\right)\cr&\qquad\times\left(\prod_{i,j}^k\frac{\sh(\mathcal{X}(\boldsymbol{x}_i)-\mathcal{X}(\boldsymbol{x}_j))\sh(\mathcal{X}(\boldsymbol{x}_i)-\mathcal{X}(\boldsymbol{x}_j)-\e_{12,13,23})}{\sh(\mathcal{X}(\boldsymbol{x}_i)-\mathcal{X}(\boldsymbol{x}_j)-\e_{1,2,3})\sh(\mathcal{X}(\boldsymbol{x}_i)-\mathcal{X}(\boldsymbol{x}_j)-\e_{123})}\right)\cr
    \end{align}
    The passage from the first to the second line applies the splitting property~\eqref{splitting} to factor the $\mathcal{J}$-factor on the enlarged diagram into a vacuum contribution and a single-box contribution at each new box location. The swapping property~\eqref{swapping prop} then converts the resulting double product into a symmetric pairwise form. Note that although $\sh(0)$ appears when $i = j$ in the second line, the overall expression remains finite.

    Substituting the integration variables $\phi_i$ for $\mathcal{X}(\boldsymbol{x}_i)$ and replacing the denominator $\mathcal{J}$-factor with $\mathcal{J}_-$ to enforce the correct pole structure, one arrives at the integral form of the DT3 invariant~\eqref{DT partition}.

\begin{figure}[ht]
    \centering
    \includegraphics[width=0.9\linewidth]{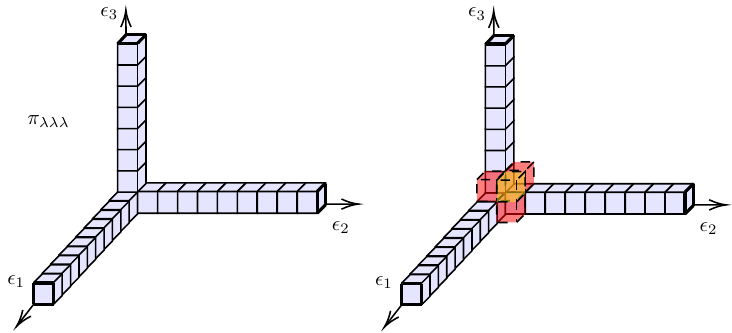}
    \caption{The left image shows the 3d Young diagram $\pi_{\lambda\lambda\lambda}$ corresponding to the vacuum configuration $(\lambda, \lambda, \lambda)$, where $\lambda=\{(1,1)\}$ is a 2d Young diagram with a single box. The right image displays the shellboxes with non‑trivial charges: red boxes carry charge $-1$, and yellow boxes carry charge $+2$.}
    \label{fig:3leg}
\end{figure}

\paragraph{3-leg case: $\pi_{\lambda\lambda\lambda}$ with $\lambda=\{(1,1)\}$.} The simplest 3-leg vacuum arises when all three legs carry a single-box boundary. A key simplification occurs because all pairwise and triple intersections in the inclusion-exclusion formula~\eqref{J 3leg} coincide with a single box: $\pi_{\lambda\emptyset\emptyset}\cap\pi_{\emptyset\lambda\emptyset} = \pi_{\lambda\emptyset\emptyset}\cap\pi_{\emptyset\emptyset\lambda} = \pi_{\emptyset\lambda\emptyset}\cap\pi_{\emptyset\emptyset\lambda} = \pi_{\lambda\emptyset\emptyset}\cap\pi_{\emptyset\lambda\emptyset}\cap\pi_{\emptyset\emptyset\lambda} = \{(1,1,1)\}$, so all intersection $\mathcal{J}$-factors are equal and the formula reduces considerably. Using~\eqref{J 3leg} and Fig.~\ref{fig:3leg}:
    \begin{align}
        \mathcal{J}\big(x\big|\pi_{\lambda\lambda\lambda}\big)=&\frac{\mathcal{J}\big(x\big|\pi_{\lambda\emptyset\emptyset}\big)\mathcal{J}\big(x\big|\pi_{\emptyset\lambda\emptyset}\big)\mathcal{J}\big(x\big|\pi_{\emptyset\emptyset\lambda}\big)\mathcal{J}\big(x\big|\pi_{\lambda\emptyset\emptyset}\cap\pi_{\emptyset\lambda\emptyset}\cap\pi_{\emptyset\emptyset\lambda}\big)}{\mathcal{J}\big(x\big|\pi_{\lambda\emptyset\emptyset}\cap\pi_{\emptyset\lambda\emptyset}\big)\mathcal{J}\big(x\big|\pi_{\lambda\emptyset\emptyset}\cap\pi_{\emptyset\emptyset\lambda}\big)\mathcal{J}\big(x\big|\pi_{\emptyset\lambda\emptyset}\cap\pi_{\emptyset\emptyset\lambda}\big)}\cr=&\frac{\sh(x-v-\e_{123})^2}{\sh(x-v-\e_{12})\sh(x-v-\e_{13})\sh(x-v-\e_{23})}
    \end{align}
    where in this configuration:
    \begin{align}
        \pi_{\lambda\emptyset\emptyset}\cap\pi_{\emptyset\lambda\emptyset}\cap\pi_{\emptyset\emptyset\lambda}=\pi_{\lambda\emptyset\emptyset}\cap\pi_{\emptyset\lambda\emptyset}=\pi_{\lambda\emptyset\emptyset}\cap\pi_{\emptyset\emptyset\lambda}=\pi_{\emptyset\lambda\emptyset}\cap\pi_{\emptyset\emptyset\lambda}=\{(1,1,1)\}
    \end{align}
    The integrand is:
    \begin{align}
         \mathcal{I}^{\text{D0-D2-D6}}_{\lambda,\lambda,\lambda;k}=(-1)^k\mathcal{I}^{\text{D0-D0}}_k\times\prod_{i=1}^k\frac{\sh(v-\e_{1,2,3}-\phi_i)\sh(\phi_i-v-\e_{123})^2}{\sh(\phi_i-v-\e_{12,13,23})\sh(v-\phi_i)^2}
    \end{align}
    After applying the JK-residue, there are three poles $\mathcal{X}(1,2,2)$, $\mathcal{X}(2,1,2)$, $\mathcal{X}(2,2,1)$, and their contributions are:
    \begin{align}
        \mathcal{Z}^{\text{D0-D2-D6}}_{\lambda,\lambda,\lambda}(1,2,2)=&-\frac{\sh(\e_1)\sh(\e_{12})\sh(\e_{13})\sh(2\e_2+\e_3)\sh(\e_2+2\e_3)}{\sh(\e_{2,3})\sh(\e_{23})\sh(\e_1-\e_2)\sh(\e_1-\e_3)}\cr\mathcal{Z}^{\text{D0-D2-D6}}_{\lambda,\lambda,\lambda}(2,1,2)=&\frac{\sh(\e_2)\sh(\e_{12})\sh(\e_{23})\sh(2\e_1+\e_3)\sh(\e_1+2\e_3)}{\sh(\e_{1,3})\sh(\e_{13})\sh(\e_1-\e_2)\sh(\e_2-\e_3)}\cr\mathcal{Z}^{\text{D0-D2-D6}}_{\lambda,\lambda,\lambda}(2,2,1)=&-\frac{\sh(\e_3)\sh(\e_{13})\sh(\e_{23})\sh(2\e_1+\e_2)\sh(\e_1+2\e_2)}{\sh(\e_{1,2})\sh(\e_{12})\sh(\e_1-\e_3)\sh(\e_2-\e_3)}
    \end{align}
Note that the three contributions are related by cyclic permutation of $(\e_1,\e_2,\e_3)$, as expected by the $S_3$ symmetry of the 3-leg vacuum.

\paragraph{2-leg case.} The computation proceeds analogously; only the intersection formula simplifies. The $\mathcal{J}$-factor is:
\begin{align}
    \mathcal{J}\big(x\big|\pi_{\lambda\mu\emptyset}\big)=&\frac{\mathcal{J}\big(x\big|\pi_{\lambda\emptyset\emptyset}\big)\mathcal{J}\big(x\big|\pi_{\emptyset\mu\emptyset}\big)\mathcal{J}\big(x\big|\pi_{\emptyset\emptyset\emptyset}\big)\mathcal{J}\big(x\big|\pi_{\lambda\emptyset\emptyset}\cap\pi_{\emptyset\mu\emptyset}\cap\pi_{\emptyset\emptyset\emptyset}\big)}{\mathcal{J}\big(x\big|\pi_{\lambda\emptyset\emptyset}\cap\pi_{\emptyset\mu\emptyset}\big)\mathcal{J}\big(x\big|\pi_{\lambda\emptyset\emptyset}\cap\pi_{\emptyset\emptyset\emptyset}\big)\mathcal{J}\big(x\big|\pi_{\emptyset\mu\emptyset}\cap\pi_{\emptyset\emptyset\emptyset}\big)}\cr=&\frac{\mathcal{J}\big(x\big|\pi_{\lambda\emptyset\emptyset}\big)\mathcal{J}\big(x\big|\pi_{\emptyset\mu\emptyset}\big)}{\mathcal{J}\big(x\big|\pi_{\lambda\emptyset\emptyset}\cap\pi_{\emptyset\mu\emptyset}\big)}
\end{align}

\subsection{D0-D8 partition function}\label{D0D8 appendix}

We compute the $k=1$ and $k=2$ contributions for a single D8-brane ($N=1$), demonstrating the $\mathcal{J}_{\geq}$-factor procedure~\eqref{Jgeq} and verifying the recursion relation~\eqref{4d recursion}. Recall that $\mathcal{J}_{\geq}(\mathcal{X}_{\mathcal{B}}(\boldsymbol{x})|\rho_{\mathcal{A}})$ selects only shell boxes with last coordinate $y_d \leq x_d$. The $k=2$ calculation additionally clarifies the sign discrepancy between the convention of \cite{Nekrasov:2018xsb} and ours: the two conventions agree for three of the four $k=2$ Young diagrams, and differ by a sign for the fourth ($\rho_4$) due to the replacement $\sh(\phi_i-\phi_j+\e_4)\to\sh(\phi_i-\phi_j-\e_4)$ in the D0-D0 sector. Throughout, we omit the label $(\4,1)$ for brevity.

\paragraph{$k=1$: $\mathcal{J}_{\geq}$ computation.} The 4d Young diagram $\{(1,1,1,1)\}$ has 15 shell boxes with charges listed in Tab.~\ref{tab:4d 1box charge}. 
   \begin{table}[ht]
        \centering
        \begin{tabular}{|c|cccc|}
           \hline  \multirow{2}{*}{$\operatorname{Q}=+1$} & $(1,1,2,2)$&$(1,2,1,2)$&$(1,2,2,1)$&$(2,1,1,2)$\\ &$(2,1,2,1)$&$(2,2,1,1)$&$(2,2,2,2)$&\\ \hline
           \multirow{2}{*}{$\operatorname{Q}=-1$} &$(1,1,1,2)$&$(1,1,2,1)$&$(1,2,1,1)$&$(1,2,2,2)$\\&$(2,1,1,1)$&$(2,1,2,2)$&$(2,2,1,2)$&$(2,2,2,1)$\\ \hline
       \end{tabular}
         \caption{The charges of the shellboxes of the 4d Young diagram $\{(1,1,1,1)\}$.}
       \label{tab:4d 1box charge}
    \end{table}
For the input $\mathcal{X}_{\4,1}(1,1,1,1)$, the $\mathcal{J}_{\geq}$ definition restricts to shell boxes with $x_4\leq 1$, reducing the 15 boxes to the 7 boxes in Tab.~\ref{tab:4d 1box charge geq}.
            \begin{table}[ht]
        \centering
        \begin{tabular}{|c|cccc|}
           \hline  $\operatorname{Q}=+1$ & $(1,2,2,1)$&$(2,1,2,1)$&$(2,2,1,1)$& \\ \hline
           $\operatorname{Q}=-1$ &$(1,1,2,1)$&$(1,2,1,1)$&$(2,1,1,1)$&$(2,2,2,1)$\\ \hline
       \end{tabular}
         \caption{The charges of the shellboxes of the 4d Young diagram $\{(1,1,1,1)\}$ when the input of $\mathcal{J}_{\geq}$ is $\mathcal{X}_{\4,\alpha}(x_1,x_2,x_3,1)$.}
       \label{tab:4d 1box charge geq}
    \end{table}
    
    Thus, we have:
    \begin{align}\label{4d YD 1box}
        \mathcal{Z}^{\text{D0-D8}}_{N=1,k=1}=\mathcal{Z}^{\text{D0-D8}}(\{(1,1,1,1)\})=&\mathcal{J}_\geq\big(\mathcal{X}_{\4,1}(1,1,1,1)\big|\{(1,1,1,1)\}_{\4,1}\big)\cr=&-\frac{\sh(\e_{12})\sh(\e_{13})\sh(\e_{23})}{\sh(\e_{1,2,3})\sh(\e_{123})}
    \end{align}

\paragraph{$k=2$: recursion relation~\eqref{4d recursion}.} The 4d Young diagram $\rho_{\4,1}=\{(1,1,1,1),(2,1,1,1)\}_{\4,1}$ has all boxes with fourth coordinate equal to $1$, so $\mathcal{J}_{\geq}$ restricts to shellboxes with $x_4\leq1$ as in Tab.~\ref{tab:4d 2box charge geq}.
     \begin{table}[ht]
        \centering
        \begin{tabular}{|c|cccc|}
           \hline  $\operatorname{Q}=+1$ & $(1,2,2,1)$&$(3,1,2,1)$&$(3,2,1,1)$& \\ \hline
           $\operatorname{Q}=-1$ &$(1,1,2,1)$&$(1,2,1,1)$&$(3,1,1,1)$&$(3,2,2,1)$\\ \hline
       \end{tabular}
         \caption{The charges of the shellboxes of the 4d Young diagram $\{(1,1,1,1),(2,1,1,1)\}$ when the input of $\mathcal{J}_{\geq}$ is $\mathcal{X}_{\4,\alpha}(x_1,x_2,x_3,1)$.}
       \label{tab:4d 2box charge geq}
    \end{table}
    
    Therefore, we have:
    \begin{align}\label{4d YD 2box 1}
        \mathcal{Z}^{\text{D0-D8}}(\{(1,1,1,1),(2,1,1,1)\}_{\4,1})=&-\frac{\sh(2\e_1+\e_2)\sh(2\e_1+\e_3)\sh(\e_{23})}{\sh(2\e_1)\sh(\e_2)\sh(\e_3)\sh(2\e_1+\e_{23})}\cr&\times\frac{\sh(\e_{12})\sh(\e_1-\e_{23})\sh(\e_{13})}{\sh(\e_1)\sh(\e_1-\e_2)\sh(\e_1-\e_3)\sh(\e_{123})}
    \end{align}
    For the other two Young diagrams $\{(1,1,1,1),(1,2,1,1)\}$ and $\{(1,1,1,1),(1,1,2,1)\}$, the corresponding contributions follow by symmetry in $\e_1$, $\e_2$, $\e_3$. However, for $\{(1,1,1,1),(1,1,1,2)\}$ the fourth coordinate of one box differs from the rest, so the full shellbox charges are listed in Tab.~\ref{tab:4d 2box charge}. Fortunately, for both inputs $\mathcal{X}_{\4,\alpha}(1,1,1,1)$ and $\mathcal{X}_{\4,\alpha}(1,1,1,2)$ the last coordinate is less than $3$, so the required $\mathcal{J}_{\geq}$ shellboxes are the same as Tab.~\ref{tab:4d 1box charge geq}, giving:
            \begin{table}[ht]
        \centering
        \begin{tabular}{|c|cccc|}
           \hline  \multirow{2}{*}{$\operatorname{Q}=+1$} & $(1,1,2,3)$&$(1,2,1,3)$&$(1,2,2,1)$&$(2,1,1,3)$\\ &$(2,1,2,1)$&$(2,2,1,1)$&$(2,2,2,3)$&\\ \hline
           \multirow{2}{*}{$\operatorname{Q}=-1$} &$(1,1,1,3)$&$(1,1,2,1)$&$(1,2,1,1)$&$(1,2,2,3)$\\&$(2,1,1,1)$&$(2,1,2,3)$&$(2,2,1,3)$&$(2,2,2,1)$\\ \hline
       \end{tabular}
         \caption{The charges of the shellboxes of the 4d Young diagram $\{(1,1,1,1),(1,1,1,2)\}$.}
       \label{tab:4d 2box charge}
    \end{table}

    Fortunately, for both inputs $\mathcal{X}_{\4,\alpha}(1,1,1,1)$ and $\mathcal{X}_{\4,\alpha}(1,1,1,2)$, the last coordinate of both boxes is less than $3$. Therefore, the shellboxes actually required are the same as those shown in Tab.~\ref{tab:4d 1box charge geq}. Thus, the contribution of this Young diagram is:
    \begin{align}\label{4d YD 2box 2}
        \mathcal{Z}^{\text{D0-D8}}(\{(1,1,1,1),(1,1,1,2)\}_{\4,1})=\frac{\sh(\e_{12})\sh(\e_{13})\sh(\e_{23})}{\sh(\e_{1,2,3})\sh(\e_{123})}\frac{\sh(\e_{12}-\e_4)\sh(\e_{13}-\e_4)\sh(\e_{23}-\e_4)}{\sh(\e_{1,2,3}-\e_4)\sh(\e_{123}-\e_4)}
    \end{align}
    Comparing the contributions of $\{(1,1,1,1)\}$~\eqref{4d YD 1box} and $\{(1,1,1,1),(2,1,1,1)\}$~\eqref{4d YD 2box 1} gives:
    \begin{align}
        \frac{\mathcal{Z}^{\text{D0-D8}}(\{(1,1,1,1),(2,1,1,1)\}_{\4,1})}{\mathcal{Z}^{\text{D0-D8}}(\{(1,1,1,1)\}_{\4,1})}=\frac{\sh(2\e_1+\e_2)\sh(\e_1-\e_2-\e_3)\sh(2\e_1+\e_3)}{\sh(2\e_1)\sh(\e_1-\e_2)\sh(\e_1-\e_3)\sh(2\e_1+\e_2+\e_3)}
    \end{align}
    According to~\eqref{4d recursion}, we obtain $\mathcal{J}_{\geq}\big(\mathcal{X}_{\4,1}(2,1,1,1)\big|\{(1,1,1,1),(2,1,1,1)\}_{\4,1}\big)$ from the charges in Tab.~\ref{tab:4d 2box charge geq}, and $\mathcal{J}_{<}\big(\mathcal{X}_{\4,1}(2,1,1,1)\big|\{(1,1,1,1)\}_{\4,1}\big)$ from the boxes with last coordinate greater than $1$ in Tab.~\ref{tab:4d 1box charge}:
    \begin{align}
        &\mathcal{J}_{\geq}\big(\mathcal{X}_{\4,1}(2,1,1,1)\big|\{(1,1,1,1),(2,1,1,1)\}_{\4,1}\big)=\frac{\sh(\e_{12})\sh(\e_1-\e_{23})\sh(\e_{13})}{\sh(\e_1)\sh(\e_1-\e_{2,3})\sh(\e_{123})}\cr&\mathcal{J}_{<}\big(\mathcal{X}_{\4,1}(2,1,1,1)\big|\{(1,1,1,1)\}_{\4,1}\big)=\frac{\sh(\e_1-\e_{24})\sh(\e_1-\e_{34})\sh(\e_4)\sh(\e_{234})}{\sh(\e_1-\e_4)\sh(\e_1-\e_{234})\sh(\e_{24})\sh(\e_{34})}
    \end{align}
    Multiplying the two expressions and applying the CY4 condition $\e_{1234}=0$ verifies the recursion relation~\eqref{4d recursion}:
    \begin{align}
        \frac{\mathcal{Z}^{\text{D0-D8}}(\{(1,1,1,1),(2,1,1,1)\}_{\4,1})}{\mathcal{Z}^{\text{D0-D8}}(\{(1,1,1,1)\}_{\4,1})}=&\mathcal{J}_{\geq}\big(\mathcal{X}_{\4,1}(2,1,1,1)\big|\{(1,1,1,1),(2,1,1,1)\}_{\4,1}\big)\cr&\times\mathcal{J}_{<}\big(\mathcal{X}_{\4,1}(2,1,1,1)\big|\{(1,1,1,1)\}_{\4,1}\big)
    \end{align}

\paragraph{$k=2$: sign rule.} The D0-D0 sector $\widetilde{\mathcal{I}}^{\text{D0-D0}}_k$ used in \cite{Nekrasov:2018xsb} differs from $\mathcal{I}^{\text{D0-D0}}_k$ in~\eqref{D0D8partition}: it replaces $\sh(\phi_i-\phi_j-\e_4)$ in the denominator by $\sh(\phi_i-\phi_j+\e_4)$:
    \begin{align}
        \widetilde{\mathcal{I}}^{\text{D0-D0}}_k=\frac{1}{k!}\prod_{i\neq j}^k\sh(\phi_i-\phi_j)\prod_{i,j}^k\frac{\sh(\phi_i-\phi_j-\e_{1,2,3}-\e_4)}{\sh(\phi_i-\phi_j-\e_{1,2,3})\sh(\phi_i-\phi_j+\e_{4})}
    \end{align}
    The corresponding full partition function is:
    \begin{align}
        \widetilde{\mathcal{I}}^{\text{D0-D8-}\overline{\text{D8}}}_{N,k}=\widetilde{\mathcal{I}}^{\text{D0-D0}}_k\times\prod_{i=1}^k\prod_{\alpha=1}^N\frac{\sh(\phi_i-w_{\4,\alpha})}{\sh(\phi_i-v_{\4,\alpha})}
    \end{align}
    For $N=1$, $k=2$ there are four poles, corresponding to:
    \begin{align}
        &\rho_1=\{(1,1,1,1),(2,1,1,1)\},\quad\rho_2=\{(1,1,1,1),(1,2,1,1)\},\cr&\rho_3=\{(1,1,1,1),(1,1,2,1)\},\quad\rho_4=\{(1,1,1,1),(1,1,1,2)\}
    \end{align}
    Only $\rho_4$ requires an additional minus sign relative to our convention, because according to the sign rule:
    \begin{align}
        \Res_{\mathcal{X}(\boldsymbol{x}\in\rho)}\left(\prod_{i=1}^k\frac{d\phi_i}{2\pi i}\right)\widetilde{\mathcal{I}}^{\text{D0-D8-}\overline{\text{D8}}}_{N,k}=(-1)^{h(\rho)}\Res_{\mathcal{X}(\boldsymbol{x}\in\rho)}\left(\prod_{i=1}^k\frac{d\phi_i}{2\pi i}\right)\mathcal{I}^{\text{D0-D8-}\overline{\text{D8}}}_{N,k}
    \end{align}
    where:
    \begin{align}
        h(\rho)=1+|\rho|+\# \{(a,d)\big|(a,a,a,d)\in\rho\text{ and } a\leq d\}
    \end{align}
    One finds $h(\rho_1)=h(\rho_2)=h(\rho_3)=4$ (even, no sign change) and $h(\rho_4)=5$ (odd, sign change). The sign difference arises because the two D0-D0 sectors differ in a specific factor: $\widetilde{\mathcal{I}}$ contains $\sh(\phi_1-\phi_2+\e_4)$ in the denominator where $\mathcal{I}$ has $\sh(\phi_1-\phi_2-\e_4)$. For the pole $(\phi_1,\phi_2)=(\mathcal{X}(1,1,1,1),\mathcal{X}(1,1,1,2))$ this matters:
    \begin{align}
        \widetilde{\mathcal{I}}^{\text{D0-D8-}\overline{\text{D8}}}_{1,2}\supset\widetilde{\mathcal{I}}=\frac{1}{\sh(\phi_{1,2}-v)\sh(\phi_1-\phi_2+\e_4)\sh(\phi_2-\phi_1+\e_4)}\cr\mathcal{I}^{\text{D0-D8-}\overline{\text{D8}}}_{1,2}\supset\mathcal{I}=\frac{1}{\sh(\phi_{1,2}-v)\sh(\phi_1-\phi_2-\e_4)\sh(\phi_2-\phi_1-\e_4)}
    \end{align}
    At the pole $\rho_4$:
    \begin{align}
        &\Res_{\mathcal{X}(\boldsymbol{x}\in\rho_4)}\widetilde{\mathcal{I}}=\frac{1}{\sh(\e_4)\sh(2\e_4)}\cr&\Res_{\mathcal{X}(\boldsymbol{x}\in\rho_4)}\mathcal{I}=\frac{1}{\sh(\e_4)\sh(-2\e_4)}=-\Res_{\mathcal{X}(\boldsymbol{x}\in\rho_4)}\widetilde{\mathcal{I}}
    \end{align}
    Since we adopt the convention~\eqref{D0D8partition}, the shell formula is formulated accordingly, and no extra sign factor needs to be tracked in our calculation.

\subsection{DT4 counting}\label{appendix DT4}

Three examples illustrate $\mathcal{J}$-factor computations in DT4 theory: (i) the four-leg case with all legs carrying a single-box 3d Young diagram, demonstrating the cutoff method and showing why charges $Q=+2$ and $Q=-3$ appear when four infinite legs meet at a shared origin; (ii) a mixed one-leg and two-surface case; and (iii) the derivation of the general DT4 integrand~\eqref{DT4 integrand} from the recursion relation~\eqref{4d recursion}, paralleling the DT3 derivation of Appendix~\ref{appendix DT}.

\paragraph{Four-leg case.} Consider the minimal 4d Young diagram $\rho_{\pi_1\pi_2\pi_3\pi_4}$ with all four legs being the single-box 3d Young diagram $\pi=\cube=\{(1,1,1)\}$:
    \begin{align}
        \rho_{\pi_1\pi_2\pi_3\pi_4}=\{(i,1,1,1)\}_{i=1}^\infty\cup\{(1,i,1,1)\}_{i=1}^\infty\cup\{(1,1,i,1)\}_{i=1}^\infty\cup\{(1,1,1,i)\}_{i=1}^\infty
    \end{align}
    In this case, it is sufficient to take the cutoff at $m>2$; we may simply set $m=3$. The resulting truncated minimal 4d Young diagram $\rho_{\pi_1\pi_2\pi_3\pi_4;3}$ is then given by:
    \begin{align}
        \rho_{\pi_1\pi_2\pi_3\pi_4;3}=\{&(1,1,1,1),(2,1,1,1),(3,1,1,1),(1,2,1,1),(1,3,1,1),\cr&(1,1,2,1),(1,1,3,1),(1,1,1,2),(1,1,1,3)\}
    \end{align}
 \begin{table}[ht]
    \centering
    \begin{tabular}{|c|cccccc|}
        \hline
        \multirow{3}{*}{$\operatorname{Q}=+1$} & $(1,1,2,4)$ & $(1,1,4,2)$ & $(1,2,1,4)$ & $(1,2,4,1)$ 
        & $(1,4,1,2)$ & $(1,4,2,1)$ \\ & $(2,1,1,4)$ & $(2,1,4,1)$ 
        & $(2,2,2,4)$ & $(2,2,4,2)$ & $(2,4,1,1)$ & $(2,4,2,2)$ \\
        & $(4,1,1,2)$ & $(4,1,2,1)$ & $(4,2,1,1)$ & $(4,2,2,2)$  & & \\
        \hline
        \multirow{4}{*}{$\operatorname{Q}=-1$} & $(1,1,1,4)$ & $(1,1,2,2)$ & $(1,1,4,1)$ & $(1,2,1,2)$
        & $(1,2,2,1)$ & $(1,2,2,4)$ \\ & $(1,2,4,2)$ & $(1,4,1,1)$ 
        & $(1,4,2,2)$ & $(2,1,1,2)$ & $(2,1,2,1)$ & $(2,1,2,4)$ \\
        & $(2,1,4,2)$ & $(2,2,1,1)$ & $(2,2,1,4)$ & $(2,2,4,1)$ 
        & $(2,4,1,2)$ & $(2,4,2,1)$ \\ & $(4,1,1,1)$ & $(4,1,2,2)$ 
        & $(4,2,1,2)$ & $(4,2,2,1)$ & & \\
        \hline
        \multirow{1}{*}{$\operatorname{Q}=+2$} & $(1,2,2,2)$ & $(2,1,2,2)$ & $(2,2,1,2)$ & $(2,2,2,1)$ & & \\
        \hline
        \multirow{1}{*}{$\operatorname{Q}=-3$} & $(2,2,2,2)$ & & & & &\\
        \hline
    \end{tabular}
    \caption{The charges of the shellboxes of the minimal 4d Young diagram $\rho_{\pi_1\pi_2\pi_3\pi_4;3}$.}
    \label{tab:4leg charges}
\end{table}
Then, using the coordinates and charges listed in Tab.~\ref{tab:4leg charges}, and discarding terms that would yield $\sh(x-v-m\,\epsilon_a+\ldots)$—namely, the shellboxes whose coordinates contain $m+1=4$—we obtain the $\mathcal{J}$-factor. Note that the $Q=+2$ and $Q=-3$ entries in Tab.~\ref{tab:4leg charges} arise because the four infinite legs all share the origin box $(1,1,1,1)$, causing multiple contributions to accumulate at a single shell box:
\begin{align}
    \mathcal{J}\big(x\big|\rho_{\pi_1\pi_2\pi_3\pi_4}\big)=\frac{\prod_{A\in\check{\4}}\sh(x-v-\e_{A})^2}{\sh(x-v-\e_{1234})^3\prod_{ab\in\6}\sh(x-v-\e_{ab})}
\end{align}
Performing the sign reversal $\sh(x) \to -\sh(-x-\e_{1234})$ on all terms except the addable boxes
\begin{align*}
    \mathfrak{A}(\rho_{\pi_1\pi_2\pi_3\pi_4})=\{(1,1,2,2),(1,2,1,2),(1,2,2,1),(2,1,1,2),(2,1,2,1),(2,2,1,1)\},
\end{align*}
gives the final $\mathcal{J}_{-\mathfrak{A}}$-factor:
\begin{align}
    \mathcal{J}_{-\mathfrak{A}}\big(x\big|\rho_{\pi_1\pi_2\pi_3\pi_4}\big)=-\frac{\prod_{a\in\4}\sh(v-x-\e_a)^2}{\sh(v-x)^3\prod_{ab\in\6}\sh(x-v-\e_{ab})}
\end{align}
The partition function of DT4 counting with minimal 4d Young diagram $\rho_{\pi_1\pi_2\pi_3\pi_4}$ is therefore:
\begin{align}
    \mathcal{I}^{\text{D0-D2-D8}}_{\pi_1\pi_2\pi_3\pi_4;k}=-\mathcal{I}^{\text{D0-D0}}_k\times\prod_{i=1}^k\frac{\prod_{a\in\4}\sh(v-\phi_i-\e_a)^2}{\sh(v-\phi_i)^3\prod_{ab\in\6}\sh(\phi_i-v-\e_{ab})}
\end{align}

\paragraph{One-leg and two-surface case.} We compute the minimal Young diagram $\rho_{\pi_1\lambda_{12}\lambda_{13}}$ with $\pi_1=\{(1,1,1)\}$, $\lambda_{12}=\{(1,1)\}$, and $\lambda_{13}=\{(1,1),(1,2)\}$. This mixed boundary configuration illustrates that the cutoff method works equally well when both leg and surface boundaries are present:
\begin{align}
    \rho_{\pi_1\lambda_{12}\lambda_{13}}=\{(i,1,1,1)\}_{i=1}^\infty\cup\{(i,j,1,1)\}_{i,j=1}^\infty\cup\{(i,1,j,1),(i,1,j,2)\}_{i,j=1}^\infty
\end{align}
We truncate at $m=4$ and discard all shellboxes whose coordinates contain $m+1=5$. The remaining shellboxes are listed in Tab.~\ref{tab:1leg 2surface}, giving:
 \begin{table}[ht]
        \centering
        \begin{tabular}{|c|cccc|}
           \hline  $\operatorname{Q}=+1$ & $(1,2,1,2)$&$(1,1,1,3)$&$(1,2,2,1)$& \\ \hline
           $\operatorname{Q}=-1$ &$(1,2,2,2)$&$(1,2,1,3)$& & \\ \hline
       \end{tabular}
         \caption{The charges of the shellboxes of the 4d Young diagram $\rho_{\pi_1\lambda_{12}\lambda_{13}}$ with $\pi_1=\{(1,1,1)\}$, $\lambda_{12}=\{(1,1)\}$, and $\lambda_{13}=\{(1,1),(1,2)\}$.}
       \label{tab:1leg 2surface}
    \end{table}
\begin{align}
    \mathcal{J}\big(x\big|\rho_{\pi_1\lambda_{12}\lambda_{13}}\big)=\frac{\sh(x-v-\e_2-2\e_4)\sh(x-v-\e_{234})}{\sh(x-v-\e_{23})\sh(x-v-2\e_{4})\sh(x-v-\e_{24})}
\end{align}
With the addable boxes $\mathfrak{A}(\rho_{\pi_1\lambda_{12}\lambda_{13}})=\{(1,2,1,2),(1,1,1,3),(1,2,2,1)\}$, the $\mathcal{J}_{-\mathfrak{A}}$-factor is:
\begin{align}
    \mathcal{J}_{-\mathfrak{A}}\big(x\big|\rho_{\pi_1\lambda_{12}\lambda_{13}}\big)=\frac{\sh(v-x+\e_4-\e_{13})\sh(v-x-\e_1)}{\sh(x-v-\e_{23})\sh(x-v-2\e_{4})\sh(x-v-\e_{24})}
\end{align}
The partition function of DT4 counting with minimal 4d Young diagram $\rho_{\pi_1\lambda_{12}\lambda_{13}}$ is therefore:
\begin{align}
    \mathcal{I}^{\text{D0-D2-D4-D8}}_{\pi_1\lambda_{12}\lambda_{13};k}=\mathcal{I}^{\text{D0-D0}}_k\times\prod_{i=1}^k\frac{\sh(v-\phi_i+\e_4-\e_{13})\sh(v-\phi_i-\e_1)}{\sh(\phi_i-v-\e_{23})\sh(\phi_i-v-2\e_{4})\sh(\phi_i-v-\e_{24})}
\end{align}

\paragraph{Derivation of the DT4 integrand~\eqref{DT4 integrand}.} Analogous to the DT3 derivation of Appendix~\ref{appendix DT}, we consider the DT4 invariant for a lattice set $\widetilde{\rho}_{\{\pi_a\},\{\lambda_{ab}\}}=\{\boldsymbol{x}_1,\dots,\boldsymbol{x}_k\}$ of $k$ boxes, arranged so that $\rho_{\{\pi_a\},\{\lambda_{ab}\}}\cup \{\boldsymbol{x}_1\}$, $\rho_{\{\pi_a\},\{\lambda_{ab}\}} \cup \{\boldsymbol{x}_1\} \cup \{\boldsymbol{x}_2\}$, $\dots$ are all valid 4d Young diagrams. Iterating the D0-D8 recursion~\eqref{4d recursion} and applying the identity~\eqref{J geq recursion} at each step gives:
    \begin{align}
    \mathcal{Z}_{\{\pi_a\},\{\lambda_{ab}\}}&(\boldsymbol{x}_1,\dots,\boldsymbol{x}_k)\cr=&\mathcal{J}_{\geq}\big(\mathcal{X}(\boldsymbol{x}_1)\big|\rho_{\{\pi_a\},\{\lambda_{ab}\}}\cup\{\boldsymbol{x}_1\}\big)\mathcal{J}_{<}\big(\mathcal{X}(\boldsymbol{x}_1)\big|\rho_{\{\pi_a\},\{\lambda_{ab}\}}\big)\cr&\times\mathcal{J}_{\geq}\big(\mathcal{X}(\boldsymbol{x}_2)\big|\rho_{\{\pi_a\},\{\lambda_{ab}\}}\cup\{\boldsymbol{x}_1\}\cup\{\boldsymbol{x}_2\}\big)\mathcal{J}_{<}\big(\mathcal{X}(\boldsymbol{x}_2)\big|\rho_{\{\pi_a\},\{\lambda_{ab}\}}\cup\{\boldsymbol{x}_1\}\big)\cr&\times\ldots\cr=&\left(\frac{\sh(-\e_{14,24,34})\sh(0)}{\sh(-\e_{1,2,3,4})}\mathcal{J}_{\geq}\big(\mathcal{X}(\boldsymbol{x}_1)\big|\rho_{\{\pi_a\},\{\lambda_{ab}\}}\big)\right)\mathcal{J}_{<}\big(\mathcal{X}(\boldsymbol{x}_1)\big|\rho_{\{\pi_a\},\{\lambda_{ab}\}}\big)\cr&\times\left(\frac{\sh(-\e_{14,24,34})\sh(0)}{\sh(-\e_{1,2,3,4})}\mathcal{J}_{\geq}\big(\mathcal{X}(\boldsymbol{x}_2)\big|\rho_{\{\pi_a\},\{\lambda_{ab}\}}\cup\{\boldsymbol{x}_1\}\big)\right)\mathcal{J}_{<}\big(\mathcal{X}(\boldsymbol{x}_2)\big|\rho_{\{\pi_a\},\{\lambda_{ab}\}}\cup\{\boldsymbol{x}_1\}\big)\cr&\times\ldots\cr=&\left(\frac{\sh(-\e_{14,24,34})\sh(0)}{\sh(-\e_{1,2,3,4})}\right)^k\mathcal{J}\big(\mathcal{X}(\boldsymbol{x}_1)\big|\rho_{\{\pi_a\},\{\lambda_{ab}\}}\big)\mathcal{J}\big(\mathcal{X}(\boldsymbol{x}_2)\big|\rho_{\{\pi_a\},\{\lambda_{ab}\}}\cup\{\boldsymbol{x}_1\}\big)\times\ldots\cr=&\left(\prod_{i,j=1}^k\frac{\sh(\mathcal{X}(\boldsymbol{x}_i)-\mathcal{X}(\boldsymbol{x}_j))\sh(\mathcal{X}(\boldsymbol{x}_i)-\mathcal{X}(\boldsymbol{x}_j)-\e_{14,24,34})}{\sh(\mathcal{X}(\boldsymbol{x}_i)-\mathcal{X}(\boldsymbol{x}_j)-\e_{1,2,3,4})}\right)\prod_{i=1}^k\mathcal{J}\big(\mathcal{X}(\boldsymbol{x}_i)\big|\rho_{\{\pi_a\},\{\lambda_{ab}\}}\big)\cr
\end{align}
where in the second equality, we use the relation~\eqref{J geq recursion}. The DT4 partition function thus factorizes into a pairwise interaction kernel and a product of vacuum $\mathcal{J}$-factors, in precise parallel with the DT3 result of Appendix~\ref{appendix DT}. Substituting $\phi_i$ for $\mathcal{X}(\boldsymbol{x}_i)$ and applying the appropriate sign reversals yields the DT4 integrand~\eqref{DT4 integrand}.

\bibliographystyle{JHEP}
\bibliography{references}

\end{document}